\documentclass[twocolumn,astrosymb]{aastex631}

\def\Sp{\textit{Spitzer}}
\global\def\tablenotemark#1{{\normalfont\textsuperscript{\currtabletypesize\it #1}}}
\received{November 8, 2024}
\revised{February 28, 2025}

\accepted{March 5, 2025 -- AJ}

\begin{document}

\title{Infrared Fluxes and Light Curves of Near-Earth Objects: The full \Sp\ Sample
}

\author[0000-0002-5599-4650]{Joseph L. Hora}
\affiliation{Center for Astrophysics $|$ Harvard \& Smithsonian, 60 Garden Street, MS-65, Cambridge, MA 02138-1516; USA}
\correspondingauthor{Joseph L. Hora}
\email{jhora@cfa.harvard.edu}

\author[0000-0002-6429-6619]{Alicia J. Allen}
\affiliation{Department of Astronomy and Planetary Science, Northern Arizona University, Flagstaff, AZ 86011, USA}

\author[0000-0003-4580-3790]{David E. Trilling}
\affiliation{Department of Astronomy and Planetary Science, Northern Arizona University, Flagstaff, AZ 86011, USA}

\author{Howard A. Smith}
\affiliation{Center for Astrophysics $|$ Harvard \& Smithsonian,
60 Garden Street, MS-65, Cambridge, MA 02138-1516; USA}
\author[0009-0005-9955-1500]{Andrew McNeill}
\affiliation{Department of Physics and Astronomy,
Bowling Green State University, 806 Ridge St, Bowling Green, OH 43403; USA}


\begin{abstract}
The IRAC camera on the Spitzer Space Telescope observed 2175 Near Earth Objects (NEOs) during its Warm Mission phase, primarily in three large surveys, and also in a small number of a dedicated projects. In this paper we present the final reprocessing of the NEO
data and determine fluxes at 3.6~\micron\ (where available) and 4.5~\micron. The observing
windows range from minutes to nearly ten hours, which means that for 39 NEOs we observe a complete lightcurve, and for these objects we present period and amplitude estimates and derive minimum cohesive strengths for the objects with well-determined periods. For an additional 128 objects we detect a significant
fraction of a complete lightcurve, and present estimated lower limits to their rotation periods. This paper presents the final and definitive Spitzer/IRAC NEO flux catalog.

\end{abstract}



\section{Introduction} \label{sec:intro}
Near Earth Objects (NEOs) are small solar system
bodies whose orbits bring them close to the Earth’s
orbit. NEOs can be used as compositional and dynamical tracers
to allow us to probe environmental conditions throughout our planetary
system and explore its history. They also provide a template for analyzing the evolution of planetary disks around other stars. NEOs are
the parent bodies of meteorites, one of our key sources of
detailed knowledge about the development of the solar
system, and so studies of NEOs are essential for understanding the origins and evolution of our planetary system and others.

The IRAC instrument \citep{2004Fazio} on the \Sp\
Space Telescope \citep{2004Werner} is a powerful NEO
characterization system. NEOs typically have daytime
temperatures $\sim$250~K, hence their thermal emission at
4.5~\micron\ is almost always significantly larger than their
reflected light at that wavelength. We can therefore employ 
a thermal model using the IR fluxes together with the optical flux estimated from the absolute magnitude $H$ to  derive NEO properties, including diameters and albedos. \citet{2010Trilling} demonstrated that this could be done reliably using only the 3.6 and 4.5~\micron\ IRAC bands available during the Spitzer Warm Mission \citep{2007Stauffer,2007Lisse}. Measuring the size distribution, albedos, and compositions for a large fraction of all known NEOs will allow us to understand the scientific, exploration, and civil-defense-related properties
of the NEO population.

After an initial pilot study to verify our observing techniques and analysis methods with the \Sp\ data \citep{Trilling08}, our team conducted
three major surveys of NEOs with \Sp/IRAC in the
Warm/Beyond Mission phases: the ExploreNEOs program \citep{2008Trillingprop}, the NEO Survey \citep{2014Trillingprop}, and the NEO Legacy Survey \citep{2018Trillingprop}.  Our
initial NEO survey results are summarized in \citet{2010Trilling,2016Trilling}.

\Sp\ completed a total of 2432 observations of 2175 unique NEOs with IRAC before the end of the mission in 2020, according to the \Sp\ solar system observation log\footnote{\url{https://irsa.ipac.caltech.edu/data/SPITZER/docs/spitzermission/observingprograms/solarsystemprograms/}}. In addition to the major surveys that observed a large number of NEOs referenced above, several targeted studies were performed during the \Sp\ Warm Mission. For example, NEOs suspected to be dormant comets were surveyed for activity \citep{2015Mommert,2016Mommert}. Observations were also made of small ($\sim$10\,m diameter) NEOs that are potential spacecraft capture targets \citep{2014mommert2009BD,2014mommert2011MD}. Several NEOs that were potential targets for a sample return mission were observed \citep{2010emery70163,2011sptz.prop80232E}. An investigation was performed of of Q-type NEOs to measure their thermal inertia to understand the possible regolith-sorting effects caused by interactions with terrestrial planets \citep{2014sptz.prop11145M}.

In another program, the Hayabusa-2 mission target 162173 Ryugu  was the target of an extensive photometric observation program \citep[Program ID\#90145;][]{muller2012,2017Mueller}. The observations include ten “point-and-shoot” measurements consisting of short standard IRAC measurements that were spaced by several days up to a few weeks, and two complete lightcurves, each using IRAC 3.6 and 4.5~\micron\ channels. The point-and-shoot observations were taken over an approximately 4 month period to cover a wide range of phase angles. Another part of the observations consisted of repeated integrations during its full period ($\sim$ 8~hr) to obtain an IR lightcurve to help to constrain the object’s shape and size. The success of these observations led us to conclude that we could extract similar lightcurves
for objects in the survey programs, which were designed
only to obtain a single flux measurement from the mosaic image averaging over all of the exposures in the observation. We found that our predicted NEO fluxes
were fairly conservative in many cases, and that we could
detect most of the NEOs in the individual IRAC exposures. We presented some initial \Sp\ lightcurve results in \citet{2018Hora}.

The Wide-field Infrared Survey Explorer (WISE;
Wright et al. 2010) has similarly used infrared observations to characterize a large sample of main-belt asteroids and NEOs. This Explorer-class mission obtained
images in four broad infrared bands at 3.4, 4.6, 12 and
22~\micron. WISE conducted its 4-band survey of the sky
starting in 2010 January, and after the cryogen was depleted later that year, it continued to operate with its
3.4 and 4.6 µm bands until 2011 February. The spacecraft was reactivated in 2013 December as NEOWISE
\citep{NEOWISE} and conducted
a sky survey in the 3.4 and 4.6 µm bands to focus on
NEO discovery and characterization. As of 2021, a total of 1845 unique NEOs have been characterized from the beginning of the cryogenic mission through year 7 of NEOWISE \citep{2021Masiero}. The WISE satellite was decommissioned in August 2024.  The WISE data can also be used to derive 
lightcurves of asteroids \citep[e.g.,][]{2015ApJ...799..191S}. However, the cadence of these observations were quite different; the WISE survey
typically provided repeated observations separated by 3~hr over a 1.5~day period, making it useful for sampling
periodicities on the order of 1 – 2 days. The \Sp\ data samples cadences from a few minutes to hours, making it ideal for small and fast-rotating NEOs, and complementary to the data that WISE provided. Also, since \Sp\ has a larger primary mirror and the observatory can track the apparent motion of the NEO, it can integrate longer  on each NEO and therefore can detect objects at the level of a few $\mu$Jy, enabling smaller and/or more distant objects to be targeted.

In this paper we present the results of an analysis of the full  sample of available \Sp\ lightcurve data. In \S\ref{sec:obs} we describe the observations and the reduction techniques using the NEOphot software. In \S\ref{sec:results} we describe the analysis used to derive periods and amplitudes of the lightcurves and presents those results. In \S\ref{sec:strengths} we estimate the cohesive strengths of a subset of the NEOs for which we have period measurements.

\section{Observations and Data Reduction}
\label{sec:obs}
\subsection{\Sp/IRAC Observations}
Observations were obtained with \Sp/IRAC during the Warm Mission \citep{2010Carey} in the ExploreNEOs program \citep[\Sp\ Program IDs 60012, 61010, 61011, 61012, 61013;][]{2008Trillingprop}, the NEO Survey \citep[Program ID 11002;]{2014Trillingprop}, the NEO Legacy Survey \citep[Program ID 13006;][]{2016Trillingprop}, and the Physical Characterization of NEOs program in the final \Sp\ cycle \citep[Program ID 14004;][]{2018Trillingprop}. 

The observations were conducted in a similar manner for these three large survey programs, taking frames while tracking the NEO motion and dithering during the observations to eliminate instrument systematics such as bad pixels or array location-dependent scattered light effects. Note that the 3.6 and 4.5~\micron\ fields of view do not overlap in the \Sp\ focal plane, so the NEO could not be observed simultaneously in both bands \citep[see Figure 2 of][]{2004Werner}.  In ExploreNEOs, we used the ``Moving Cluster'' target mode with custom offsets to perform the dithers, alternating between the 3.6 and 4.5~\micron\ fields of view. For the other programs, we used the ``Moving Single'' target mode and used a large cycling dither pattern with the source in the 4.5~\micron\ field of view only.  The observations and initial results from the large NEO surveys are described more fully in \citet{2010Trilling} and \citet{2016Trilling} and references therein. A first look at the lightcurves that could be extracted from the \Sp\ data was presented by \citet{2018Hora}, who provide additional details on the observation planning and design in the larger programs.

In this paper, we also include observations of NEOs performed in several other small \Sp\ programs which are listed in Table~\ref{tab:programs}. The \Sp\ Science Center pipeline-processed data for all of these programs can be retrieved from the \Sp\ Heritage Archive\footnote{\url{https://irsa.ipac.caltech.edu/applications/Spitzer/SHA/}} \citep{SHA}. The full NEO dataset was downloaded from the SHA in 2022 February (two years after the end of the \Sp\ mission), having been reprocessed by pipeline version S19.2.0.
In general, each NEO was observed using one Astronomical Observing Request (AOR), although some objects were observed at two or more different dates in separate AORs.

\begin{deluxetable}{ccccccc}
\caption{\Sp\ NEO programs}\label{tab:programs}
\tablehead{Program ID & Reference}
\startdata
60012, 61010, 61011,\\
61012, 61013 & \citet{2008Trillingprop}\\
11002 & \citet{2014Trillingprop}\\
13006 & \citet{2016Trillingprop}\\
14004 & \citet{2018Trillingprop}\\
11097 & \citet{2014sptz.prop11097R}\\
10109 & \citet{2013sptz.prop10109M}\\
10132 &  \citet{2013sptz.prop10132T}\\
11145 & \citet{2014sptz.prop11145M}\\
12043 &  \citet{2015sptz.prop12043M} \\
13102 & \citet{2016sptz.prop13102M}\\
13164 & \citet{2017sptz.prop13164M}\\
14025 &  \citet{2018sptz.prop14025M}\\
70054 & \citet{2010sptz.prop70054M}\\
70163 & \citet{2010emery70163}\\
80084 & \citet{2011sptz.prop80084M}\\
80232 & \citet{2011sptz.prop80232E}\\
90145 & \citet{2012sptz.prop90145M}\\
90256 &\citet{2013sptz.prop90256T}\\ 
\enddata
\end{deluxetable}

\subsection{IRAC Data Reduction}
\begin{figure}
    \centering
    \includegraphics[width=0.99\linewidth]{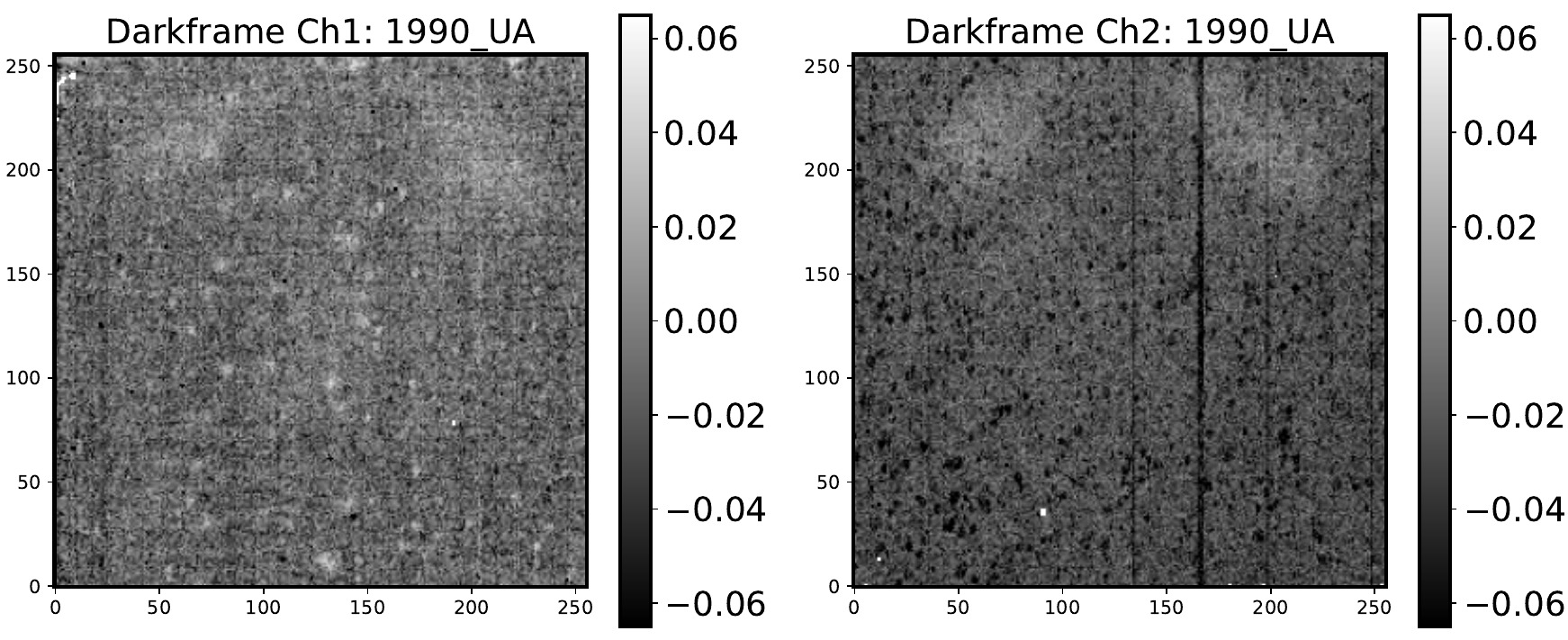}
    \caption{The 3.6~\micron\ (left) and 4.5~\micron\ (right) residual dark frames calculated from the 1990~UA observations.}
    \label{fig:darks}
\end{figure}

\begin{figure*}
    \centering
    \includegraphics[width=0.93\linewidth]{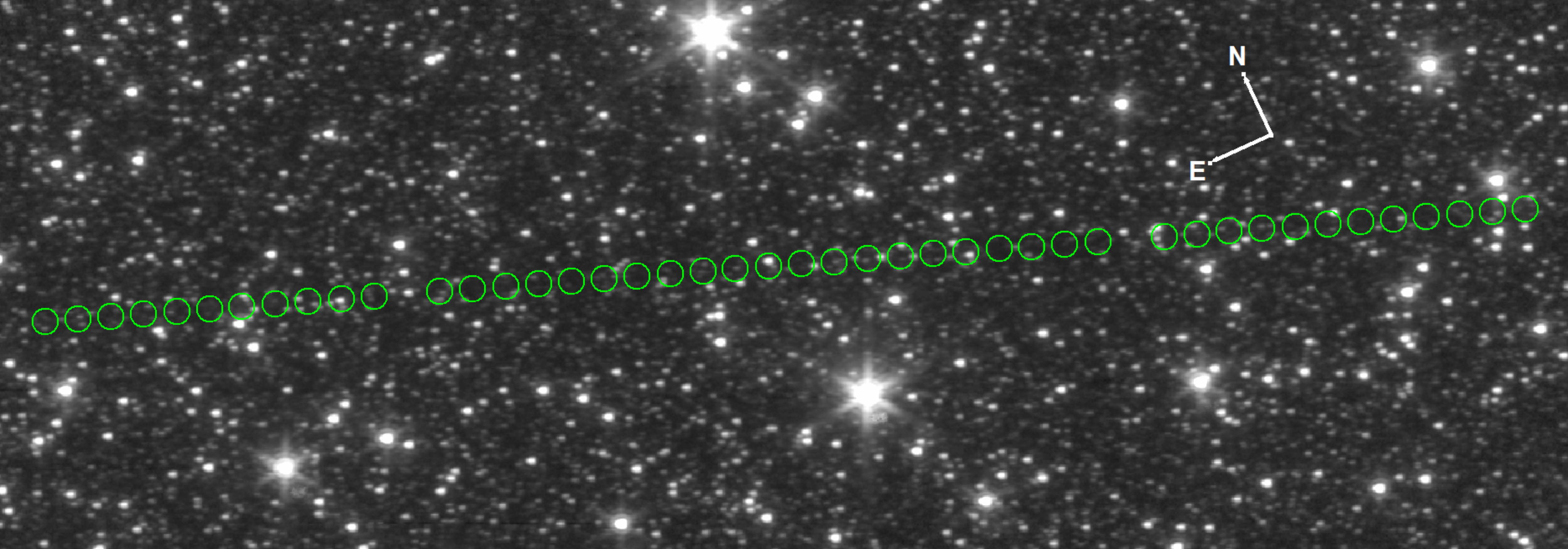}
    \caption{Mosaic of the 4.5~\micron\ sky background from the example 1990~UA observations. The stars appear slightly trailed because \Sp\ is tracking on the NEO during these observations. The green circles indicate the average position of the NEO during various individual IRAC frames, with the NEO moving from right to left starting with frame 277 (see Figure~\ref{fig:patches}). The missing positions in the sequence are frames that were rejected due to bright stars, cosmic rays, or the NEO landing on bad pixels on the array. The image is approximately 15\farcm5 wide, and centered near R.A., Decl. of 291\fdg223, 10\fdg649 (J2000), with the orientation shown in the image.}
    \label{fig:skymos}
\end{figure*}

\begin{figure*}
    \centering
    \vskip -10pt
    \includegraphics[width=0.63\linewidth]{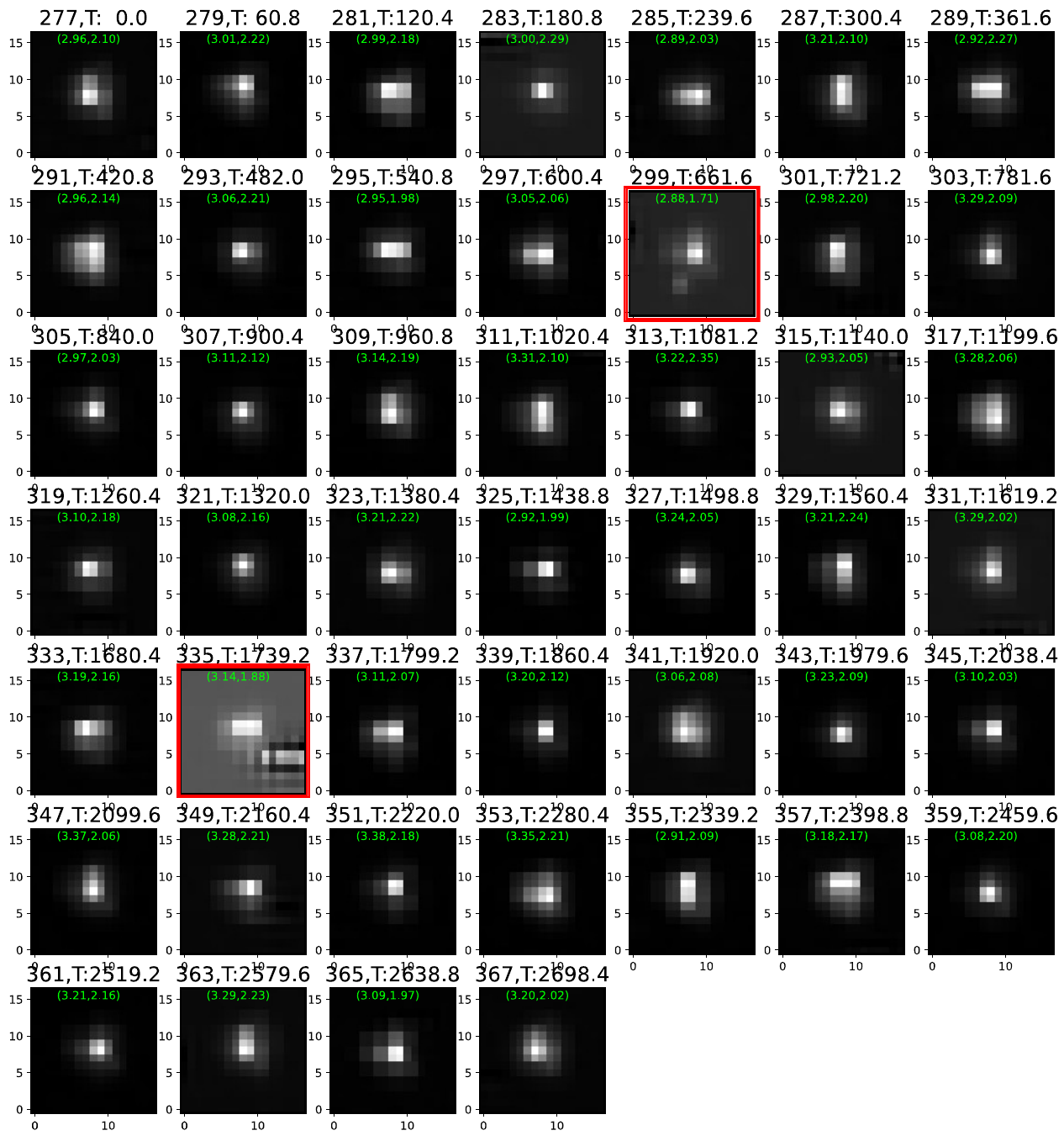}
    \caption{IRAC 4.5~\micron\ image patches from individual sky-subtracted cBCD frames displayed by NEOphot. The black labels above each frame show the frame number and relative time in seconds from the first frame in this series. The green label underneath show the relative offset in pixels of the source centroid relative to the expected NEO position.The images outlined in red are the frames that were rejected due to artifacts. An example of an artifact is visible in frame 335, where the NEO is passing near a residual from a bright star subtraction.}
    \label{fig:patches}
\end{figure*}

\begin{figure*}
    \centering
    \includegraphics[width=0.99\linewidth]{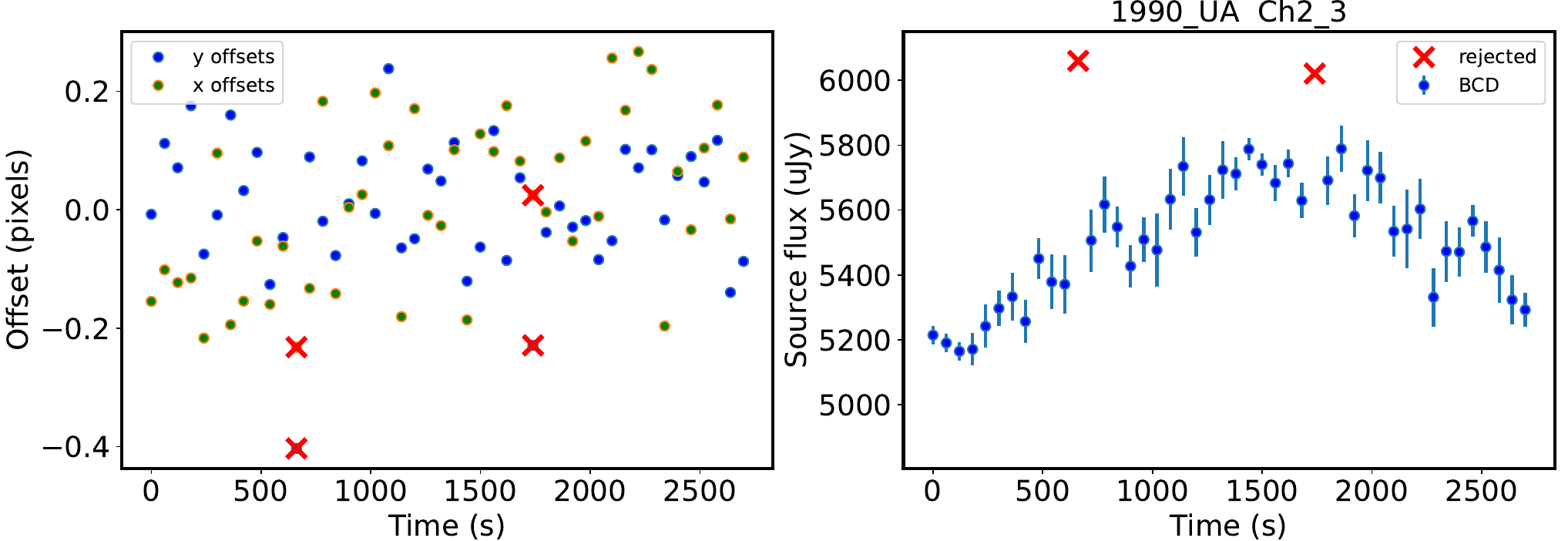}
    \caption{Interactive plots from the NEOphot program of the x and y offsets of the NEO relative to the nominal position (left) and the NEO photometry (right) for a portion of the 1990~UA dataset. The time is relative to the first frame in the dataset being examined. The results for two frames have been rejected from the analysis, marked by red 
    X's. The same rejected points are indicated in both graphs, showing that the frames were outliers in terms of the flux measured and their position in the frame.}
    \label{fig:1Dplots}
\end{figure*}

\begin{figure}
    \centering
    \includegraphics[width=0.99\linewidth]{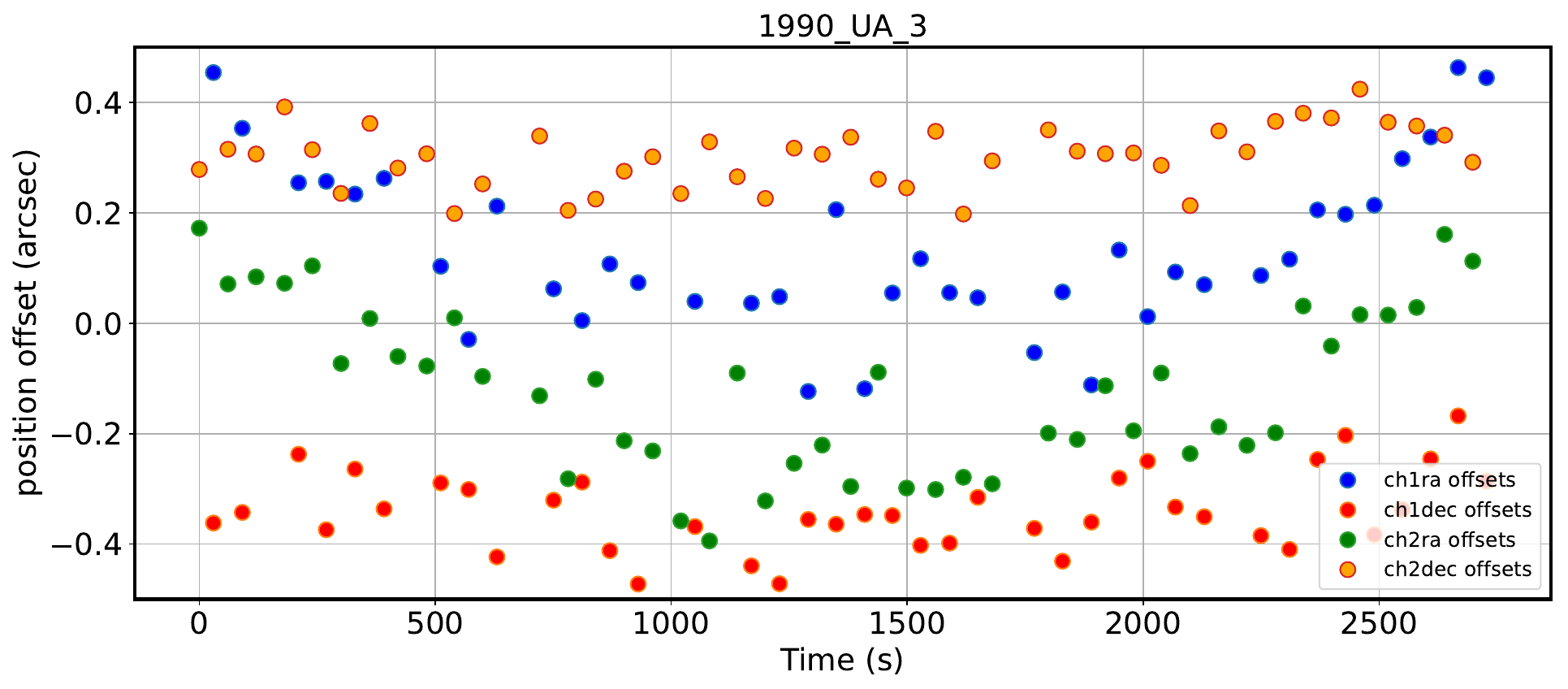}
    \caption{Plot showing the Right Ascension (R.A.) and Declination (Decl.) offsets in arcsec relative to the expected position of the NEO for both the 3.6 and 4.5~\micron\ images, which were taken in alternating on-source and sky frames. The Decl. offset is relatively stable in both channels, whereas the R.A. offset in both channels changes by about 0\farcs4 during the observation, with a minimum near the halfway point. This behavior may be due to the change of apparent motion of the NEO from \Sp's viewpoint, where the R.A. non-sidereal rate changed by $\sim$0\farcs4/minute during the observation, whereas the Decl. rate was relatively unchanged.}
    \label{fig:poserr}
\end{figure}

\begin{figure}
    \centering
    \includegraphics[width=0.99\linewidth]{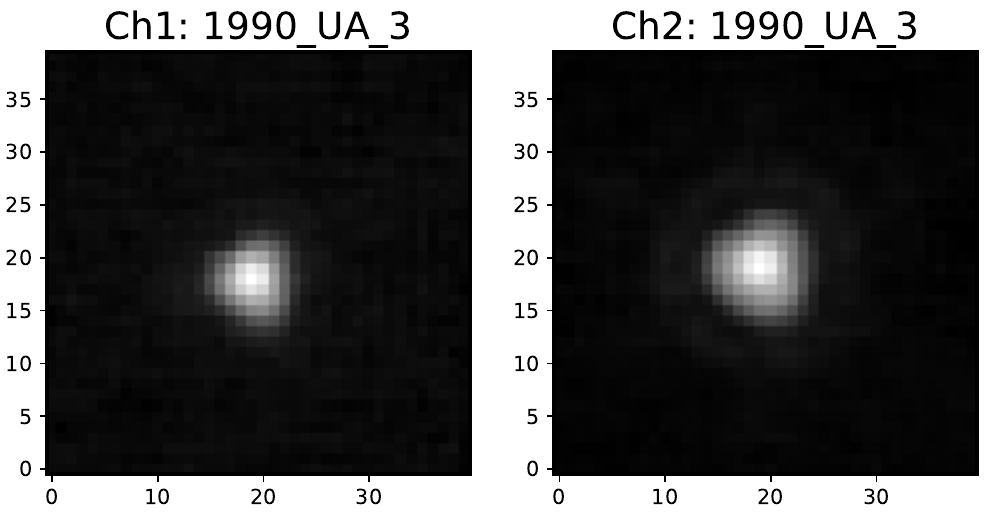}
    \caption{A plot from the NEOphot notebook that shows the 3.6~\micron\ (left) and 4.5~\micron\ (right) mosaics constructed from the frames shown in Figure~\ref{fig:patches}, excluding the rejected frames.  The axes are labeled in units of mosaic pixels (0\farcs6/pixel). The NEOs in this study were not spatially resolved, so these images effectively show the shape of the IRAC PSF.}
    \label{fig:NEOmos}
\end{figure}

\begin{figure}
    \centering
    \includegraphics[width=0.99\linewidth]{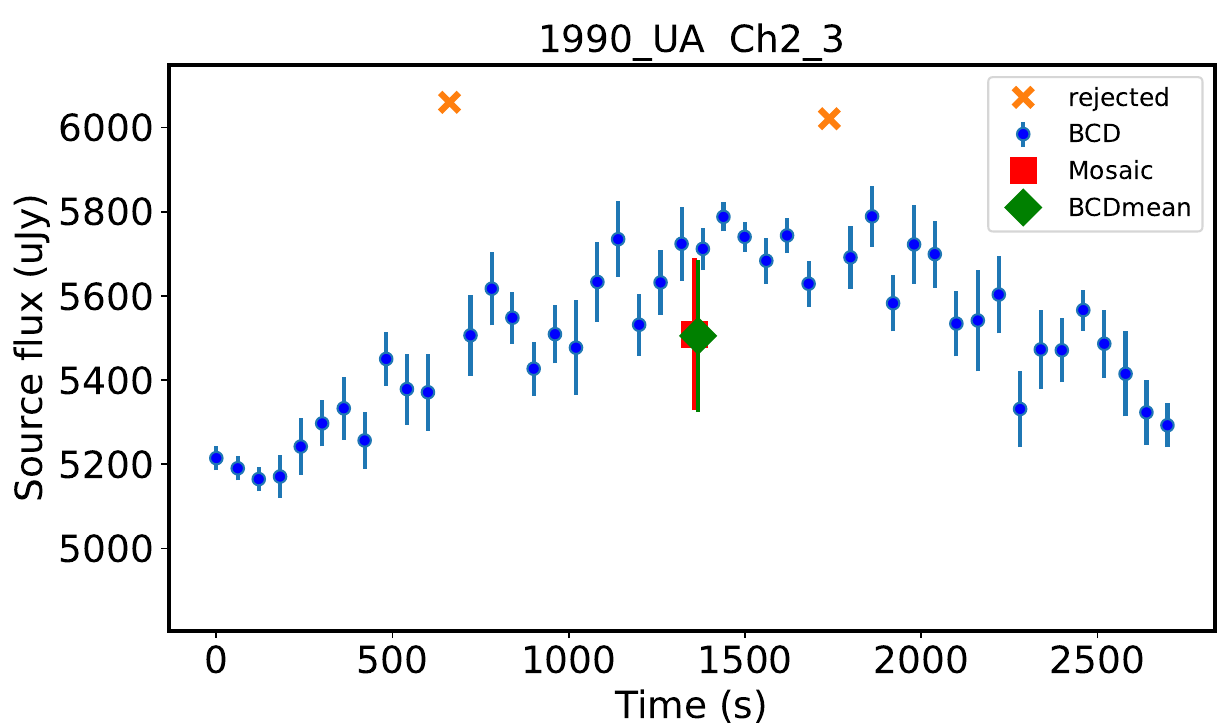}
    \caption{A plot produced by NEOphot of the accepted photometry for this part of the 1990~UA dataset,
    as final confirmation of the results. The plot shows the lightcurve (blue circles with error bars), rejected photometry (orange Xs), the mosaic photometry (red square with error bars) and the median of the lightcurve photometry (green diamond with error bars set to the standard deviation of the blue points). In general, the mosaic and mean of the BCD photometry should be within 1$\sigma$ of each other, and both should be consistent with the individual BCD photometry points.}
    \label{fig:finalphot}
\end{figure}

The data were reduced using the NEOphot software, which is a Jupyter notebook written for this project that produces mosaics and single-frame photometry for moving objects from IRAC data. The notebook uses many functions in the Astropy package \citep{astropy:2013,astropy:2018,astropy:2022}  to manipulate images and performs the aperture photometry using photutils \citep{larry_bradley_2023_7946442}. The NEOphot software is available on github\footnote{\url{https://github.com/jhora99/NEOphot}}. In the examples of the reduction steps shown below, a portion of the observations of 1990~UA are used (AOR 42169088) in the various plots and images. The plots and images in Figures~\ref{fig:darks} -- \ref{fig:finalphot} are from the NEOphot notebook and are typical of those produced when an object is analyzed.

The input data to NEOphot are the ``corrected Basic Calibrated Data" frames (cBCDs; *.cbcd.fits) produced by the IPAC pipeline. The first step is to calculate and subtract from all frames a residual dark frame which NEOphot generates from all  frames in the AOR. There is often a residual pattern in the dark frames which are not fully corrected in the IPAC processed cBCD frames. The median of the frame stack was calculated, after rejecting pixels above or below threshold values in order to reject stars and bad pixels in individual frames. These residual darks (Figure~\ref{fig:darks}) were then subtracted from all frames in the AOR to remove the dark pattern from the images.

The expected NEO position is given in the cBCD FITS headers. However, for some objects, this position was off by several arcsec. Therefore, NEOphot has an option to query Horizons\footnote{\url{https://ssd.jpl.nasa.gov/horizons/}} \citep{1996DPS....28.2504G,2015IAUGA..2256293G} to recalculate the position of the NEO at the observation time for each frame obtained by \Sp\ and update the NEO position information in the FITS headers to allow them to be masked off in the sky frames, and for the photometry routine to locate the objects.

NEOphot then calculates an image of the sky during the NEO observation. In this step, the images are registered to the position on the sky defined in the WCS of the FITS file headers. The NEO is masked out in each frame in this step as to not affect the calculation of the sky mosaic. For most observations, the NEO moved by many arcsec during the AOR, so a sky frame could be calculated that includes every point of the NEO's path throughout the observation (Figure~\ref{fig:skymos}). However, for some objects the motion during the AOR was on the order of or less than the \Sp\ point spread function (PSF), so a sky frame could not be calculated and subtracted. Therefore, if there happened to be a star directly behind the NEO during the \Sp\ observations, the flux for those objects would be overestimated. We have marked these cases in the results given in Table~\ref{tab:mosflux}.

The sky mosaics are made by first reprojecting the images to a common WCS and calculating the mean value at each sky pixel from the frames after masking bad pixels and the NEO location. This mosaic and all others in NEOphot use a pixel scale of 0\farcs6/pixel. Bad pixels are masked by comparing each frame to a median image of the combined frames and removing pixels that are $>3\sigma$ away from the median value, where $\sigma$ is the local noise estimate. Due to the undersampled IRAC images, the flux in a pixel close to the center of a star can vary greatly if the star is well-centered on a pixel compared to being near the corner of a pixel. NEOphot therefore calculates a local noise estimate to keep the central bright pixel of a star from being rejected, similar to the method used in the IRACproc software \citep{2006Schuster}.

Once the sky frame is calculated, it is subtracted from each cBCD frame to produce an image containing only the NEO (an ``NEO frame''), although still having bad pixels due to cosmic ray impacts, bad array pixels, and some residuals from incomplete subtraction of bright stars. Aperture photometry is then performed on each NEO frame and the results are displayed to the user in the form of a set of images showing a small region around the NEO (see Figure~\ref{fig:patches}). The aperture size and background annulus inner and outer radius are user-selectable parameters; in our processing we used an aperture radius of 3\farcs6 and inner and outer annulus radii of 7\farcs2 and 9\farcs6, respectively. To derive the calibration factors to convert from ADU to Janskys using these radii, the same photometry techniques were used on three IRAC flux calibration stars HD165459, HD184837, and 1812095 (2MASS J18120957+6329423). The fluxes for these stars were taken from \citet{Reach05}, and the data for these stars were obtained at various times during the \Sp\ Warm Mission and processed with the same pipeline version (S19.2.0). Ten separate AORs for each calibration star were processed, and the derived calibration factors for the observations were averaged to give the conversion factors used for the NEO photometry. The standard deviation of the calibration star photometry was 0.6\% for the 3.6~\micron\ band and 0.3\% for the 4.5~\micron\ band, much smaller than the $\sim$2\% relative photometric accuracy expected for photometry from the cBCD frames \citep[e.g., see][]{Hora2008}.

The program also displays two 1-D plots: the x and y offset in pixels from the expected position versus time, and the flux versus time obtained from the images (Figure~\ref{fig:1Dplots}). Outliers from the curve in either position or flux due to bad pixels in the cBCD can easily be seen in these plots. These 1-D plots are interactive, allowing the user to click on individual measurements to remove outliers from the lightcurve. Once the outliers have been eliminated, the program creates a plot showing the relative offset of the centroid of the NEO in each of the frames (Figure~\ref{fig:poserr}) for both the 3.6 and 4.5~\micron\ channels, if available. This plot allows the user to confirm that the positions found were consistent with the expected position of the NEO in each frame.

After this step, a mosaic is made of all the NEO frames, excluding those rejected in the previous step (Figure~\ref{fig:NEOmos}). The user can choose whether to recenter the frames based on the NEO image, or use the expected NEO position in the cBCD headers to stack the frames. For most frames recentering is not necessary, but for some AORs the position of the NEO stored in the FITS header is not correct so the images must be shifted slightly to line up properly. After constructing the mosaic, photometry is performed and the result shown in a 1-D plot of the cBCD photometry along with the mosaic photometry and the median of the cBCD photometry (Figure~\ref{fig:finalphot}). The mosaic photometry is used in the thermal modeling along with the absolute magnitude $H$ obtained from the JPL small body database\footnote{\url{https://ssd.jpl.nasa.gov/sb/}} or from PanSTARRS observations (Allen et al., in preparation). The final results are written to text files that record the light curves and mosaic photometry. All of the images shown here are saved for each object, and the plots are also saved as pdf files.


\section{Results}
\label{sec:results}
The fluxes derived for all NEOs processed using the methods described above are given in Table~\ref{tab:mosflux}. These fluxes are the result of aperture photometry performed on the mosaics constructed from all of the frames in the AORs. In some cases where the S/N of the individual frames was low due to short integration times and low NEO fluxes, the mosaic flux is a more reliable measure of the average flux of the NEO during the observation period and is less susceptible to effects such as bad pixels, cosmic rays, or background stars that can affect the photometry in individual cBCD frames.

The 3.6~\micron\ fluxes are given in the table if the object was observed in that band. When present, the 3.6 and 4.5~\micron\ frames were taken alternating between the two channels since they have different fields of view. The column marked ``4.5~\micron\ Amplitude Ratio" gives the ratio of the peak-to-peak range of the lightcurve fluxes divided by the mosaic flux. This value was calculated for objects where the minimum flux was greater than 50~$\mu$Jy in the 4.5~\micron\ channel and the flux was also greater than 3$\times$ the median flux uncertainty for that measurement. This prevents anomalously large amplitude values being reported for objects that are low S/N and the amplitude uncertainty is high.

Most of the NEO fluxes reported previously by \citet{2010Trilling} and \citet{2016Trilling} differ only by a few percent or less from the new fluxes that we derive here. Some of these effects are due to the updated IRAC calibration and the improved background subtraction using sky frames constructed from the masked cBCDs, which was not previously done in every case. Larger differences are due to cases where bad cBCDs (e.g., cases where the NEO position in the frame was affected by bad pixels, cosmic rays, bright star residuals, etc.) were not rejected in the previous reduction method prior to constructing the mosaic frame. In this current reduction, we visually inspected the cBCD frames and rejected those with artifacts before constructing the mosaic, thereby minimizing their effect on the mosaics and the aperture photometry.

The fluxes determined here, along with updated absolute magnitude values derived from PanSTARRS measurements, will be used to derive albedo and diameter estimates using the NEATM thermal model. This will be presented in a future paper \citep{2023allen,2025allen}.

In addition to the mosaic photometry, the full dataset of BCD photometry for every NEO in our sample is available in the supplemental data to this paper.

\begin{deluxetable*}{lrlccrrrrrrc}
\tabletypesize{\scriptsize}
\caption{IRAC Fluxes from Mosaics for all NEOs Observed}\label{tab:mosflux}
\tablehead{& & & Start& & 3.6~\micron & 3.6~\micron&4.5~\micron & 4.5~\micron & 4.5~\micron \\Object &  & & MJD & Duration & Flux & Flux unc. & Flux & Flux unc.& Amplitude \\
Designation & SPKID & Name & (d) & (h) & ($\mu$Jy) & ($\mu$Jy) &  ($\mu$Jy)&  ($\mu$Jy) & Ratio & Notes{\scriptsize \tablenotemark{a}}}
\startdata
A898 PA  &  20000433 &433 Eros         &      55068.25899 &0.130& 12675.67 &  88.507  &40479.61 & 193.71 &0.04 &  \nodata\\
1929 SH  &  20001627 &1627 Ivar         &     55363.35810 &0.135&  1263.77 &  18.85&   2769.74 &  31.89 &0.08  & \nodata\\
1932 EA1 &  20001221 &1221 Amor         &     57887.82225 &0.332&     \nodata&  \nodata &    327.06 &   3.10 &0.17   &\nodata\\
1932 HA  &  20001862 &1862 Apollo       &     57992.46300 &2.136&    \nodata & \nodata  &     19.04 &   0.62 &   \nodata    &\nodata\\
1936 CA  &  20002101 &2101 Adonis       &     58060.80713 &0.678 & \nodata  &  \nodata&     75.62 &   2.14 &0.44   &\nodata\\
1937 UB  &  20069230 &69230 Hermes      &     55387.84496 &0.131 &  641.61 &  12.81&   2884.92 &  12.81 &0.03   &\nodata\\
1947 XC  &  20002201 &2201 Oljato       &     58243.92796& 0.346 &  \nodata  &  \nodata  &    405.00 &   3.99 &0.21   &\nodata\\
1948 EA  &  20001863 &1863 Antinous     &     55076.53238& 0.132 & 5249.02 &  40.14&  23497.25 &  75.60 &0.03   &\nodata\\
1948 OA  &  20001685 &1685 Toro         &     55436.33325& 0.136 &  289.56 &  11.15&    741.87 &   4.97 &0.06& 1 \\
1949 MA  &  20001566 &1566 Icarus       &     55457.11490& 0.588 &  305.31 &   5.72&   1189.47 &  12.01 &0.09 &  \nodata\\
\enddata
\tablenotetext{a}{A ``1" indicates objects with small relative motion during the observation.
A ``2'' indicates objects that passed near a bright star during the observation.
}
\tablecomments{Table~\ref{tab:mosflux} is published in its entirety in the machine-readable format. A portion is shown here for guidance regarding its form and content.}
\end{deluxetable*}

\subsection{Deriving periods}

The 4.5~\micron\ photometry was used for searching for rotational periods since it exists for every object in the dataset and is of higher signal-to-noise (S/N) because of the thermal emission that dominates that band compared to the 3.6~\micron\ band which has a larger contribution from reflected sunlight and lower thermal flux. In those cases where the rotation was well-detected in both bands, the periods derived from the 3.6~\micron\ band closely match those from the 4.5~\micron\ period.

The AORs for the NEO programs were designed to obtain a single flux measurement at sufficient S/N, and many of the NEOs were observed with only a few on-source frames and fairly short timescales which is inadequate to determine reliable periods. In searching for signs of rotational variations, we excluded sources that had fewer than 18 samples or a duration of less than 0.5~hr.
\begin{figure}
    \centering
    \includegraphics[width=0.99\linewidth]{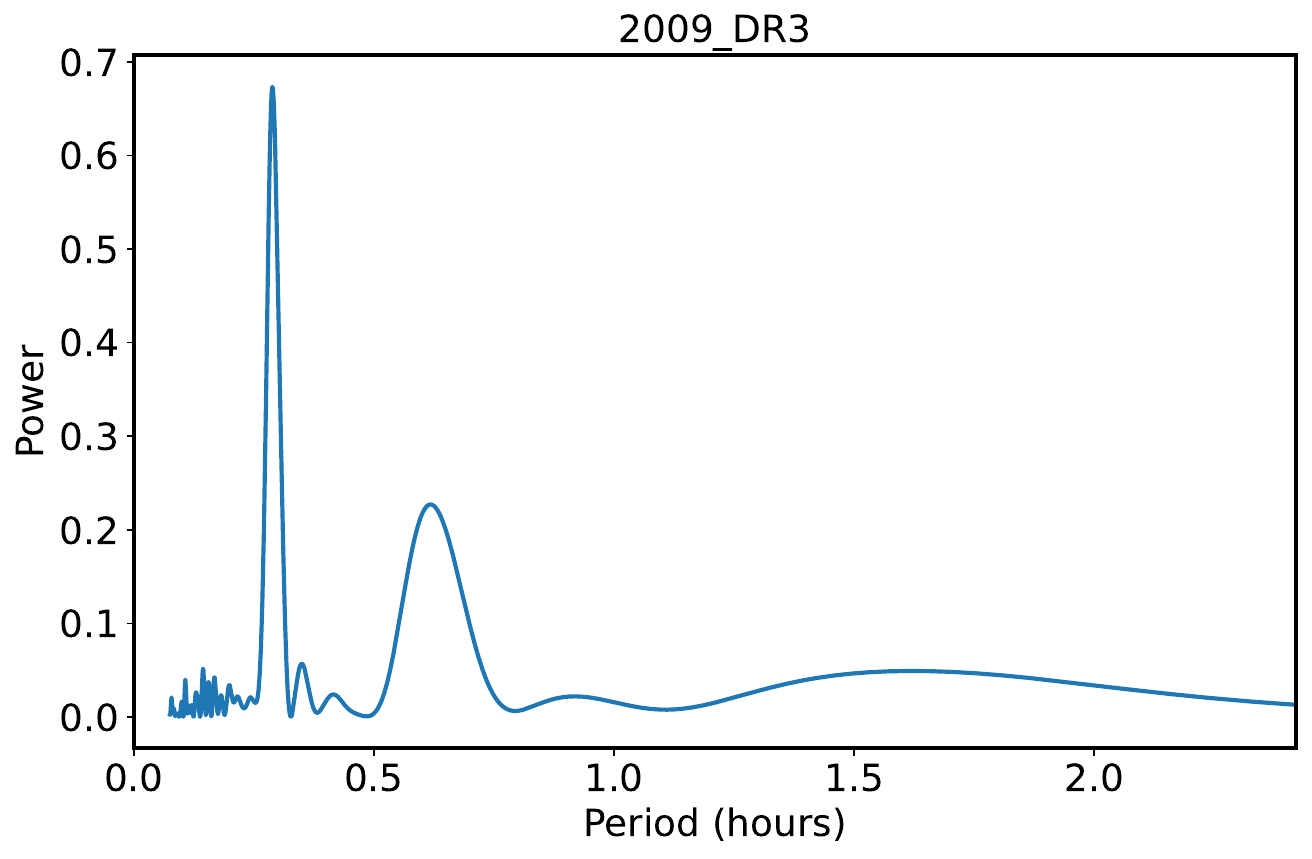}
    \caption{ The periodogram produced by the L-S
    algorithm for the NEO 2009~DR3. The observations used the 100~s frame time, and with dithering one frame was obtained every $\sim$110~s; or 0.031~hr (with some frames not used due to bad pixels or issues with subtracting the background field). The highest
    peak in the periodogram is at 0.28857~hr, and doubling this value gives a rotational period of 0.577$\pm$0.003~hr (see its lightcurve in Figure~\ref{fig:lc6}). No evidence of significant power is seen at the data sampling period.}
    \label{fig:Pgram}
\end{figure}

We used the Lomb-Scargle (L-S) periodogram \citep{1976Ap&SS..39..447L,1982ApJ...263..835S} as implemented in the Astropy timeseries subpackage to determine the periodicity of the NEO timeseries data. The \citet{2008baluev} method was used to estimate the false alarm probability (FAP) of the solution, where larger values indicate a higher chance of an incorrect period determination. The peak in the periodogram with the highest power was selected, and this period was doubled to estimate the rotational period of the NEO. An example is shown in Figure~\ref{fig:Pgram} for the NEO 2009~DR3.

In the full Spitzer sample, the successful period results sampled a range of 0.8 to 11.1 periods for individual NEOs. Lightcurves and phase plots for the NEOs where at least one full period was found are shown in Appendix~\ref{appdx:lc}.  

\subsubsection{Lightcurves with $\sim$1 Period Sampled}
Because of the relatively short duration of most of the Spitzer observations, a number of the NEOs had $\sim$1 or slightly higher number of periods sampled. This will lead to higher systematic uncertainties in the rotational periods for these objects. For NEOs whose lightcurves do not sample more than two full periods, there is always a chance that the objects have a more complex shape and have triple peaked lightcurves or other complex behavior that was not fully sampled by the data presented here. Therefore, NEOs with less than two sampled periods in this dataset should in some respects be considered lower limits to their periods. However, if those effects are not present, the rotational periods can be accurately determined, as shown in Figure~\ref{fig:2005GL95hr}. In Table~\ref{tab:fullperiods} we list the NEOs and periods determined where more than two full periods were sampled and the FAP is less than 10\%. Table~\ref{tab:uncon_periods} lists NEOs for which either less than two full periods were sampled, or the FAP was $>$10\%. These objects are subject to the caveats detailed above and should be considered less reliable than the NEO periods in Table~\ref{tab:fullperiods}.

\subsubsection{Estimating uncertainties}
The period uncertainties were estimated by adding random noise to each data point in the individual flux measurements of the lightcurve based on the photometric uncertainty, and running the L-S period determination on the altered data. This was repeated 1000 times for each source in Table~\ref{tab:fullperiods} and \ref{tab:uncon_periods}, and the standard deviation of the derived periods is reported in the Uncertainty column.  This should be taken as the contribution to the uncertainty from the photometric measurements at each point in time, but as described in the previous section, there may be other systematic errors that result in errors larger than the values given in the table. 
\begin{figure}
    \centering
    \includegraphics[width=0.99\linewidth]{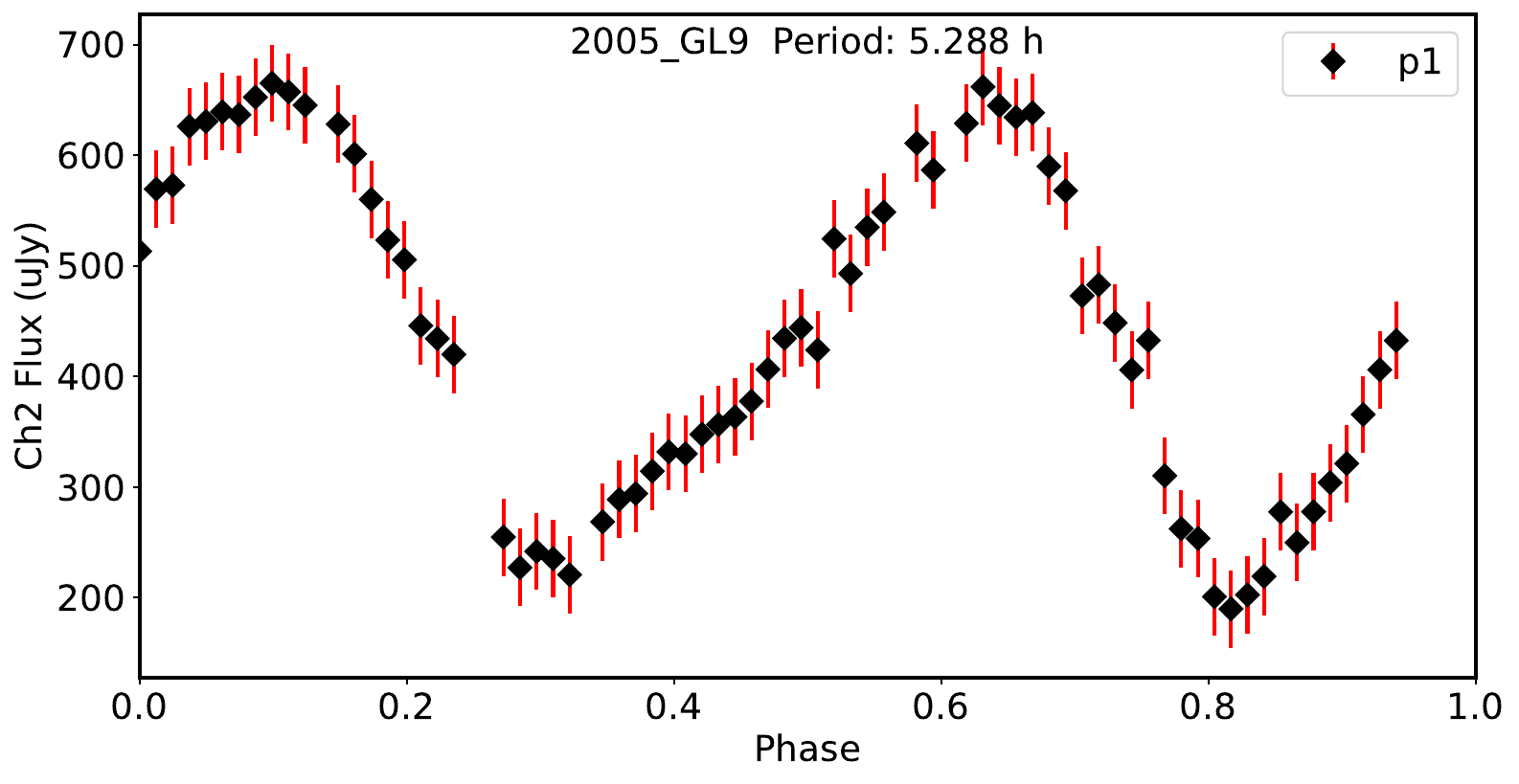}
    \caption{The phase plot for the NEO 2005~GL9 when using only the first 5~hr of data. The rotational period determined was 5.288$\pm$0.094~hr, compared to 5.149$\pm$0.026~hr determined when using the full 9~hr dataset.}
    \label{fig:2005GL95hr}
\end{figure}

As an example of the possible effects this may have on the period determination, we take the Spitzer observations of 2005~GL9 where the lightcurve length is $\sim$1.75 rotational periods of length 5.199$\pm$0.025~hr. If we reanalyze the data using only the first 5~hr of observations, we obtain a period of  5.288$\pm$0.062~hr, which is shown in Figure~\ref{fig:2005GL95hr} (compare to Figure~\ref{fig:lc5}). 

\subsubsection{Incomplete lightcurves}
NEOs that were seen to have significant flux variations but did not appear to have sampled a full period were designated as lower limits. In these cases the length of the observation period is given as the lower limit, although in some cases it might be possible to estimate the full period if, for example, one assumes a sinusoidal variation. The objects that vary with incomplete periods are given in Table~\ref{tab:incomplete}.

\begin{deluxetable*}{lcccrrcr}
\caption{NEO Periods Determined from IRAC Observations}\label{tab:fullperiods}
\tablehead{Object & Period& Uncertainty &  & Number of &  & MJD Start & Duration\\
Designation & (h) & (h) & FAP & Periods & Frames & (d) & (h)}
\startdata
    1996 FG3 &  3.558 &  0.001 & 0.00E+00 &   2.76 &    4654 &     55737.377000 &    9.816 \\
  2000 DP107 &  2.802 &  0.001 & 0.00E+00 &   3.63 &    5115 &     55897.548561 &   10.175 \\
   2001 UV16 &  0.483 &  0.002 & 3.82E-06 &   6.27 &      85 &     57161.411979 &    3.033 \\
  2002 AJ129 &  3.977 &  0.036 & 2.57E-14 &   2.25 &      99 &     55726.388994 &    8.940 \\
   2007 BX48 &  0.782 &  0.026 & 4.56E-02 &   3.09 &      60 &     57947.409105 &    2.421 \\
   2008 EY68 &  0.447 &  0.003 & 5.64E-06 &   5.42 &      66 &     58291.938392 &    2.420 \\
    2009 DR3 &  0.577 &  0.003 & 2.17E-08 &   4.20 &      64 &     58022.097453 &    2.421 \\
    2011 LL2 &  0.103 &  0.000 & 5.13E-04 &  17.56 &      55 &     58382.383701 &    1.811 \\
    2011 XA3 &  0.725 &  0.011 & 3.12E-05 &   2.50 &      53 &     57860.410521 &    1.810 \\
   2012 BF86 &  0.164 &  0.001 & 6.81E-03 &  11.06 &      48 &     57872.918427 &    1.815 \\
    2012 WK4 &  0.129 &  0.001 & 1.36E-03 &   2.68 &      26 &     58164.824117 &    0.346 \\
    2013 VO5 &  0.377 &  0.001 & 4.64E-12 &   6.42 &      69 &     58191.807763 &    2.420 \\
     2015 XC &  0.545 &  0.008 & 7.74E-02 &   3.21 &      54 &     57826.763754 &    1.749 \\
\enddata
\end{deluxetable*}

\begin{deluxetable*}{lcccrrcr}
\caption{NEO Under-Constrained Periods ($<$2 Full Periods Sampled or $>$10\% FAP)}\label{tab:uncon_periods}
\tablehead{Object & Period& Uncertainty &  & Number of &  & MJD Start & Duration\\
Designation & (h) & (h) & FAP & Periods & Frames & (d) & (h)}
\startdata
1982 XB &  0.652 &  0.050 & 1.31E-01 &   2.68 &      43 &     57167.050929 &    1.750 \\
1986 JK &  2.720 &  0.287 & 8.31E-01 &   1.09 &      69 &     57062.269246 &    2.974 \\
     1989 WD &  2.873 &  0.059 & 5.50E-26 &   0.95 &      77 &     57520.329365 &    2.730 \\
     1990 UA &  3.088 &  0.012 & 6.85E-83 &   1.72 &     292 &     55750.720817 &    5.313 \\
    1991 CB1 &  5.632 &  0.257 & 3.93E-07 &   0.93 &     248 &     55724.298690 &    5.254 \\
     1998 PG &  1.113 &  0.070 & 9.97E-01 &   2.08 &      67 &     57345.876309 &    2.311 \\
    1999 JE1 &  6.420 &  0.310 & 3.77E-08 &   1.39 &      94 &     55765.110523 &    8.941 \\
    1999 JU3 &  5.801 &  0.027 & 7.48E-66 &   1.37 &     403 &     56333.838000 &    7.944 \\
   2001 KO20 &  1.227 &  0.129 & 9.96E-01 &   1.97 &      59 &     58014.963331 &    2.422 \\
   2001 XR30 &  0.391 &  0.028 & 1.00E+00 &   4.41 &      44 &     57278.675843 &    1.721 \\
     2002 SV &  2.233 &  0.080 & 1.00E+00 &   1.36 &      79 &     57400.359146 &    3.031 \\
   2002 TW55 &  0.118 &  0.015 & 9.29E-01 &   2.79 &      45 &     58145.059415 &    0.328 \\
   2003 EO16 &  5.656 &  0.221 & 7.78E-01 &   1.60 &     105 &     55728.364000 &    9.072 \\
   2004 KK17 &  4.699 &  0.153 & 3.71E-07 &   1.10 &      55 &     55743.594207 &    5.169 \\
   2004 PS92 &  1.769 &  0.299 & 1.00E+00 &   1.37 &      62 &     57692.226007 &    2.420 \\
   2004 TK14 &  0.735 &  0.023 & 1.00E+00 &   3.30 &      57 &     57769.475032 &    2.425 \\
    2005 GL9 &  5.149 &  0.026 & 1.53E-63 &   1.75 &     219 &     55727.772870 &    9.031 \\
    2005 HC3 &  2.534 &  0.089 & 1.15E-02 &   1.17 &      76 &     57486.682971 &    2.974 \\
     2006 GU &  1.426 &  0.147 & 5.27E-04 &   1.27 &      50 &     57324.009213 &    1.812 \\
  2007 DU103 &  2.964 &  0.292 & 3.42E-02 &   1.02 &      83 &     57300.517128 &    3.034 \\
    2008 UF7 &  2.814 &  0.119 & 2.11E-03 &   1.08 &      83 &     57054.277971 &    3.036 \\
    2009 HU2 &  0.604 &  0.014 & 1.00E+00 &   3.90 &      60 &     58051.629606 &    2.358 \\
  2009 WD106 &  2.490 &  0.131 & 4.53E-02 &   1.22 &      82 &     57122.693453 &    3.034 \\
   2010 RG42 &  2.566 &  0.152 & 4.52E-01 &   1.17 &      71 &     58432.832030 &    3.007 \\
   2011 EP51 &  1.573 &  0.196 & 1.00E+00 &   1.87 &      78 &     57562.672216 &    2.945 \\
    2012 AD3 &  1.912 &  0.143 & 6.24E-01 &   1.59 &      73 &     57057.446607 &    3.032 \\
\enddata
\end{deluxetable*}

\begin{deluxetable*}{lcccrcc}
\caption{NEO Period Lower Limits ($<$1 Full Period Sampled)}\label{tab:incomplete}
\tablehead{&Period & &Approximate\\Object & Lower Limit &  & Number of &  & MJD Start & Duration\\
Name & (h) & FAP & Periods & Frames & (d) & (h)}
\startdata
1971 UA & 1.4 & 1.25E-09 & 0.25 & 24 & 55201.241826 & 0.349 \\
1978 DA & 5.2 & 3.25E-14 & 0.45 & 56 & 57919.492040 & 2.327 \\
1994 LW & 13.0 & 7.59E-26 & 0.70 & 113 & 55772.963104 & 9.092 \\
1998 MZ & 2.4 & 8.36E-08 & 0.75 & 48 & 57277.992492 & 1.814 \\
1998 SJ2 & 2.0 & 4.38E-11 & 0.90 & 49 & 58746.951334 & 1.816 \\
1998 ST4 & 5.4 & 1.16E-16 & 0.45 & 55 & 58271.522096 & 2.420 \\
1998 XA5 & 4.4 & 3.35E-03 & 0.55 & 55 & 58228.589408 & 2.421 \\
1999 LD30 & 3.2 & 2.42E-17 & 0.75 & 61 & 58383.662326 & 2.424 \\
1999 LU7 &	1.8 &	2.91E-07 & 0.40& 22 & 57061.293281 & 0.708 \\
\enddata
\tablecomments{Table~\ref{tab:incomplete} is published in its entirety in the machine-readable format. A portion is shown here for guidance regarding its form and content.}
\end{deluxetable*}

\subsection{Comments on Individual NEOs}
A majority of the NEOs in our sample have no prior rotation period measurements. In this section we comment on specific NEOs and their lightcurves and derived periods for cases with prior measurements listed in the Light Curve DataBase \citep[LCDB;][]{2009LCDB} and where the \Sp\ measurements sampled one full period or more, according to our L-S analysis. We also comment on measurements that have low S/N and therefore may be less certain that the L-S FAP may indicate. References to the LCDB are to the 2023 February version retrieved from the JPL Small-Body Database Lookup web page\footnote{\url{https://ssd.jpl.nasa.gov/tools/sbdb_lookup.html}}. Several objects have the note in the LCDB that the ``Results are based on less than full coverage, so that the period may be wrong by 30 percent or so", which we will abbreviate below as the LFC (Less than Full Coverage) note.  

\subsubsection{1982 XB}
The LCDB value for the period of 1982~XB is 9.012~hr, with a LFC note. The value derived from \Sp\ data is significantly shorter, 0.647~hr. However, the flux and amplitude are low compared to the measurement uncertainty, with several discrepant points, which likely makes the \Sp\ determination less certain.

\subsubsection{1986 JK}
There is no previous rotation period listed in the LCDB. The amplitude of the variation in the \Sp\ lightcurve is low compared to the measurement uncertainty, and the L-S period is just slightly shorter than the length of the observation. It is possible that the full period was not sampled, even though the L-S FAP is low.

\subsubsection{1989 WD}
The \Sp-derived period of 2.873~hr is close to the LCDB period of 2.89111~hr. The \Sp\ observations did not quite cover the full period length, so that likely led to a slight difference from the previously-derived value. 

\subsubsection{1996 FG3}
The NEO 1996~FG3 is a binary asteroid known to undergo mutual eclipses, with an orbital period of approximately 16~hr. This object was targeted in \Sp\ program 70054 \citep{2010sptz.prop70054M} to obtain a thermal IR lightcurve of the binary system to measure its thermal inertia and constrain the surface properties of the NEO. A similar analysis has been performed on this object by \citet{2024Jackson} using data from WISE/NEOWISE. The thermal effects of the eclipse apparently  affected the flux during the phases sampled here, however we were still able to obtain a \Sp-derived period of 3.558~hr. This is close to the LCDB value of 3.5942~hr, and that determined by \citet{2015Icar..245...56S} of 3.595195$\pm$0.000003~hr. 

\subsubsection{1998 PG}
The LCDB period for this object is 2.5163~hr with a LFC note, which is slightly more than double the \Sp-derived value of 1.101~hr. The \Sp\ observation was approximately 2.5~hr, so it would have sampled the full lightcurve if it was the longer period. The phased \Sp\ lightcurve seems consistent with the derived shorter period.

\subsubsection{1999 JU3}
The \Sp-derived period for 1999~JU3 (162173 Ryugu, the Hayabusa2 target) of 7.633~hr is consistent with the 7.63262~hr period reported by \citet{watanabe2019}. 

\subsubsection{2000 DP107}
The NEO 2000~DP107 is a binary asteroid with an orbital period of approximately 1.755~d \citep{2002Margot}. 
The \Sp-derived period of 2.776~hr is consistent with the value of 2.7745~hr obtained from radar observations by \citet{2015Naidu}. The \Sp\ observations sampled over three periods in just over 10~hr, and slight differences in the lightcurve shape are seen, as well as a brightening trend, perhaps partly due to the increasing illumination percentage during the measurement period, or thermal inertia effects.

\subsubsection{2002 AJ129}
The \Sp-derived period of 3.915~hr is consistent with the value of 
3.9333~hr reported in the LCDB.

\subsubsection{2005 GL9}
The \Sp-derived period of 5.198~hr is consistent with the LCDB value of 5.131~hr. This NEO lightcurve has one of the largest amplitudes observed by \Sp, with amplitude/median flux $>$1.

\subsubsection{2005 HC3}
The \Sp-derived period of 2.478~hr is much shorter than the LCDB value of 14.40~hr. The \Sp\ observations were only $\sim$3~hr long and fairly low S/N, so the period derived by \Sp\ could be spurious, although the amplitude of the variations was significant and appear real.

\subsubsection{2011 LL2}
The \Sp-derived value of 0.10315$\pm$0.00015~hr is consistent with the LCDB value of 0.103154 (as reported by \citealt{Hergenrother2012}) to within the measurement uncertainties.

\subsubsection{2011 XA3}
The \Sp-derived value of 0.724~hr is consistent with the 0.73~hr period in the LCDB.

\subsubsection{2012 BF86}
\begin{figure}
    \centering
    \includegraphics[width=0.99\linewidth]{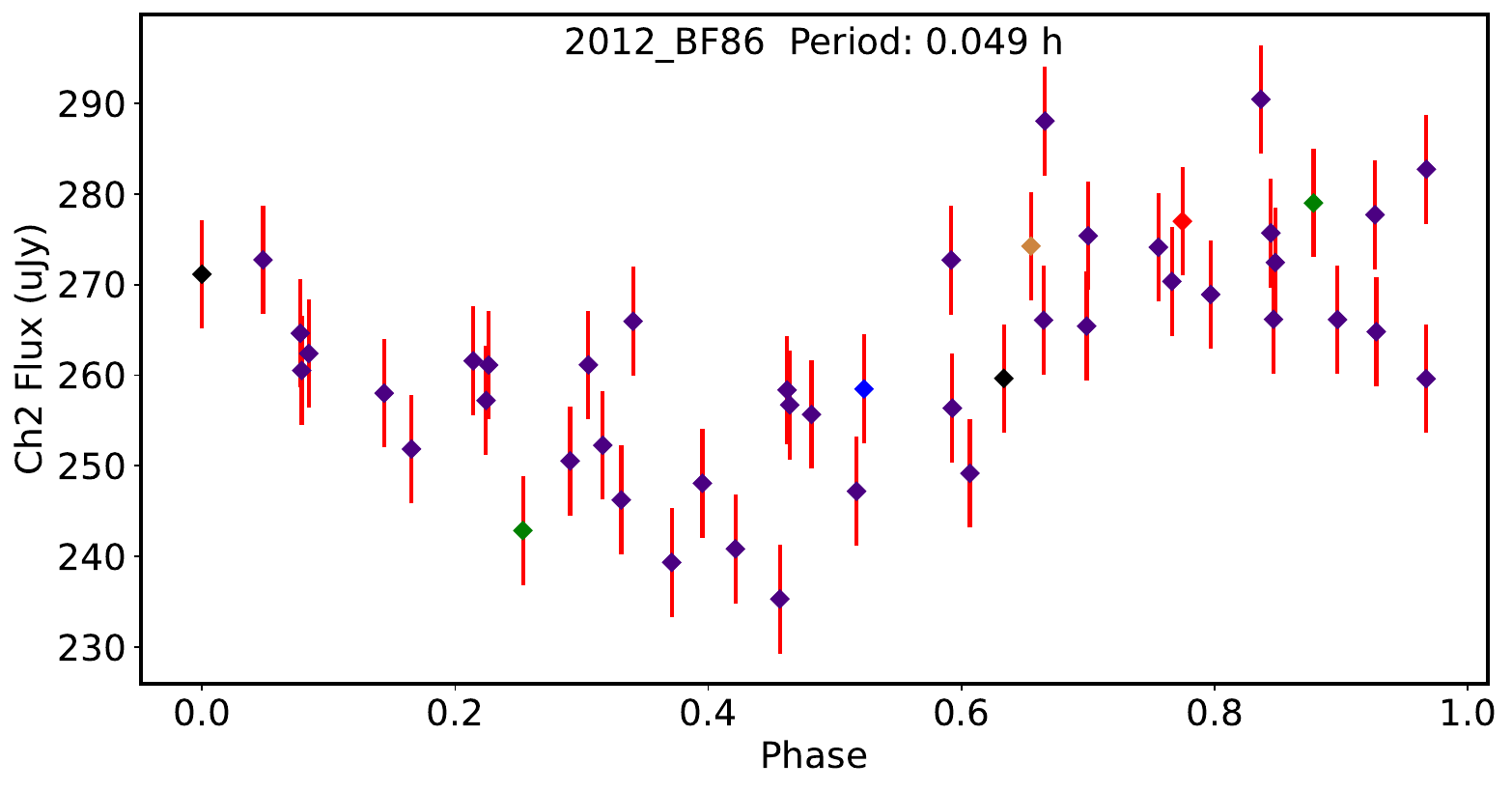}
    \caption{The \Sp\ lightcurve for 2012~BF86, forced to the period of 0.0491~hr as obtained by \citet{2018Thirouin}. }
    \label{fig:BF86}
\end{figure}

The \Sp-derived period of 0.164~hr differs from the LCDB value of 0.0491~hr with a LFC note, as reported by \citet{2018Thirouin}. Their Discovery Channel Telescope observations were performed with 15~s integrations over a period of 71~minutes. The \Sp\ observations were obtained with 100~s frames at a median cadence of 110~s, over a total period of 109~minutes. Therefore the \Sp\ analysis is likely less able to pick out the shorter periods, obtaining 1.6 samples per period (assuming the 0.0491~hr rotation period) compared to the \citet{2018Thirouin} observations which obtained over 11 samples per period. However, forcing this shorter period on the \Sp\ lightcurve, shown in Figure~\ref{fig:BF86}, shows only a single peak and so is perhaps consistent with half of the actual period. More observations are required to obtain the true period for this object.

\subsubsection{2015 XC}
\begin{figure}
    \centering
    \includegraphics[width=0.99\linewidth]{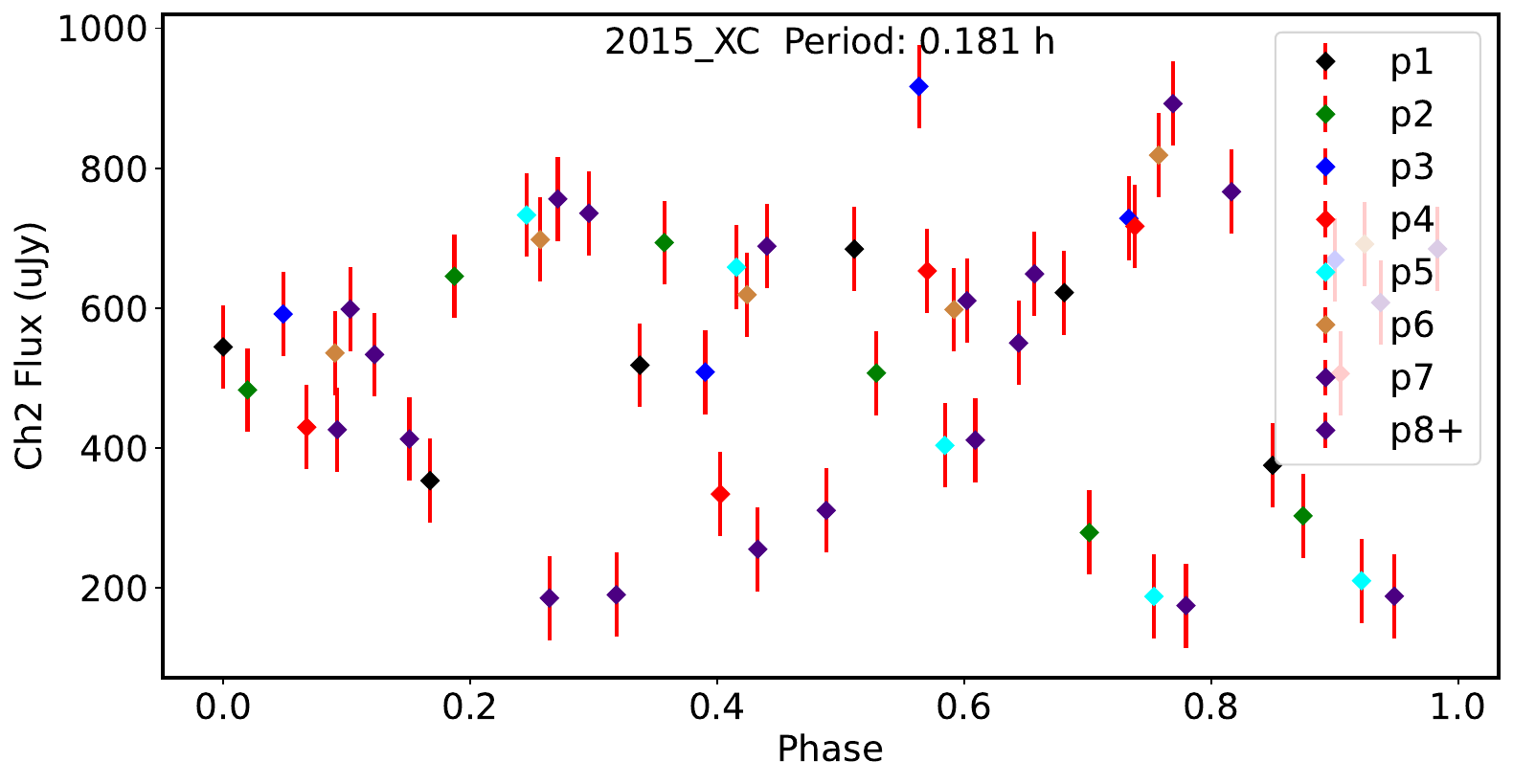}
    \includegraphics[width=0.99\linewidth]{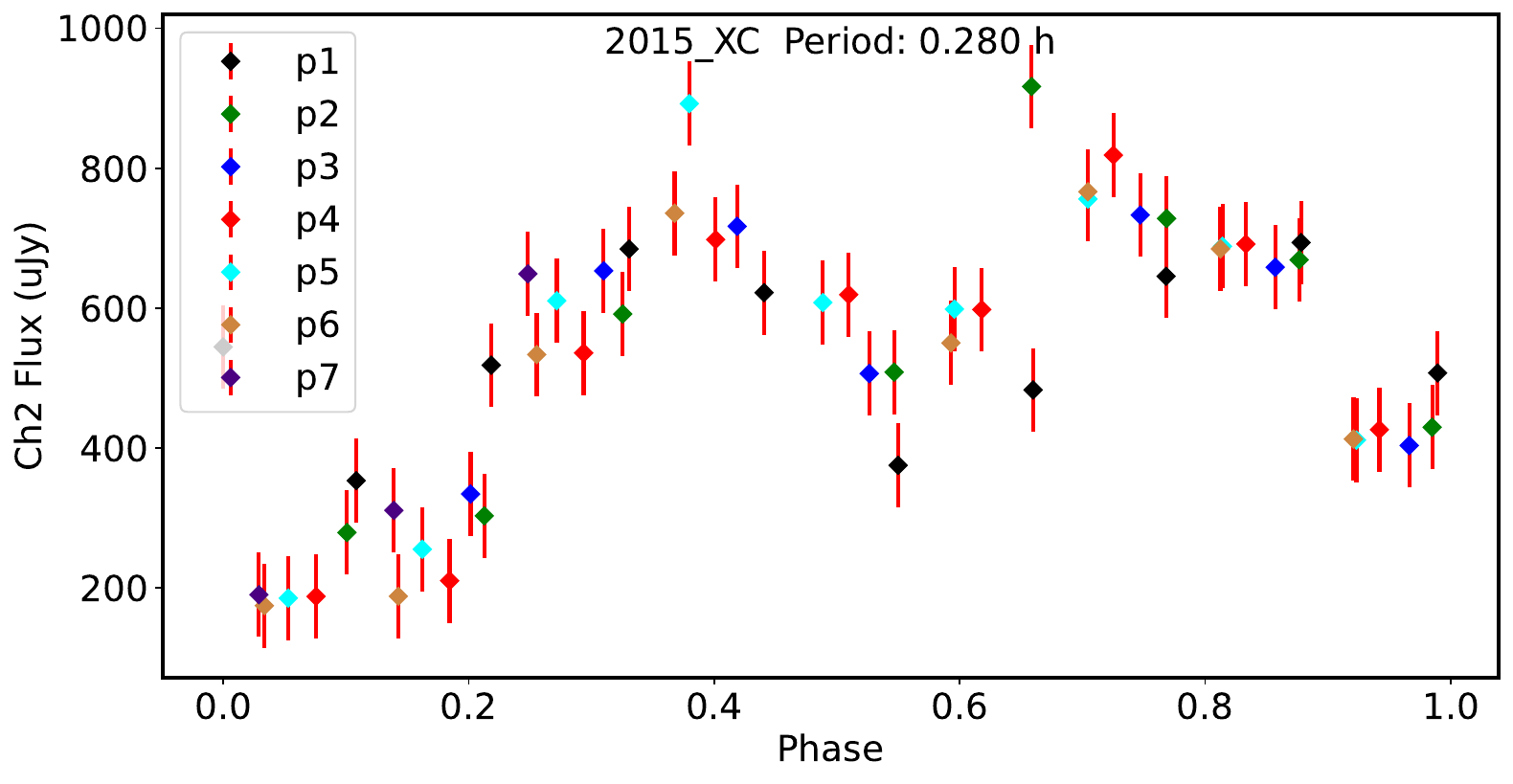}
    \includegraphics[width=0.99\linewidth]{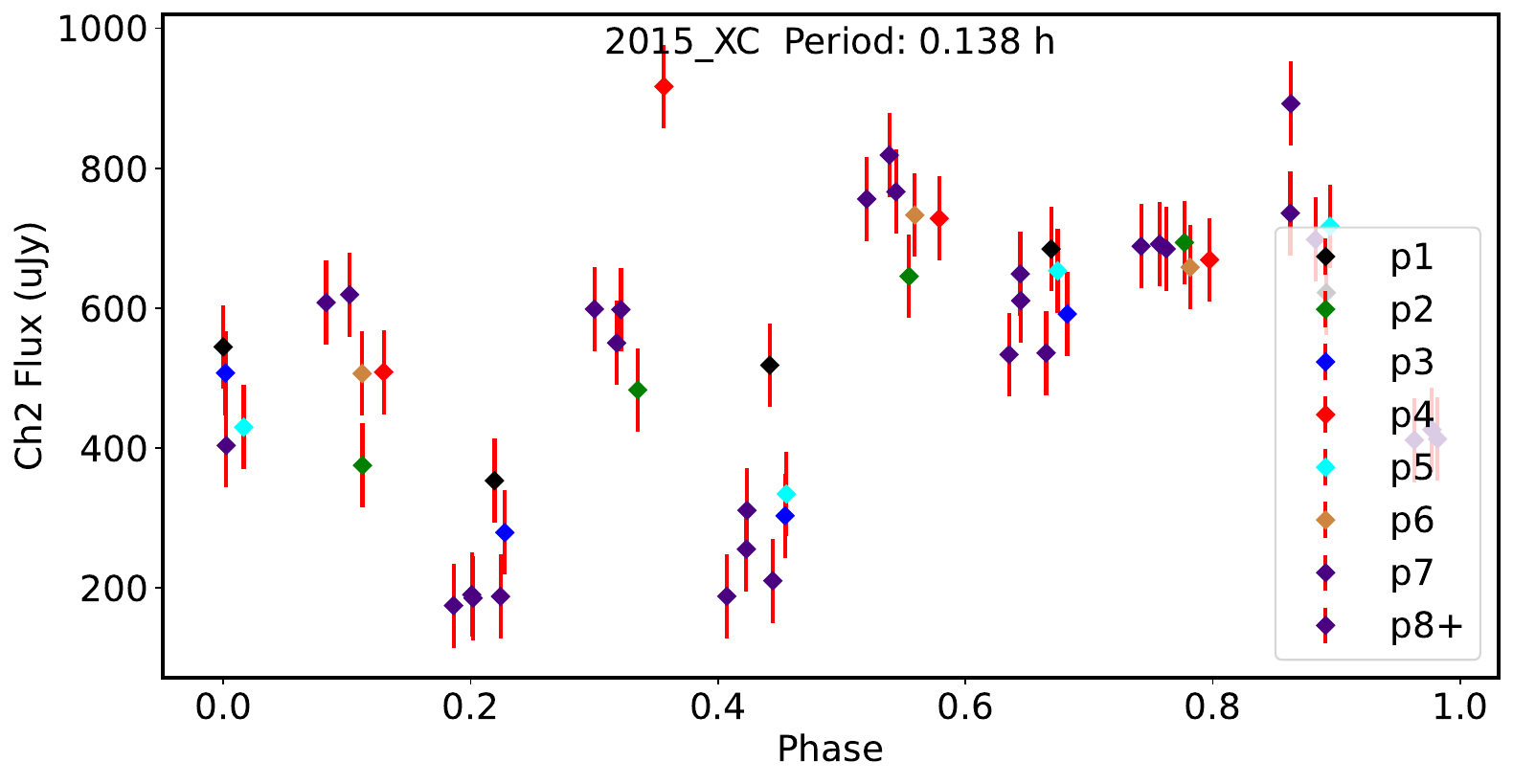}
    \caption{Phased plots at various alternate periods for 2015~XC. The top and middle plots show the Spitzer data phased according to periods found by P. Pravec as given by \citet{2016Warner}. The lower plot shows the phased plot using a period of 0.13831~hr, which was the second highest peak in the power spectrum of the \Sp\ lightcurve.}
    \label{fig:2015XC}
\end{figure}

This object was reported as a tumbling NEO by \citet{2016Warner} who derived a period of 0.541~hr using the ``Float'' mode in {\it MPO Canopus}, and found other periods of 0.181099~hr and 0.27998~hr. The first period is consistent with the \Sp-derived period of 0.545~hr. Figure~\ref{fig:2015XC} shows phased lightcurve plots for the alternate periods, consistent with the 0.280~hr period and inconsistent with the 0.181~hr period. The second highest peak in the power spectrum from the L-S fit was at 0.13831~hr, shown in the lower plot of the figure.

\section{Internal strengths}
\label{sec:strengths}
\begin{figure}
    \centering
    \includegraphics[width=0.99\linewidth]{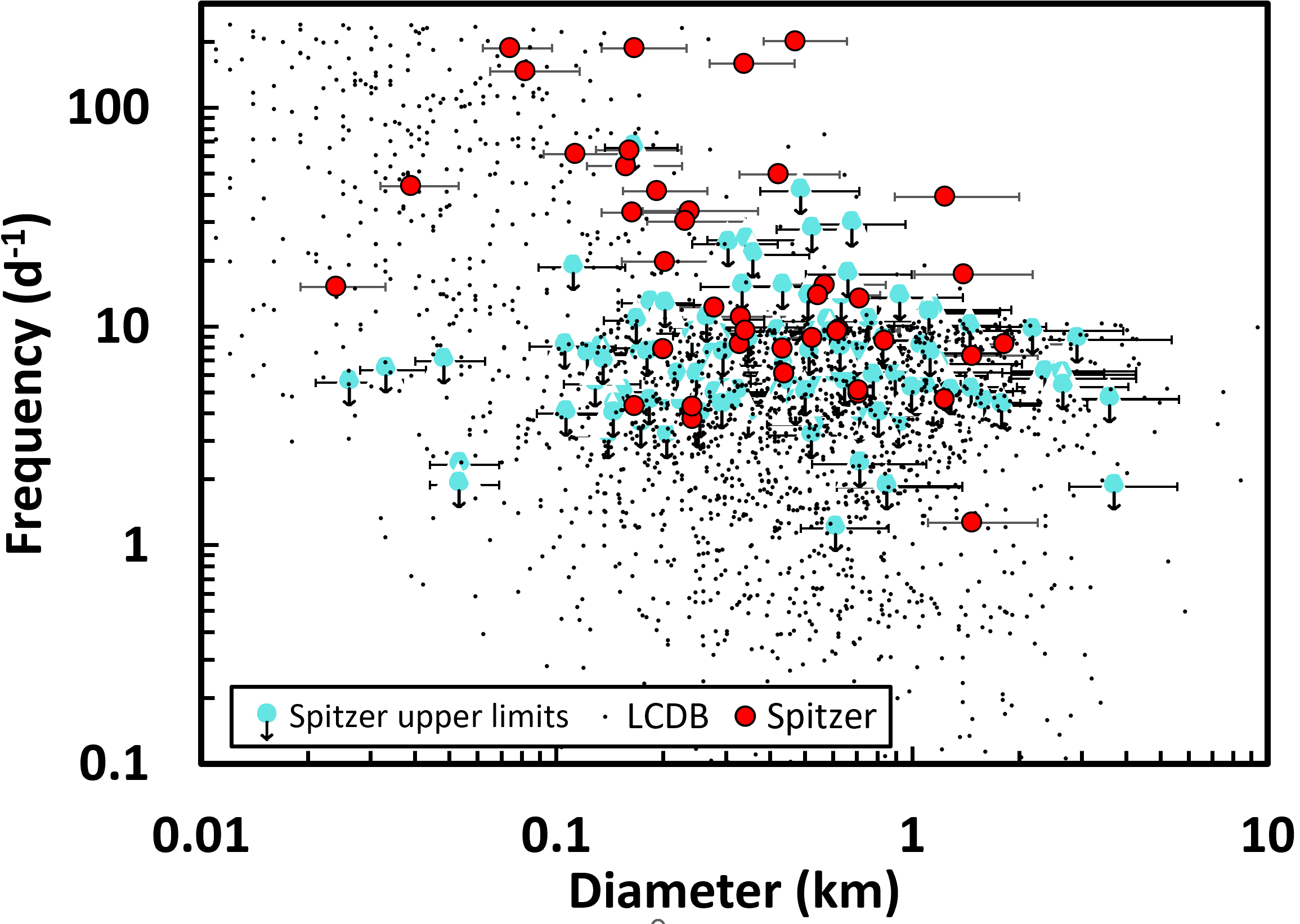}
    \caption{NEO periods as a function of derived diameter. Black points are data from the Lightcurve Database \citep{2009LCDB}. Red points are \Sp-determined periods and diameters from this work, and blue points are upper limits to the periods derived from \Sp\ lightcurves that did not cover more than one period. The 1~$\sigma$ diameter uncertainties are shown for the \Sp\ observations, and the upper limits are shown with a downward pointing arrow for those \Sp\ points where less than one period was observed. For the objects with \Sp-determined periods, the statistical uncertainties are smaller than the plotted points.}
    \label{fig:PvsDplot}
\end{figure}
The NEOs for which we have determined periods or lower limits to the periods are shown in a frequency versus diameter plot in Figure~\ref{fig:PvsDplot}, along with the set of objects from the LCDB.
The ``spin barrier'' represents a theoretical lower limit to the rotation period of an asteroid assuming it to be a strengthless rubble pile of aggregate material. This is the critical spin rate at which the object would undergo rotational fission. While there is a relative dearth of objects in the size range $0.2<D<10$ km with rotational periods exceeding this so-called barrier at $P<2.2$ h, such objects have been and continue to be discovered (\citealt{2009LCDB}, \citealt{2024Strauss}). Objects with rotation periods shorter than this limit are known as super-fast rotators (SFRs). Objects rotating with such periods and exhibiting no evidence of ongoing rotational fission must have some additional strength beyond that of simple self-gravity acting to resist break-up.

A monolithic object, perhaps a single coherent fragment from collisional disruption of a larger parent body, would have much greater internal strength than a rubble pile, allowing it to rotate at rates beyond the assumed critical spin rate. As the collisional lifetimes of objects of this size range are much shorter than the 
lifetime of the solar system it is likely that they should have been collisionally broken down into rubble piles rather than remaining coherent \citep{deelia2007}. A possibility is that they could be a fragment from a relatively recent collision but to date no such identification has been made. It is perhaps more feasible to find a coherent NEO than a Main Belt Object due to the near-zero collisional probability of an object in near-Earth space.

An alternative hypothesis is that the SFR may resist rotational fission if it has some additional cohesive strength in its internal structure. This is more in line with the literature values for strength derived to date, which are generally of order 100--1000s Pa. To constrain the potential cohesive strengths of these rotating ellipsoids we use a simplified form of the Drucker-Prager model, a three-dimensional model estimating the stresses within a geological material at its critical rotation state \citep{holsapple2007, alejano2012}. From this we derive lower limits on the internal strength required, beyond those of self-gravity and intra-aggregate friction, for objects with rotation periods approaching or exceeding the spin barrier.

Due to the phase angle amplitude effect which causes lightcurve amplitudes to appear greater due to increased shadowing at larger phase angles \citep{zappala1990} we must correct our measured amplitudes following the method previously used in \cite{mcneill2019}. This prevents overestimation of the strength required to resist fission.

Eleven of the NEOs with FAP$<$10\% from Tables~\ref{tab:fullperiods} and \ref{tab:uncon_periods}  were found to have a combination of period and amplitude necessitating some additional cohesive strength beyond those of self-gravity and friction. Without a formal taxonomic classification for these objects we use the albedo for these objects as a stand-in to assign a density of $1700$ or $2500$~kg/m$^3$ for albedo lower or higher respectively than $0.10$. The computed strength values are given in Table\ \ref{table:strength}. One of the objects had less than two fully sampled periods, so is segregated at the bottom of the table and should be considered less reliable than the other NEOs in the table that had greater than two periods sampled.

\begin{deluxetable}{lcr}
\caption{Summary of Objects Requiring Cohesive Strength to Resist Fission}\label{table:strength}
\tablehead{Object   & $\rho$ & Cohesive Strength \\
    & (kg m$^3$) & (Pa)}
\startdata
2000 DP107 & 1700                    & $11_{-4}^{+2}$         \\
2001 UV16  & 2500                    & $945_{-176}^{+231}$    \\
2007 BX48  & 2500                    & $44_{-10}^{+14}$       \\
2008 EY68  & 1700                    & $51_{-6}^{+13}$       \\
2009 DR3   & 1700                    & $46_{-11}^{+9}$        \\
2011 LL2   & 2500                    & $2011_{-398}^{+728}$   \\
2011 XA3   & 2500                    & $75_{-12}^{+15}$       \\
2012 BF86  & 2500                    & $98_{-17}^{+27}$       \\
2012 WK4   & 2500                    & $192_{-33}^{+58}$     \\
2013 VO5   & 2500                    & $151_{-43}^{+36}$      \\
\hline
Less than 2 periods sampled:\\
2006 GU    & 1700                    & $203_{-40}^{+43}$      \\
\enddata
\end{deluxetable}


\section{Conclusions}
We have developed a data processing pipeline and reduced in a uniform way the set of NEO observations made by \Sp/IRAC which used the moving object mode (i.e., tracking the NEO's apparent motion as viewed by \Sp) during the warm mission. We conclude that:
\begin{itemize}
\item We present the final and definitive \Sp\ NEO flux catalog from the major survey programs performed and from smaller projects that targeted NEOs.
\item The NEOs that IRAC targeted were generally bright enough at Band 2 to be detected in single frames, enabling light-curve analyses of multi-frame observations. The observations obtained with the IRAC camera were both sensitive enough to obtain important and reliable infrared photometry of NEOs and to obtain new insights from infrared results. These results also highlight the advantages of pointed observations which can measure a continuous lightcurve from a single sequence, versus survey results.
\item We find 39 \Sp\ NEOs with enough time samples lasting one or more periods to retrieve their lightcurves, with 13 having two or more periods sampled and the L-S FAP $<$10\% Another 128 NEOs had only incomplete lightcurves and lower limits were derived.  The remainder of the sample had a small number of observations where only a mean flux could be determined.
\item The lightcurve-diameter distribution of the full-period NEOs  resembles that for previously published NEOs.
\item The shortest lightcurve of our set is 0.1192~hr; altogether in this set we found 25 ``super-fast rotators" with periods under 2.2~hr. The longest period was 6.39~hr. Some of of the lightcurves are complex with multiple peaks during one rotation, and deserving of further analysis.

\item For all of the NEOs we have constructed mosaics and performed photometry to measure the mean flux during the period of observation. These values will be used along with optical magnitudes to fit a thermal model to the NEO and derive estimated diameters and albedos. The lightcurve measurements will allow estimates of the uncertainty of the thermal modeling based on single flux values. This work is ongoing \citep{2023allen} and will be presented in a subsequent paper \citep{2025allen}.
\item Eleven of the NEOs with periods determined with FAP$<$10\%  were found to have a combination of period and amplitude necessitating some additional cohesive strength
beyond those of self-gravity and friction. We estimated the lower limits of the cohesive strengths required for these NEOs.
\end{itemize}

\begin{acknowledgments}
We acknowledge support for this program from the NASA YORPD program under grant 80NSSC22K0243. This work is based on observations made with the
\Sp\ Space Telescope, which was operated by the Jet
Propulsion Laboratory, California Institute of Technology under a contract with NASA. Support for this work
was provided by NASA through an award issued by
JPL/Caltech.

This research has made use of NASA’s Astrophysics Data
System and the arXiv preprint server.
\end{acknowledgments}

%

\vspace{5mm}
\facilities{Spitzer(IRAC)}


\software{astropy \citep{astropy:2013,astropy:2018,astropy:2022}, matplotlib \citep{Hunter2007}, Jupyter notebook \citep{Kluyver2016jupyter}, photutils \citep{larry_bradley_2023_7946442}, numpy \citep{harris2020array}}



\bibliography{references}{}
\bibliographystyle{aasjournal}

\appendix
\section{NEO Lightcurves and phase curves} \label{appdx:lc}
\begin{figure*}
    \centering
    \includegraphics[width=0.495\linewidth]{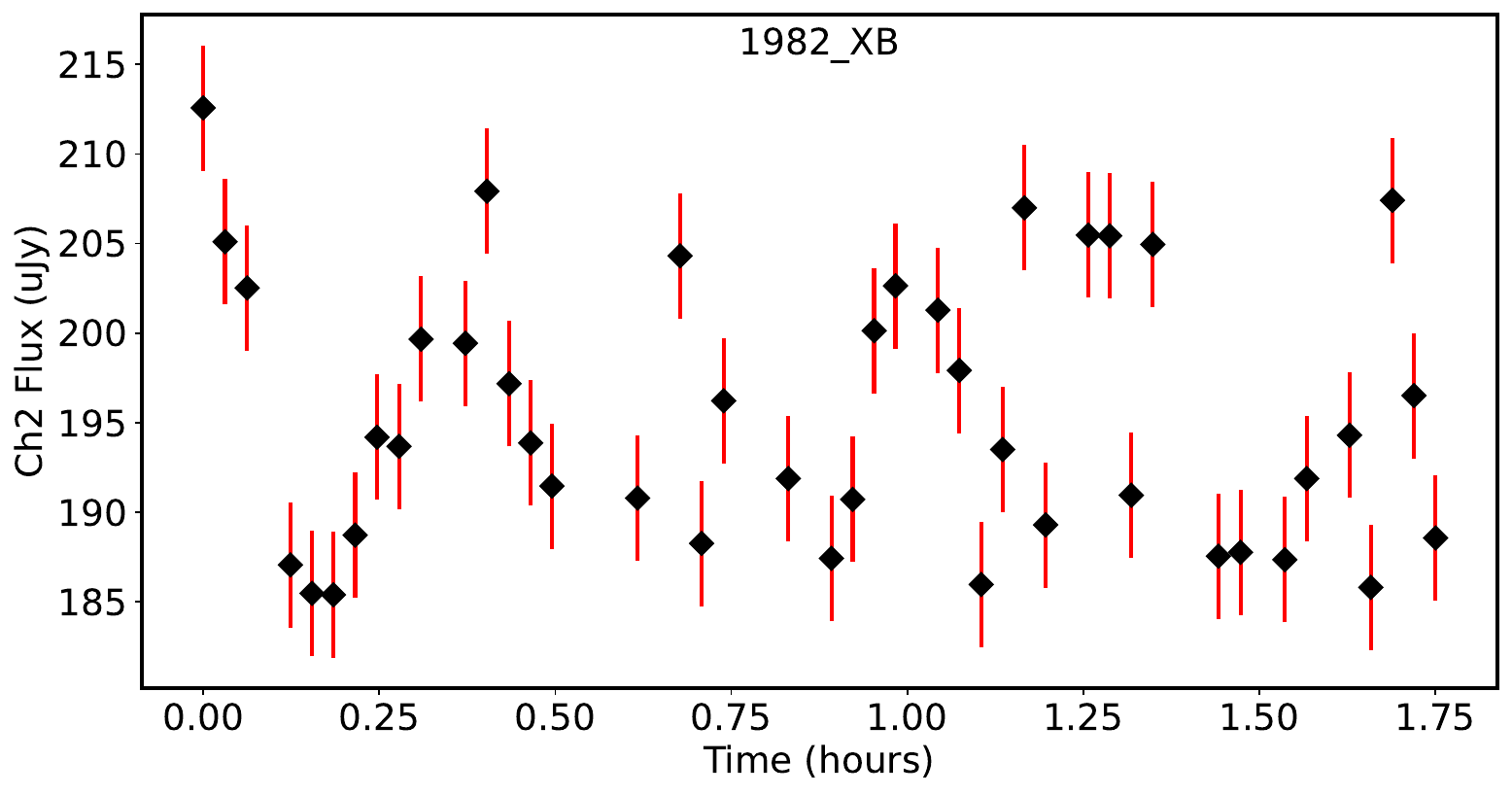}
    \includegraphics[width=0.495\linewidth]{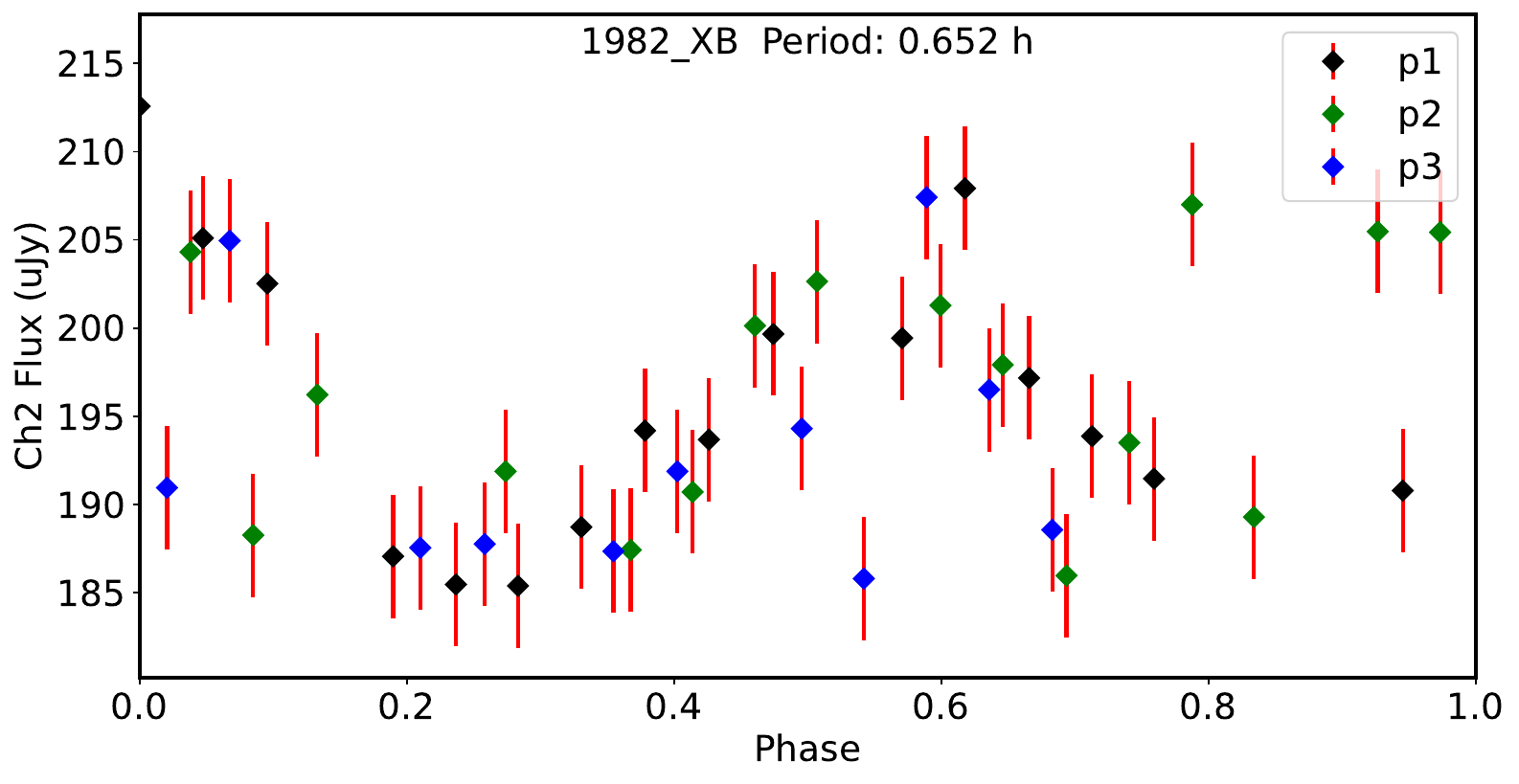}
    \includegraphics[width=0.495\linewidth]{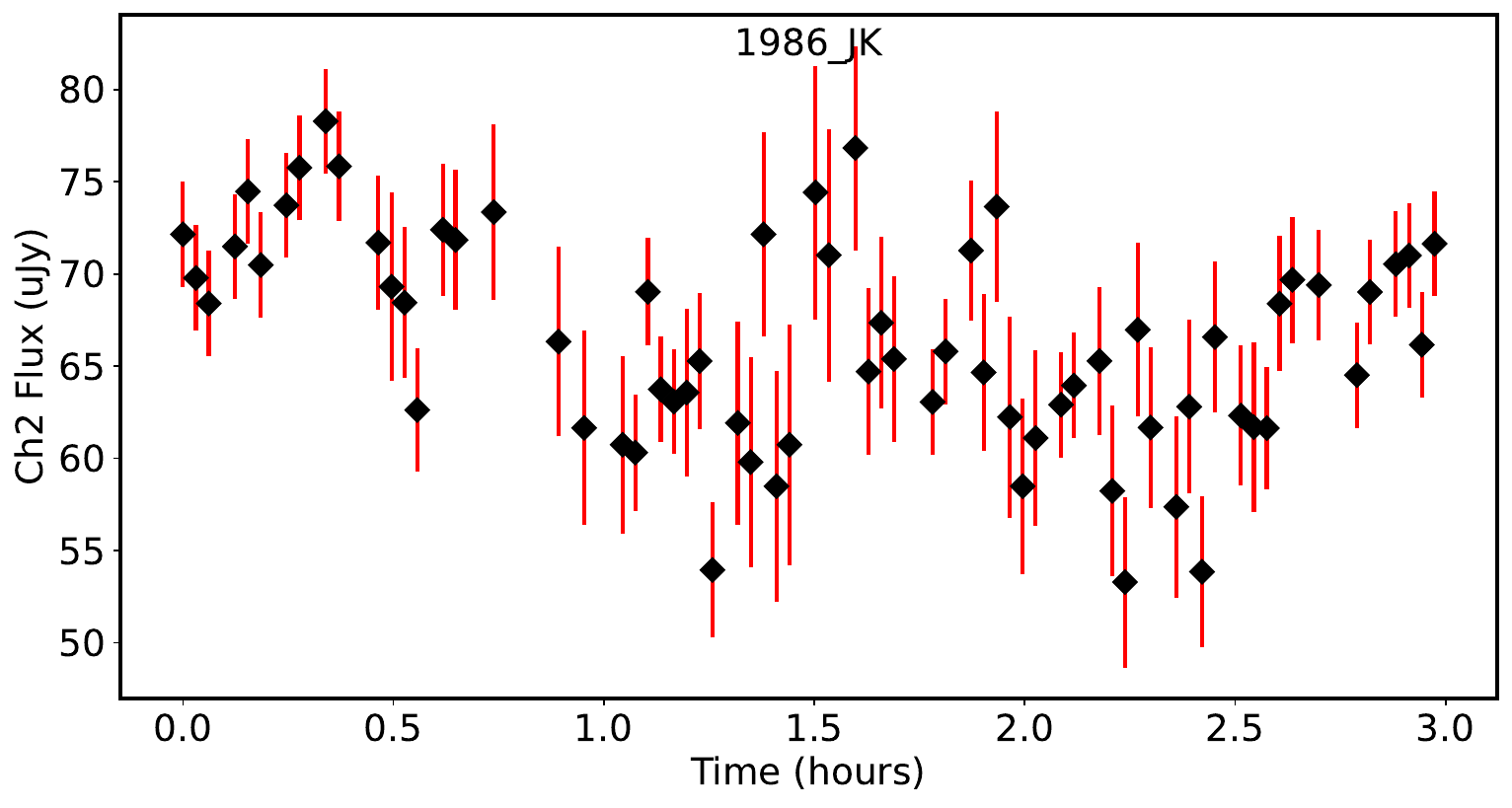}
    \includegraphics[width=0.495\linewidth]{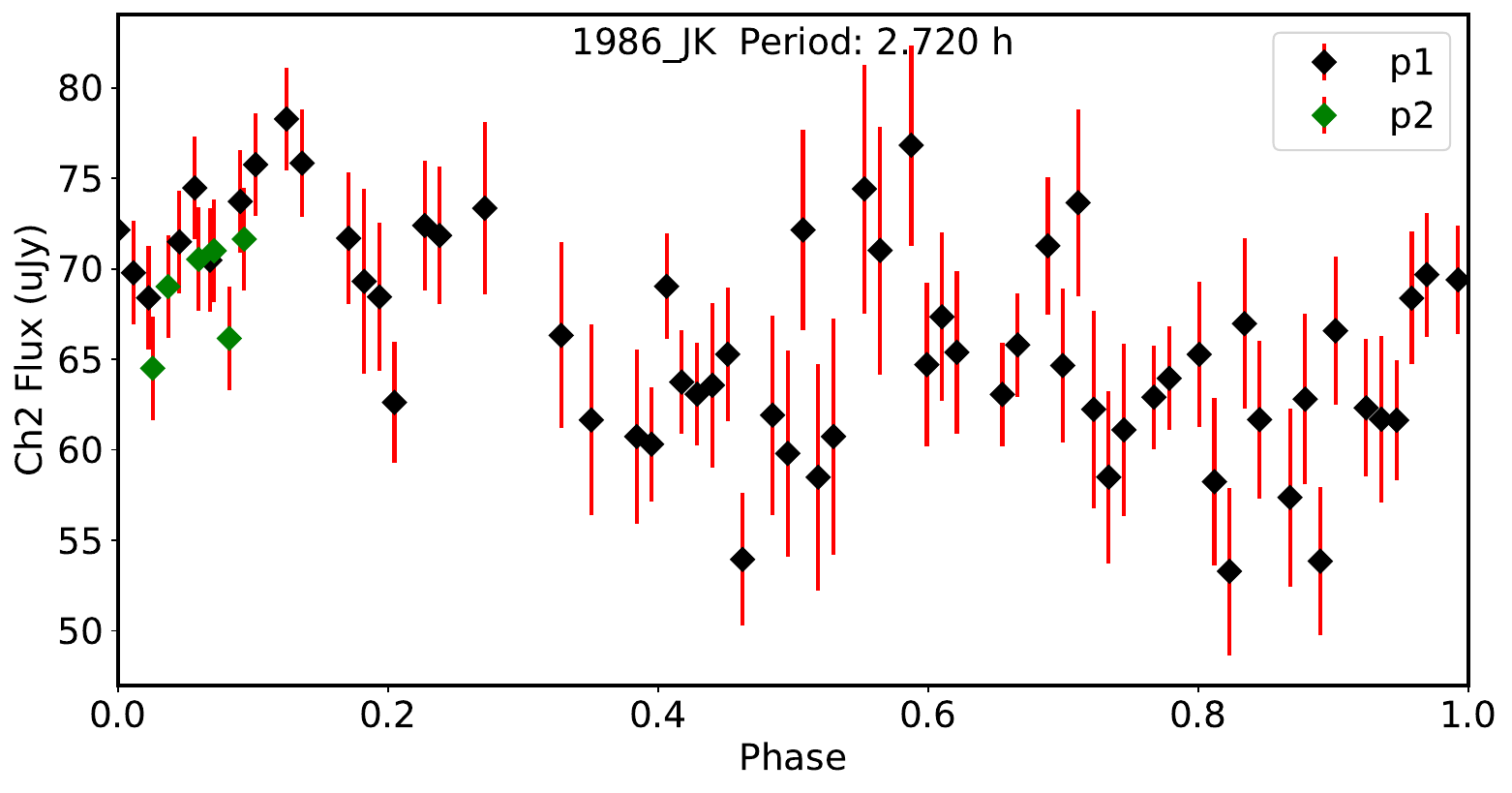}
    \includegraphics[width=0.495\linewidth]{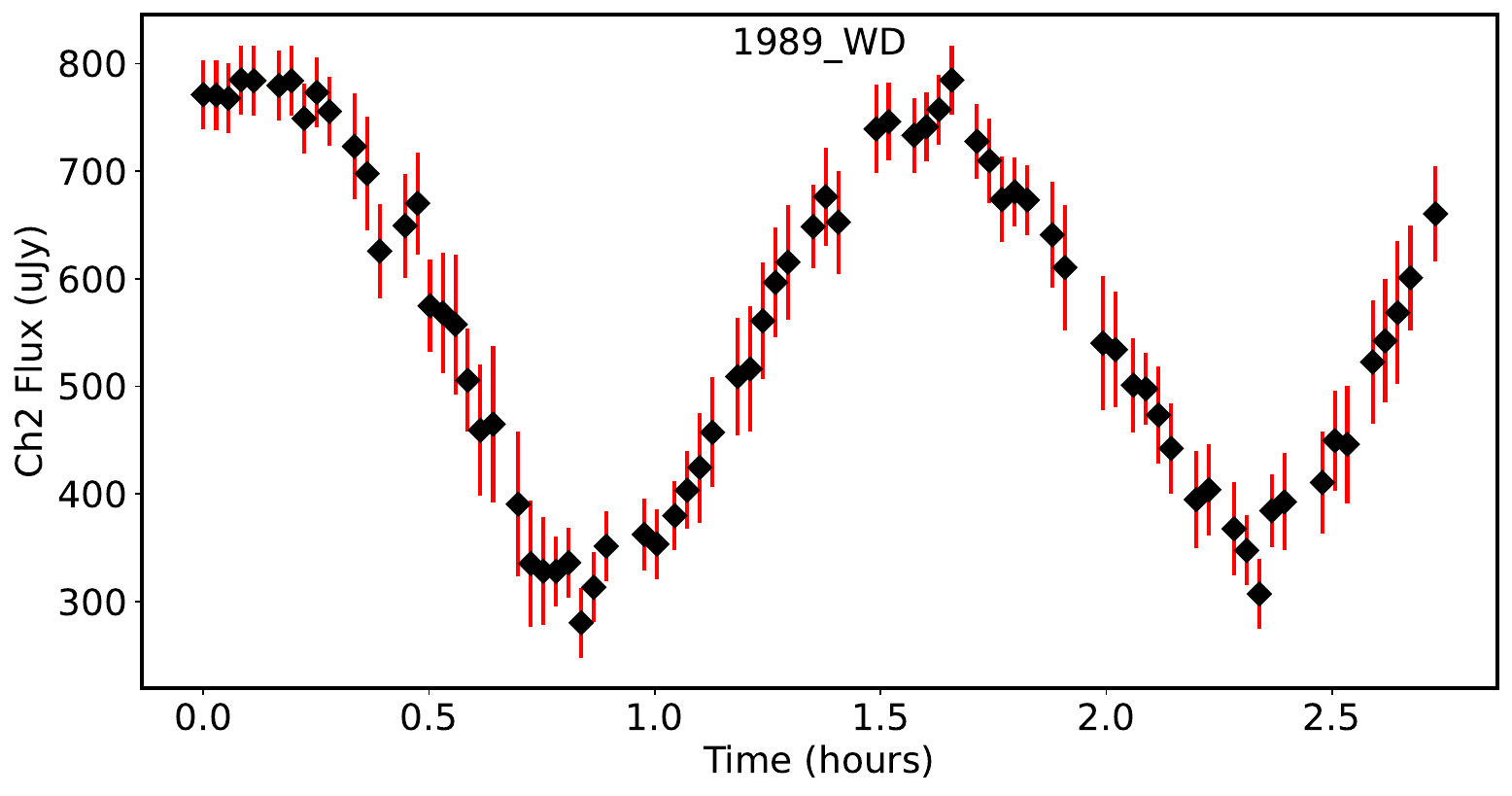}
    \includegraphics[width=0.495\linewidth]{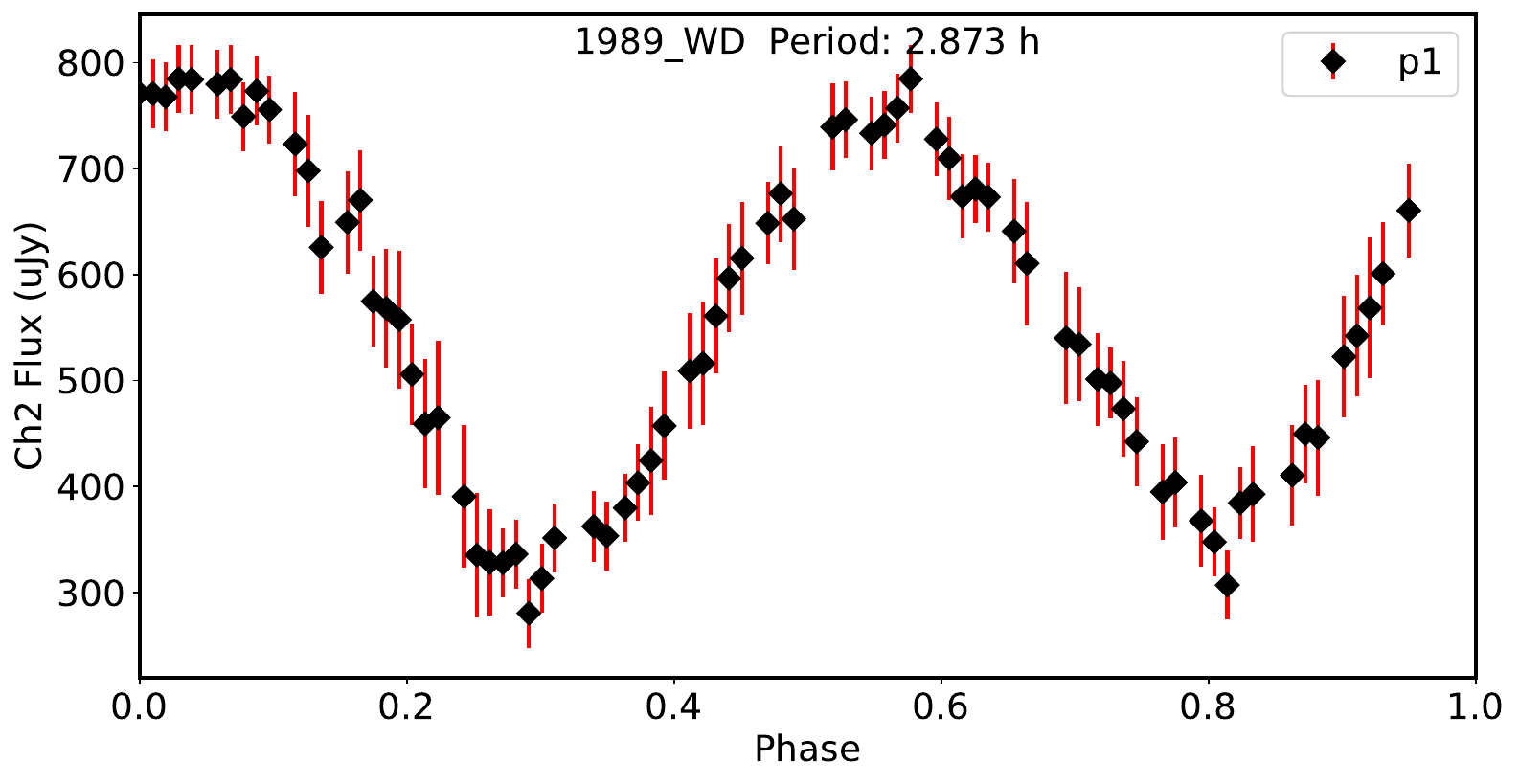}
    \includegraphics[width=0.495\linewidth]{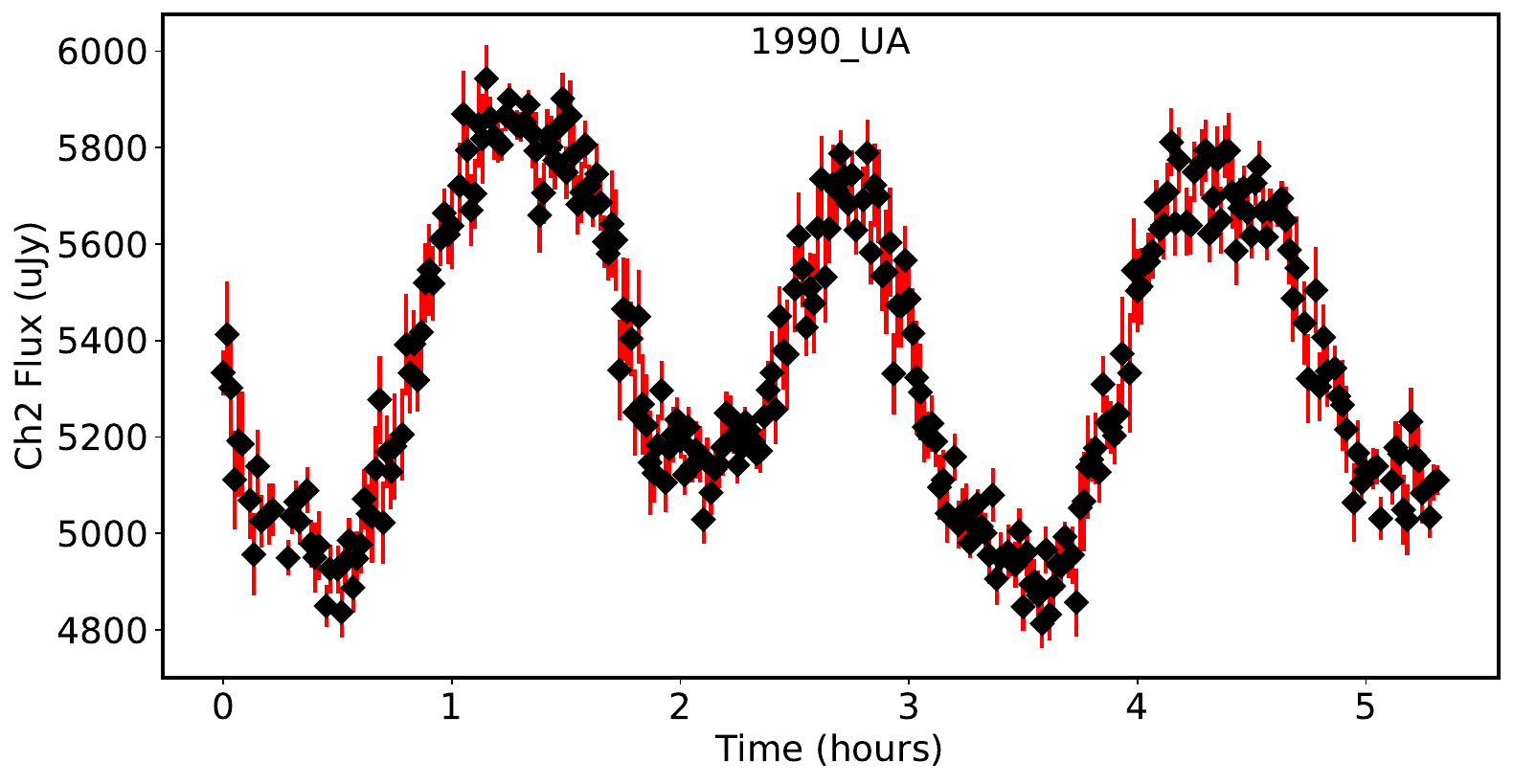}
    \includegraphics[width=0.495\linewidth]{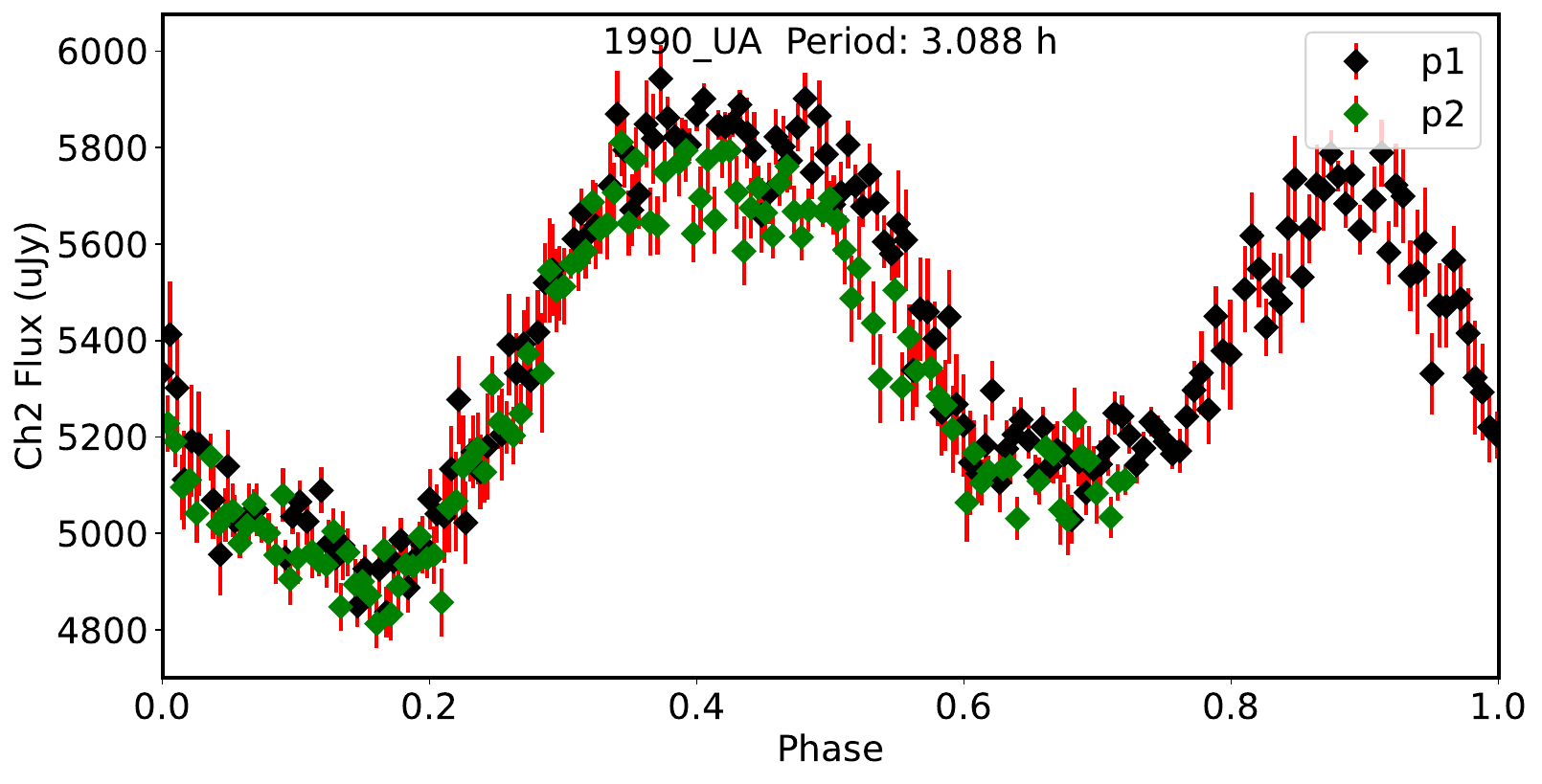}
    \caption{Lightcurves (left column) and phased lightcurves (right column) for sources with one or more periods sampled and periods determined. The periods are plotted with different colors in the plots on the right. }
    \label{fig:lc1}
\end{figure*}

\begin{figure*}
    \centering
    \includegraphics[width=0.495\linewidth]{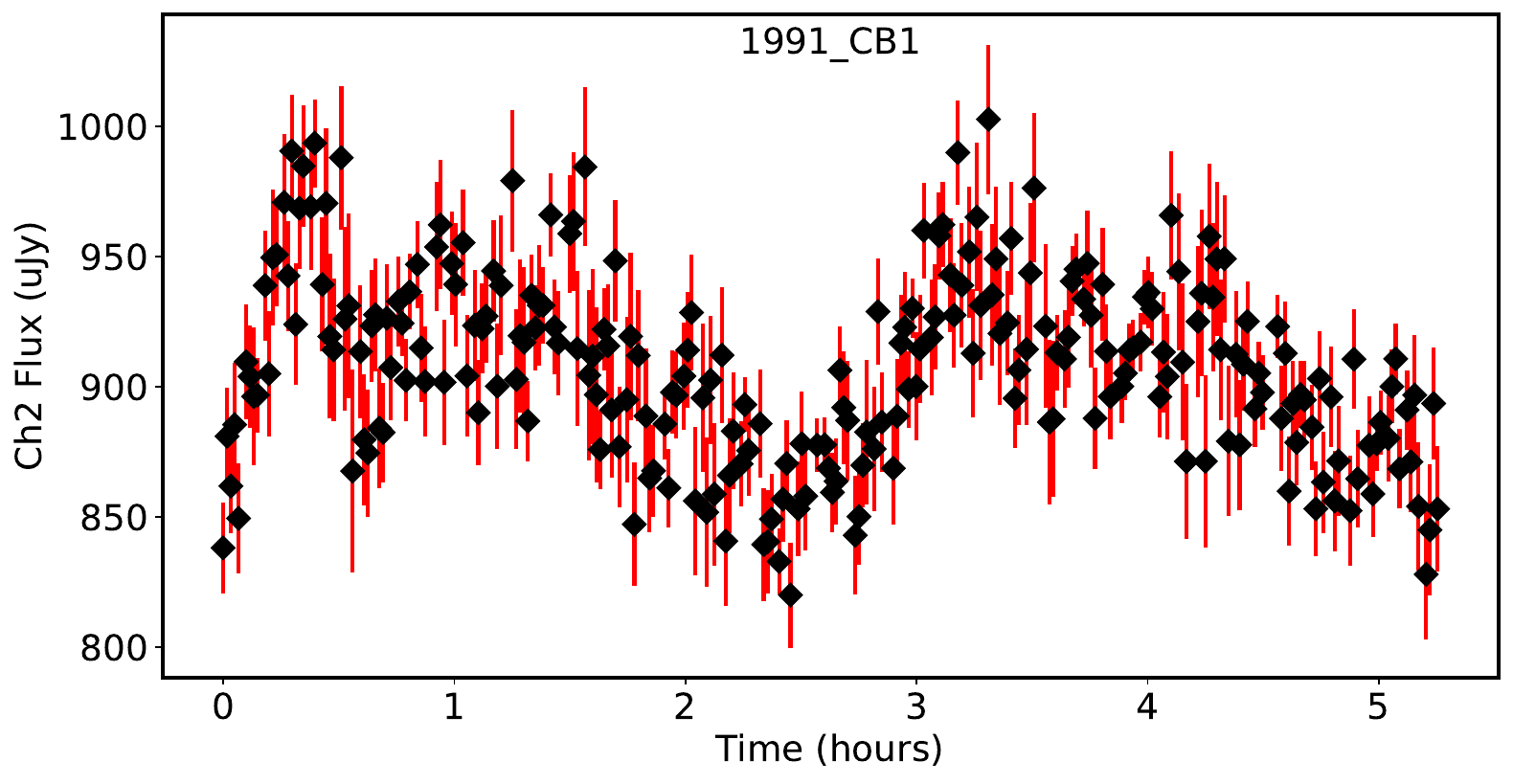}      \includegraphics[width=0.495\linewidth]{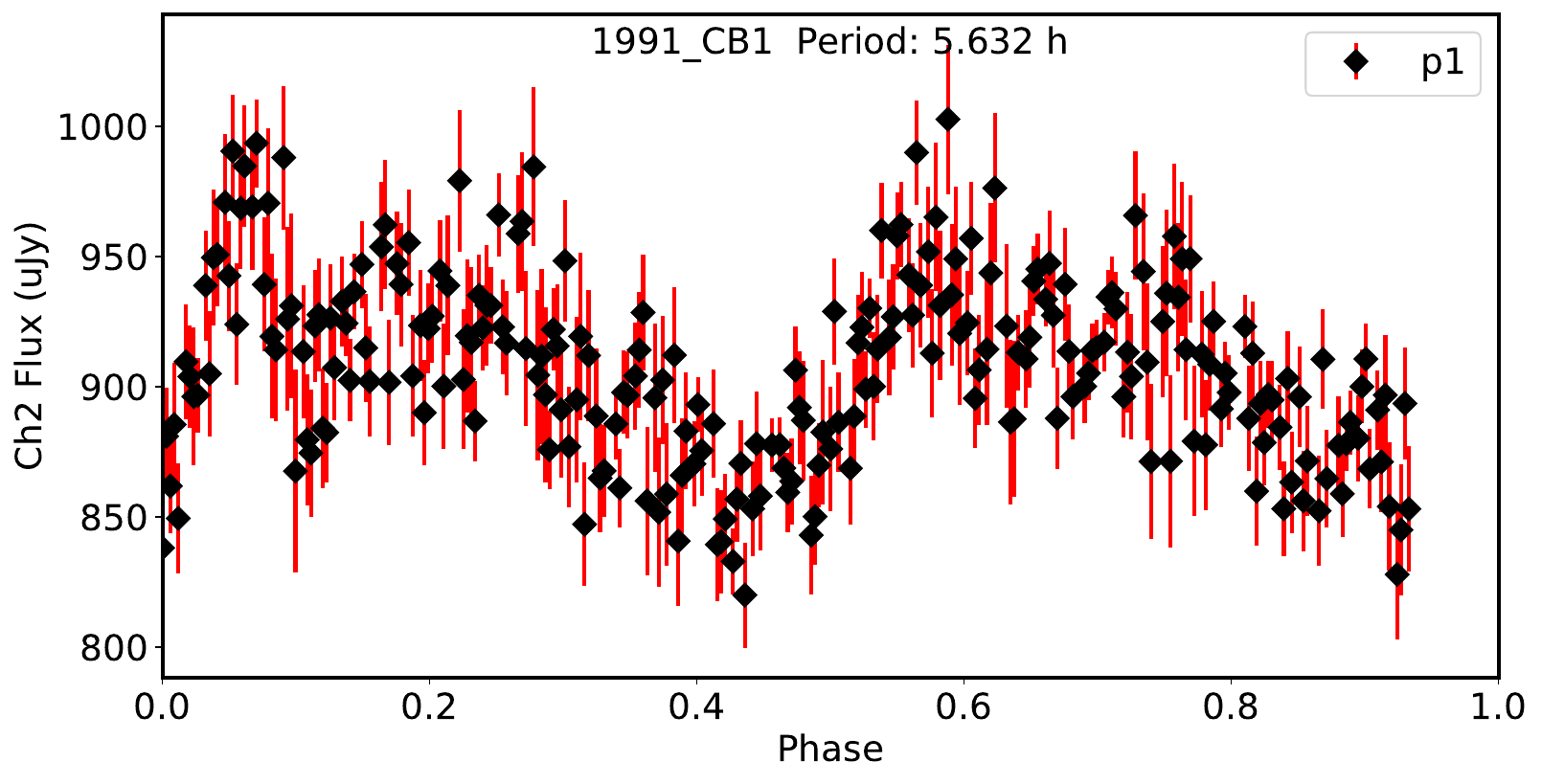} 
   \includegraphics[width=0.495\linewidth]{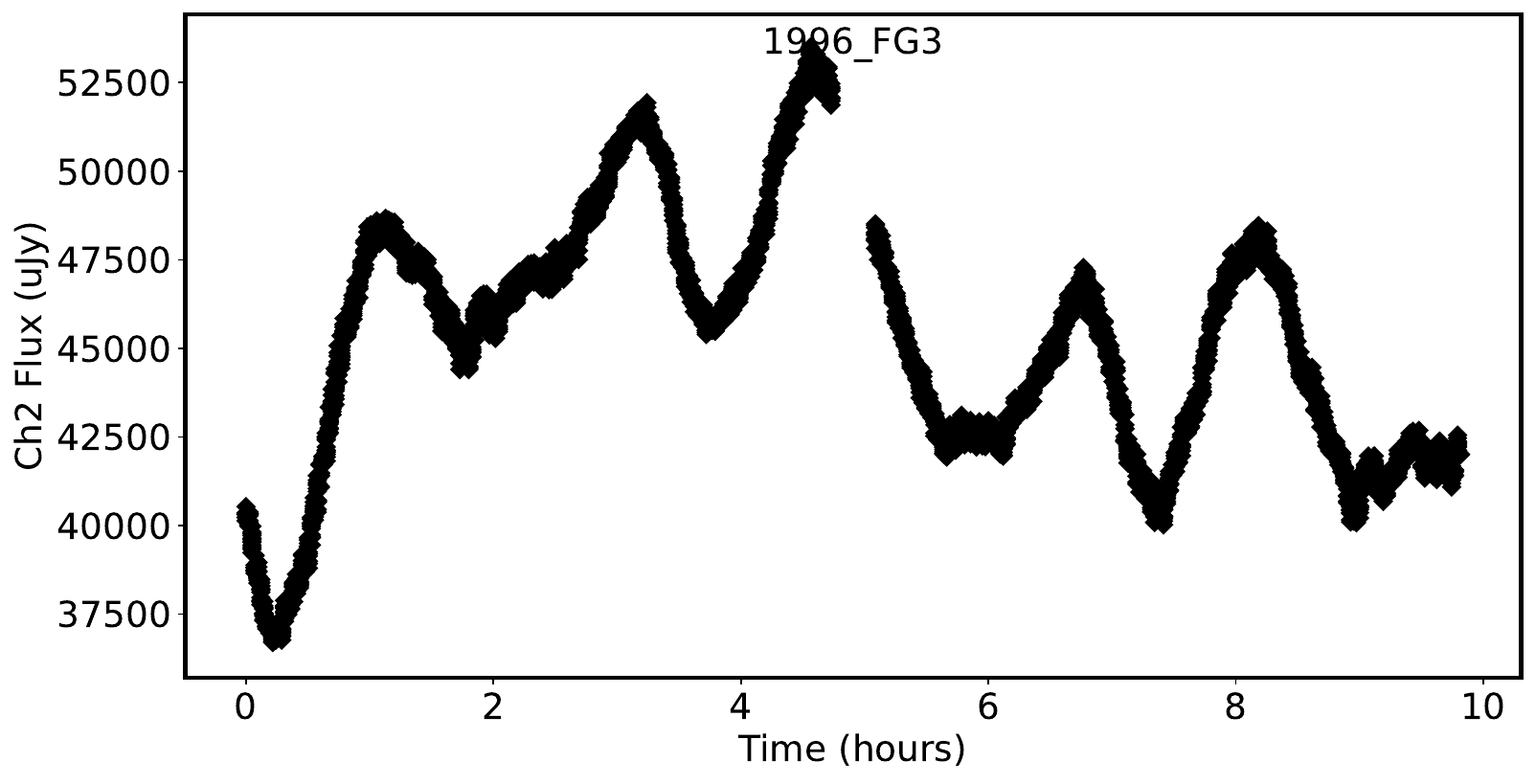}
    \includegraphics[width=0.495\linewidth]{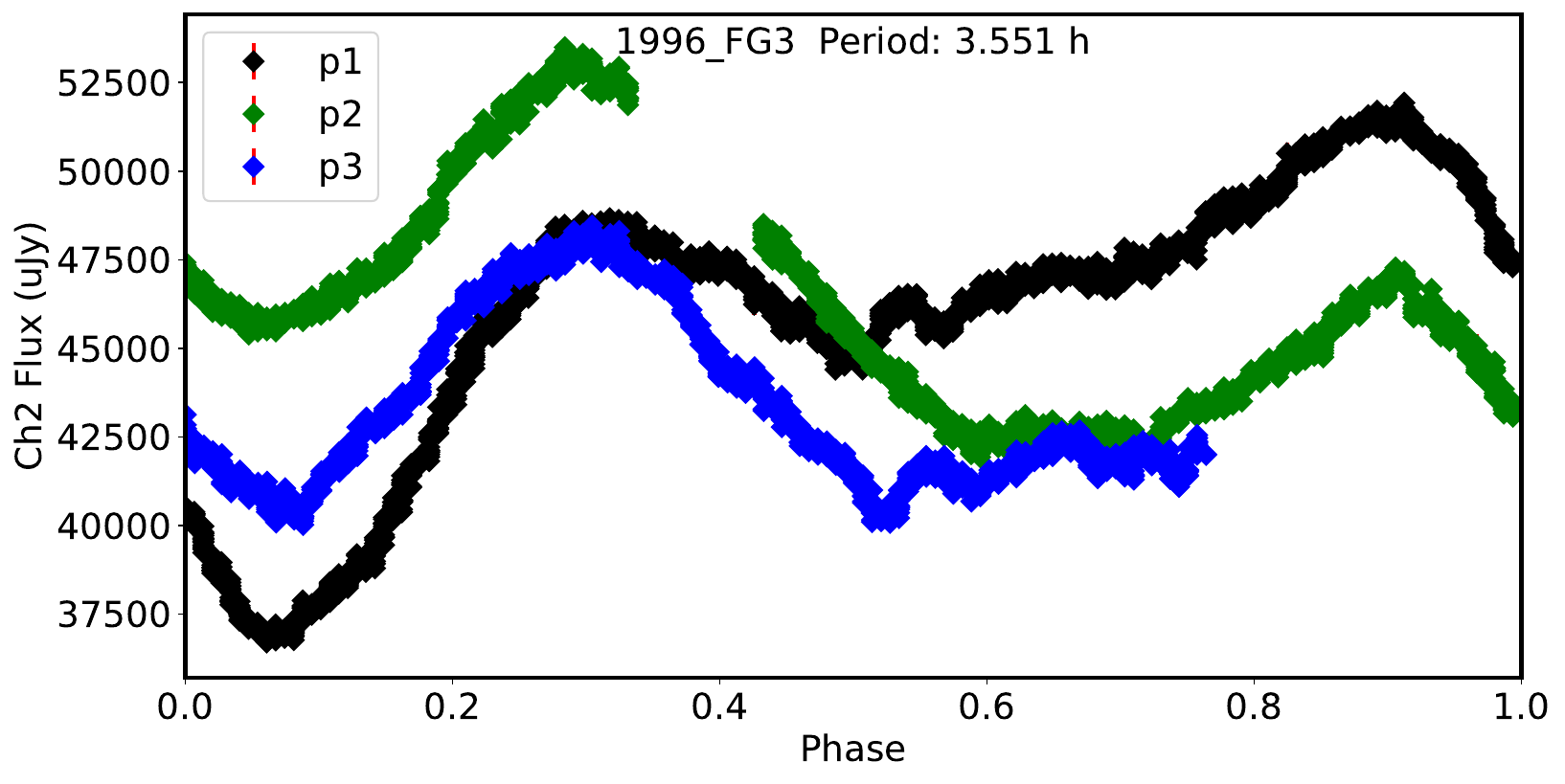}
   \includegraphics[width=0.495\linewidth]{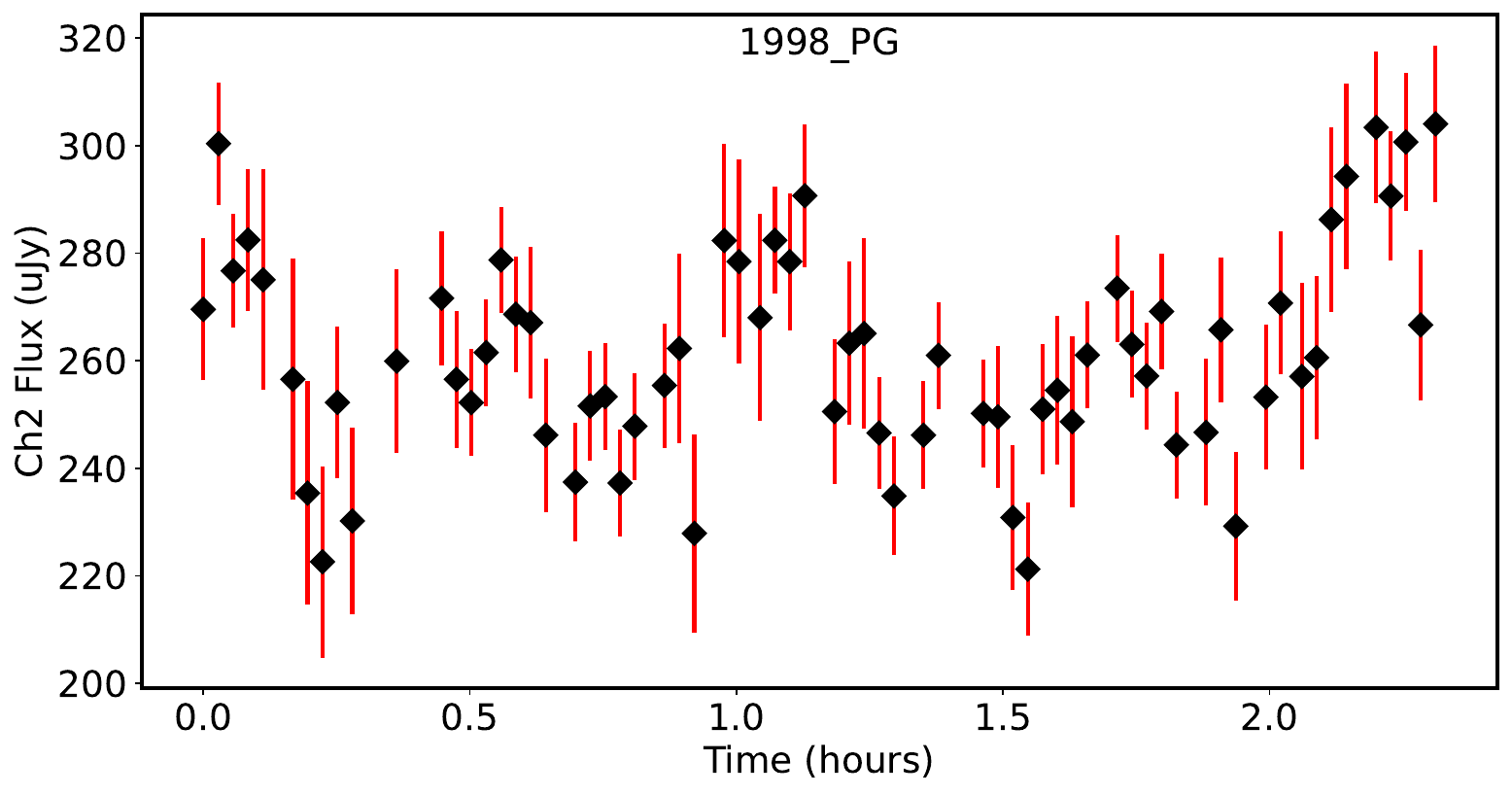}
    \includegraphics[width=0.495\linewidth]{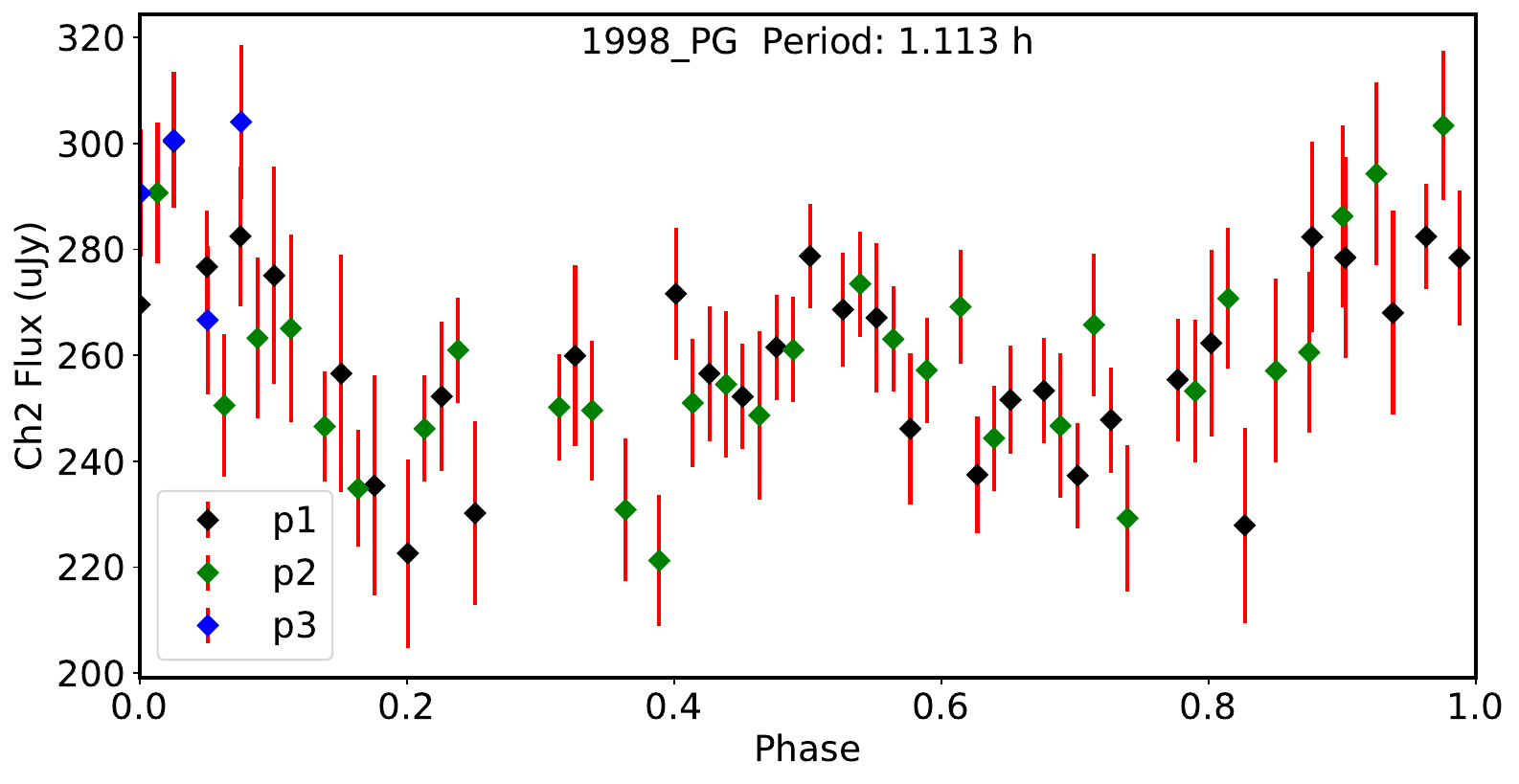}
    \includegraphics[width=0.495\linewidth]{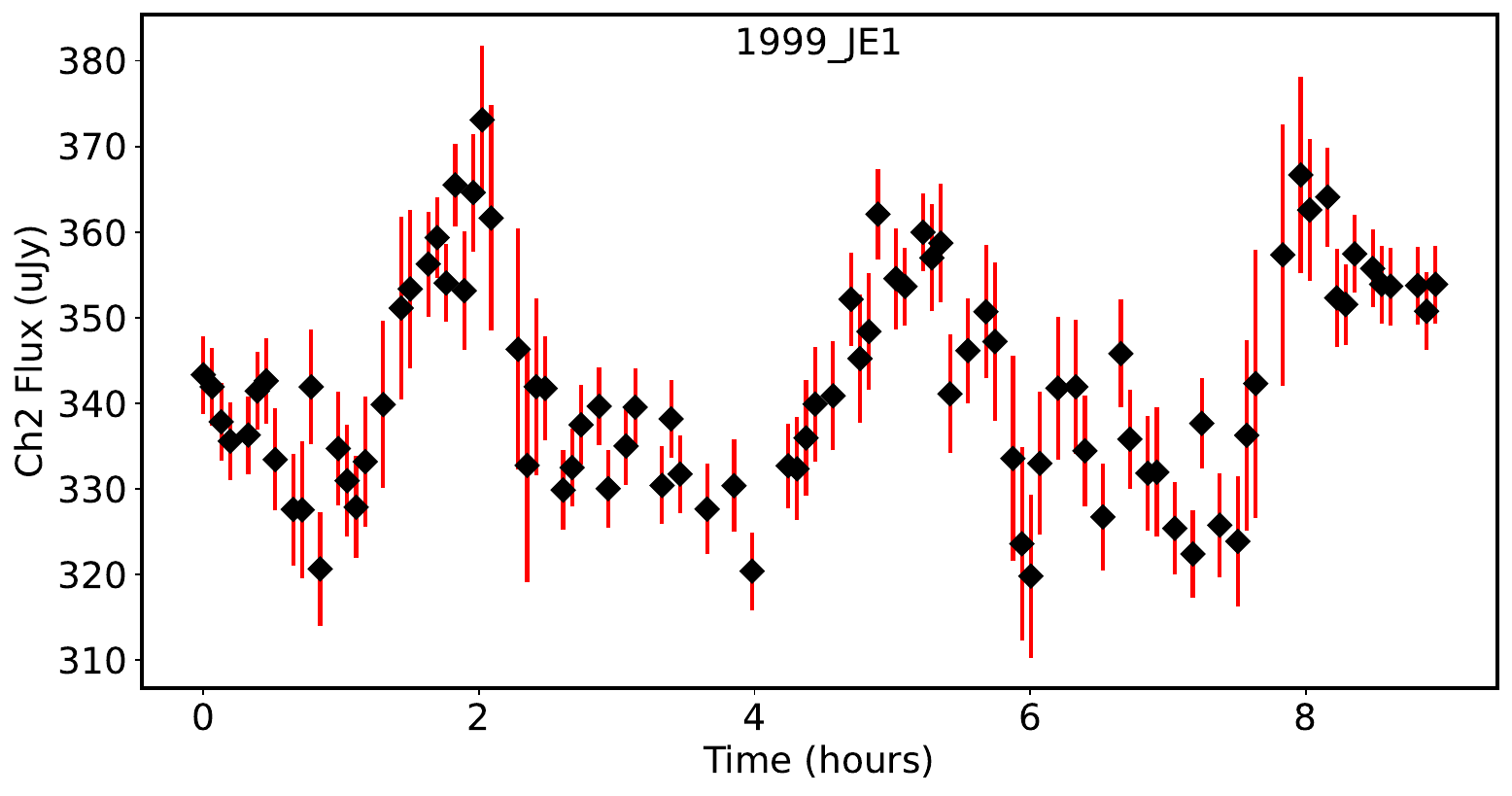}
    \includegraphics[width=0.495\linewidth]{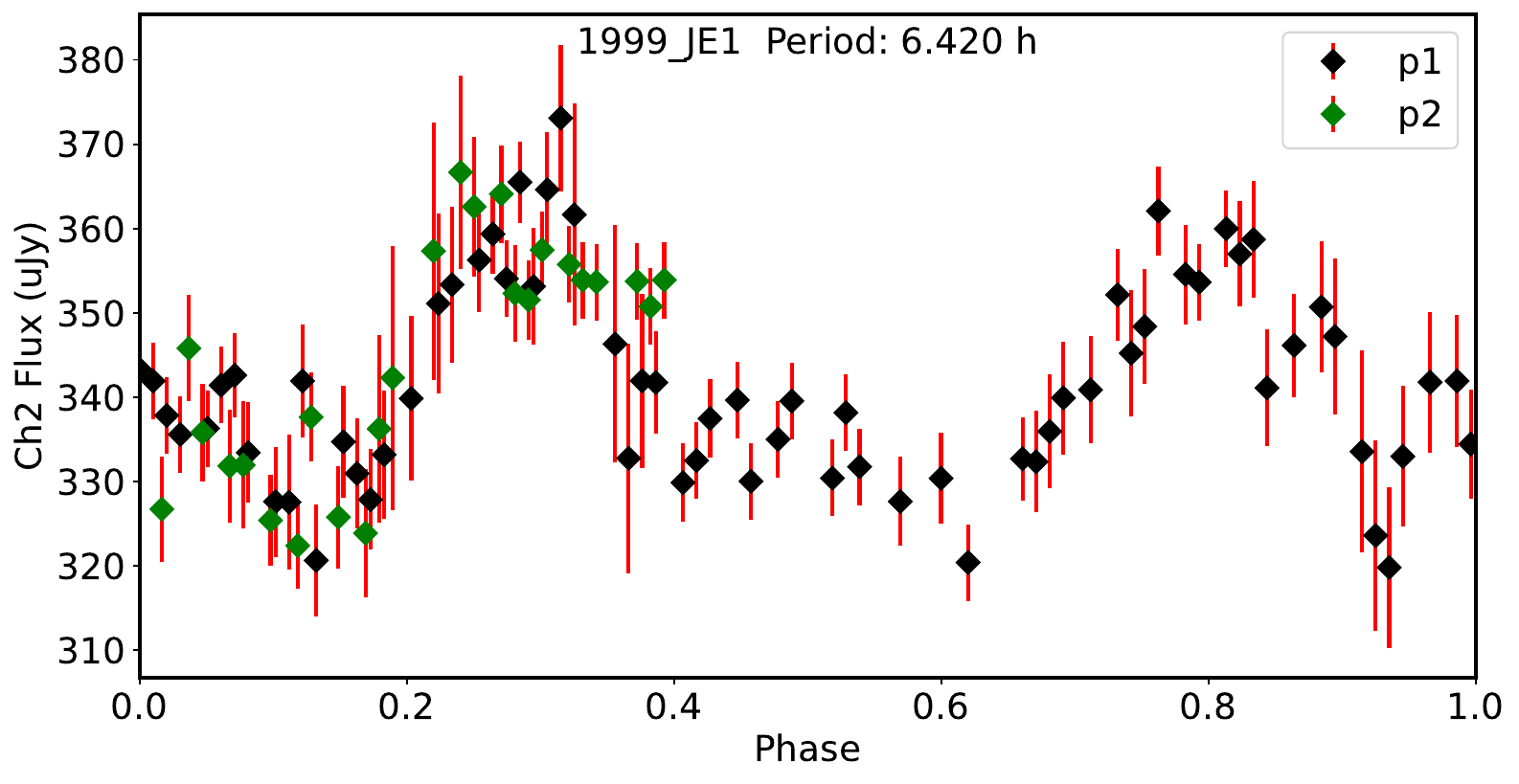}    \includegraphics[width=0.495\linewidth]{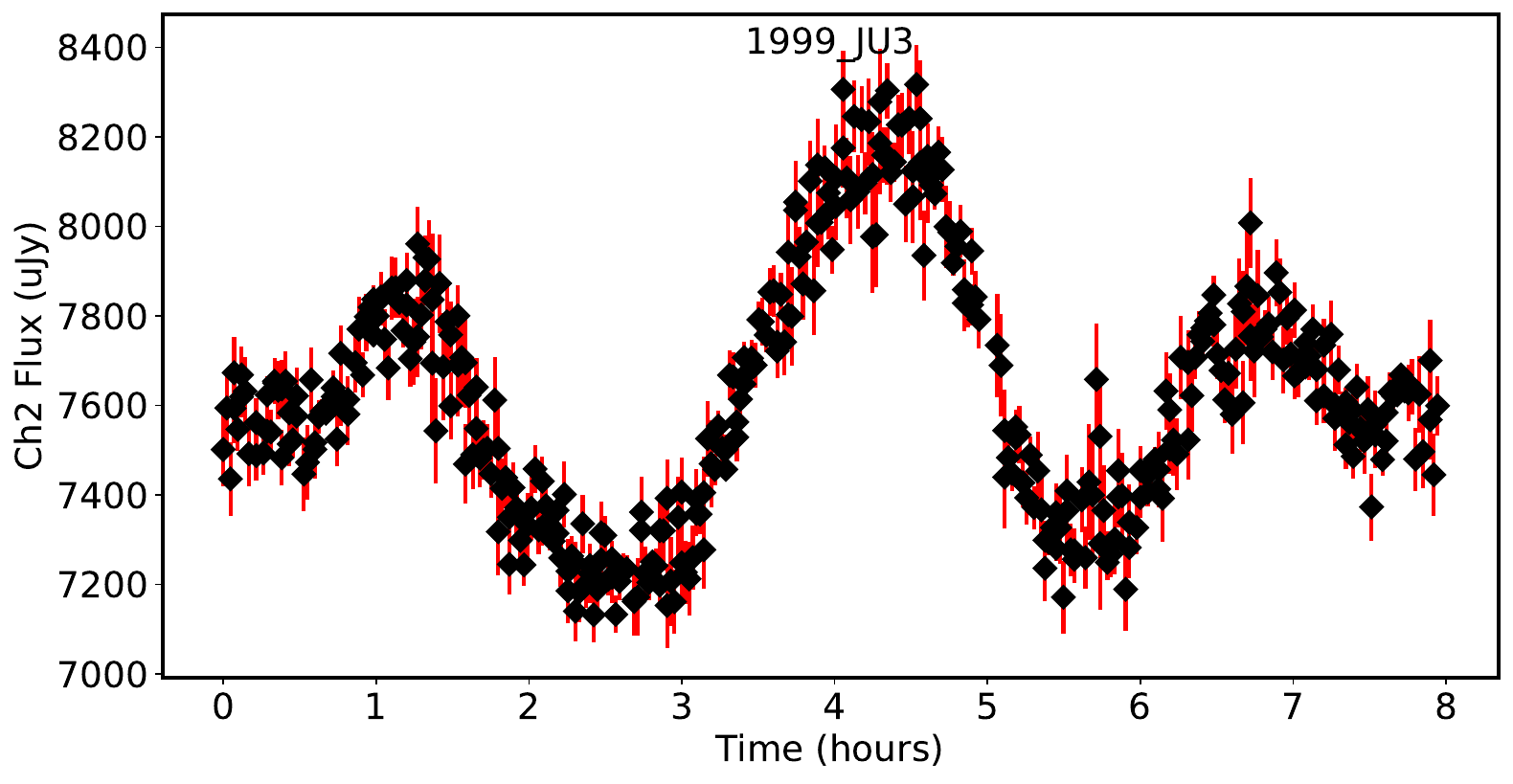}
    \includegraphics[width=0.495\linewidth]{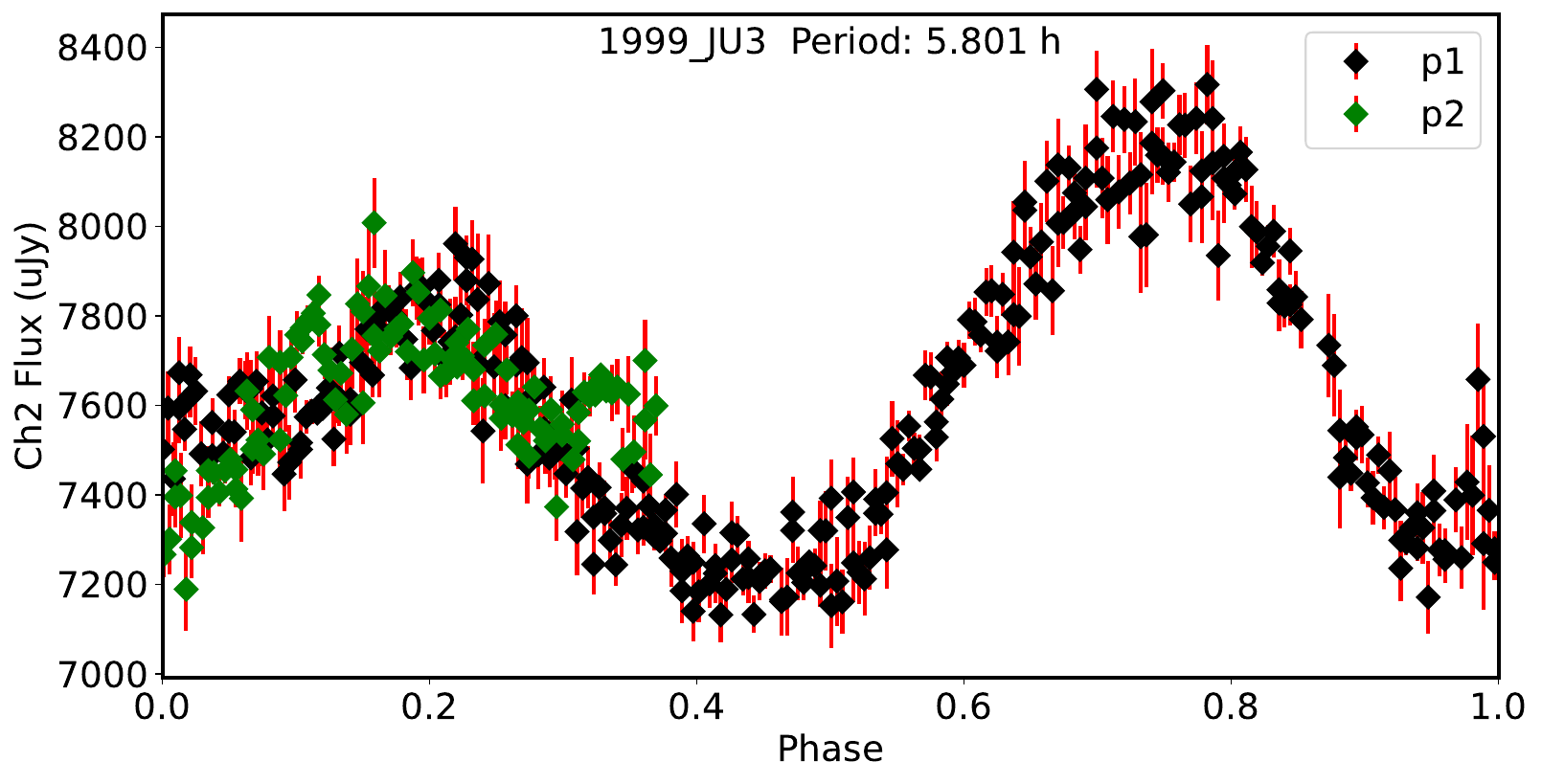}        \caption{Lightcurves (left column) and phased lightcurves (right column) for sources with one or more periods sampled and periods determined. The periods are plotted with different colors in the plots on the right.}
    \label{fig:lc2}
\end{figure*}

\begin{figure*}
    \centering        
\includegraphics[width=0.495\linewidth]{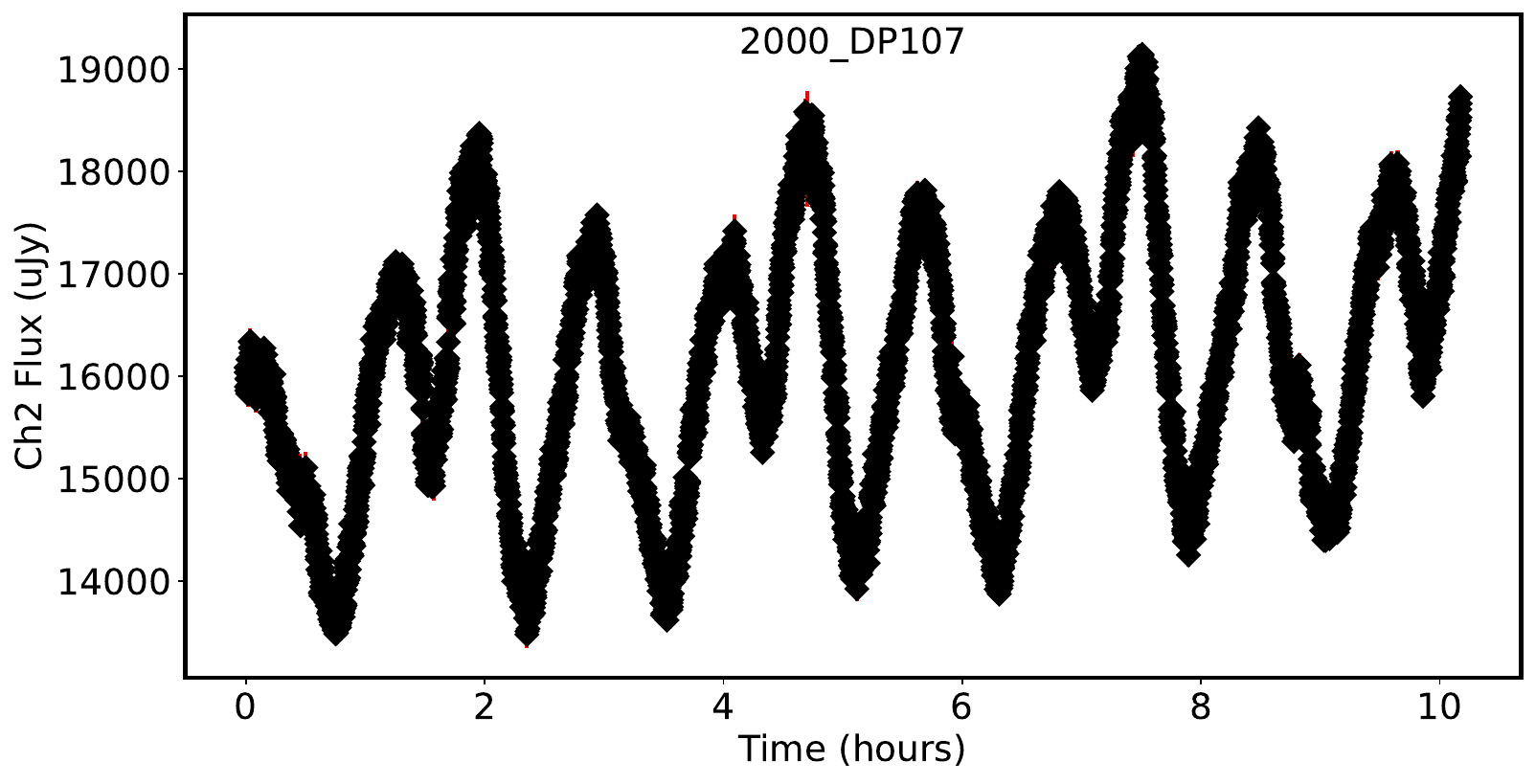}
    \includegraphics[width=0.495\linewidth]{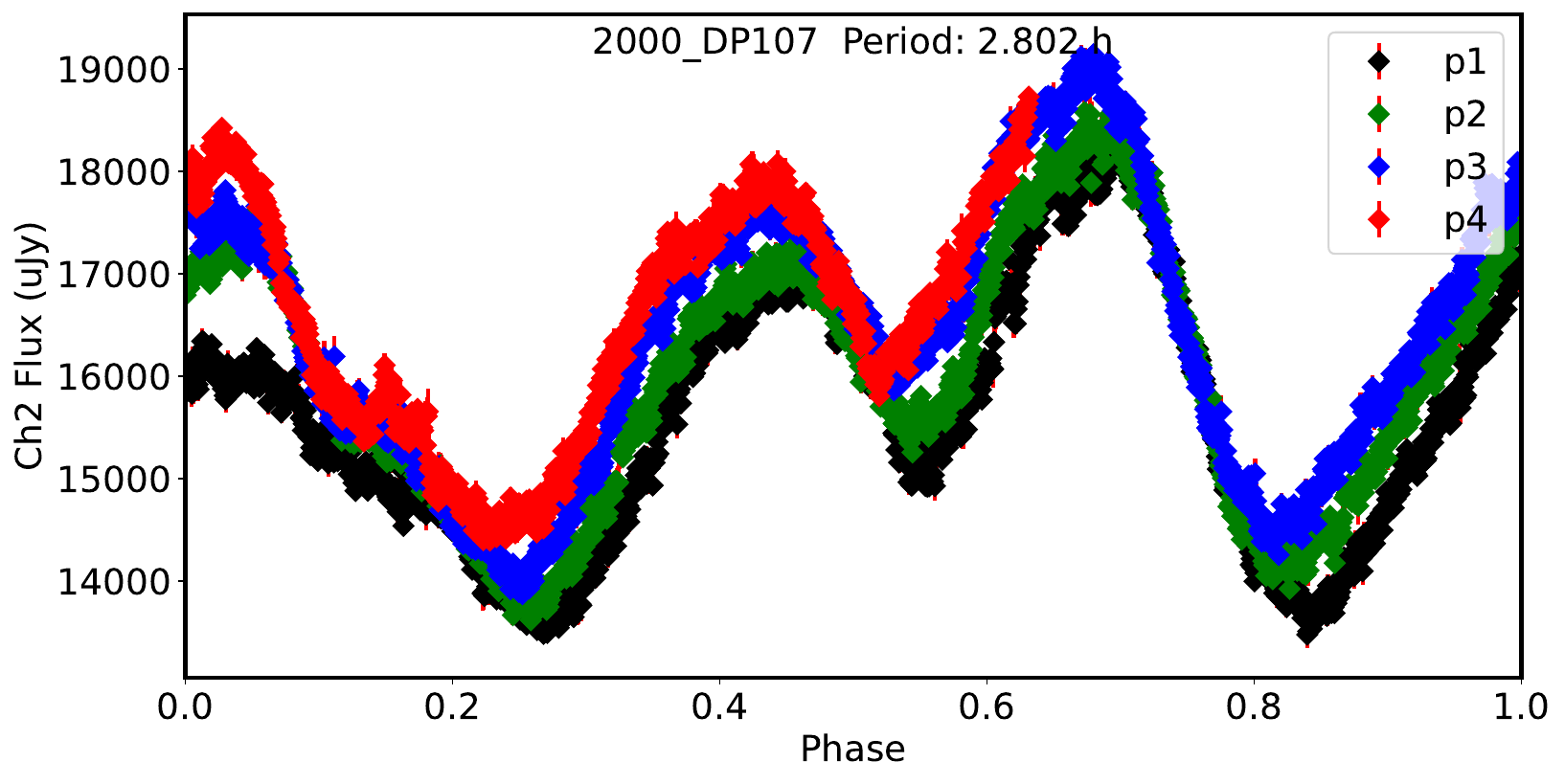}
    \includegraphics[width=0.495\linewidth]{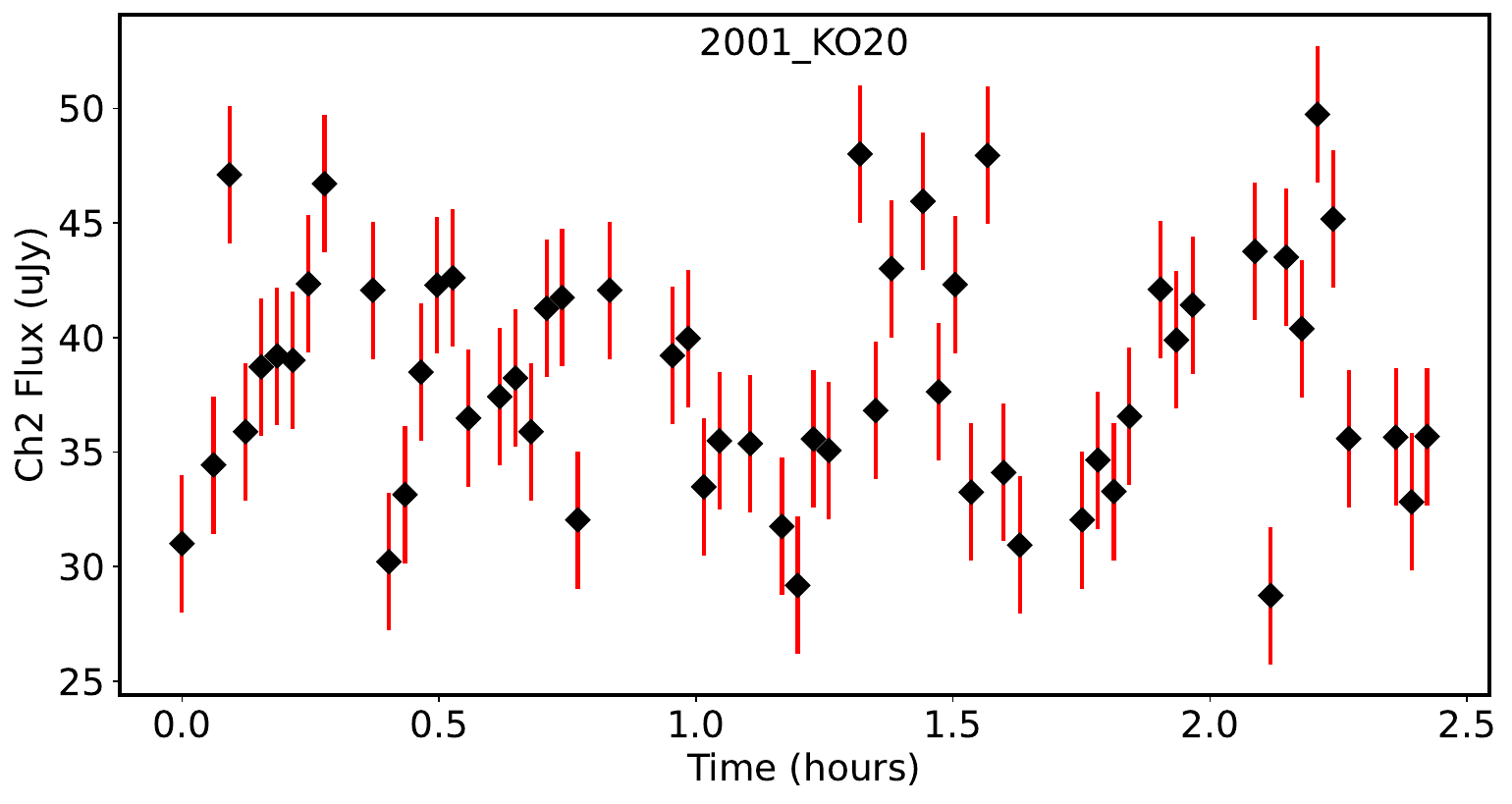}
    \includegraphics[width=0.495\linewidth]{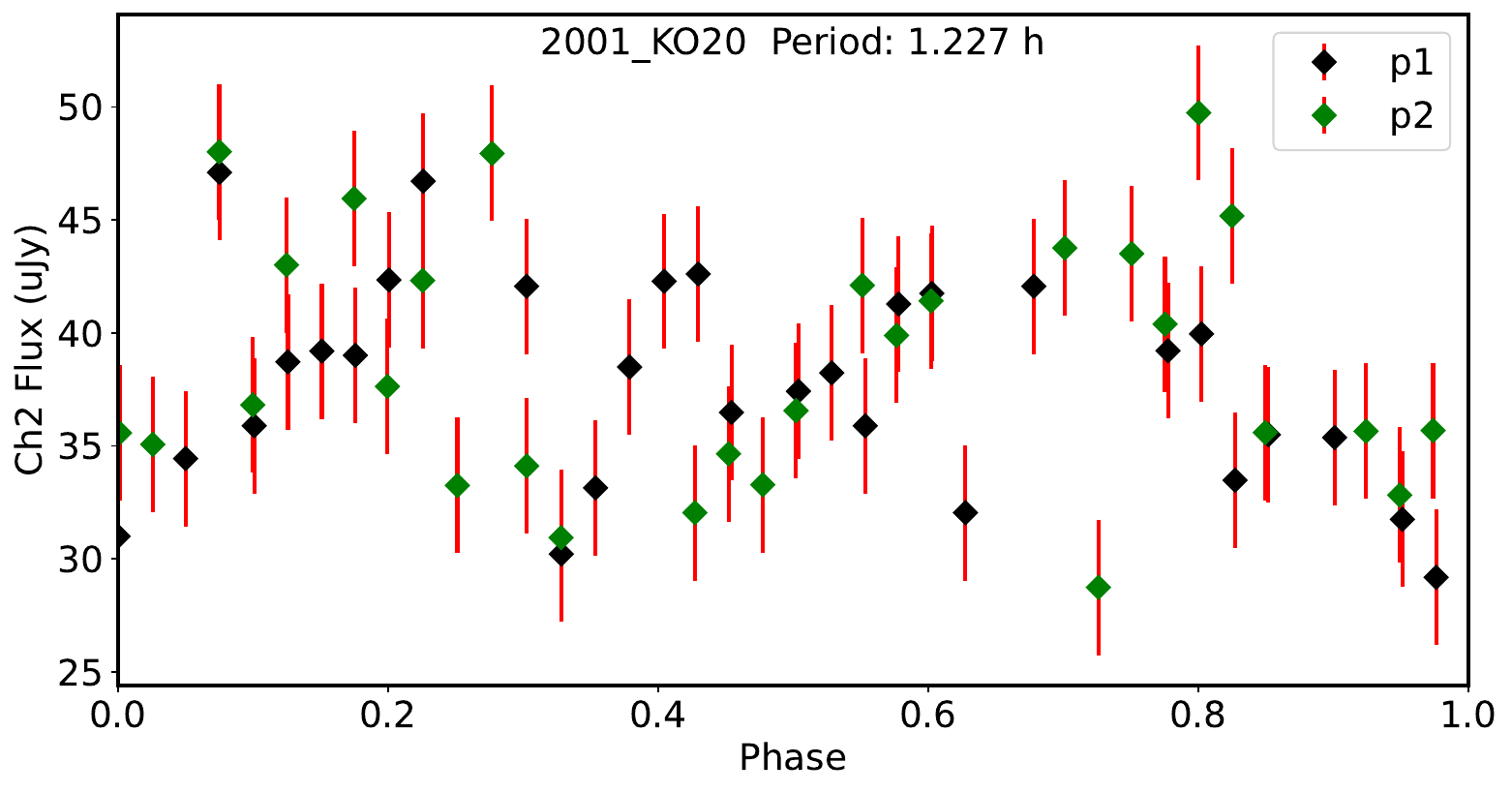}
    \includegraphics[width=0.495\linewidth]{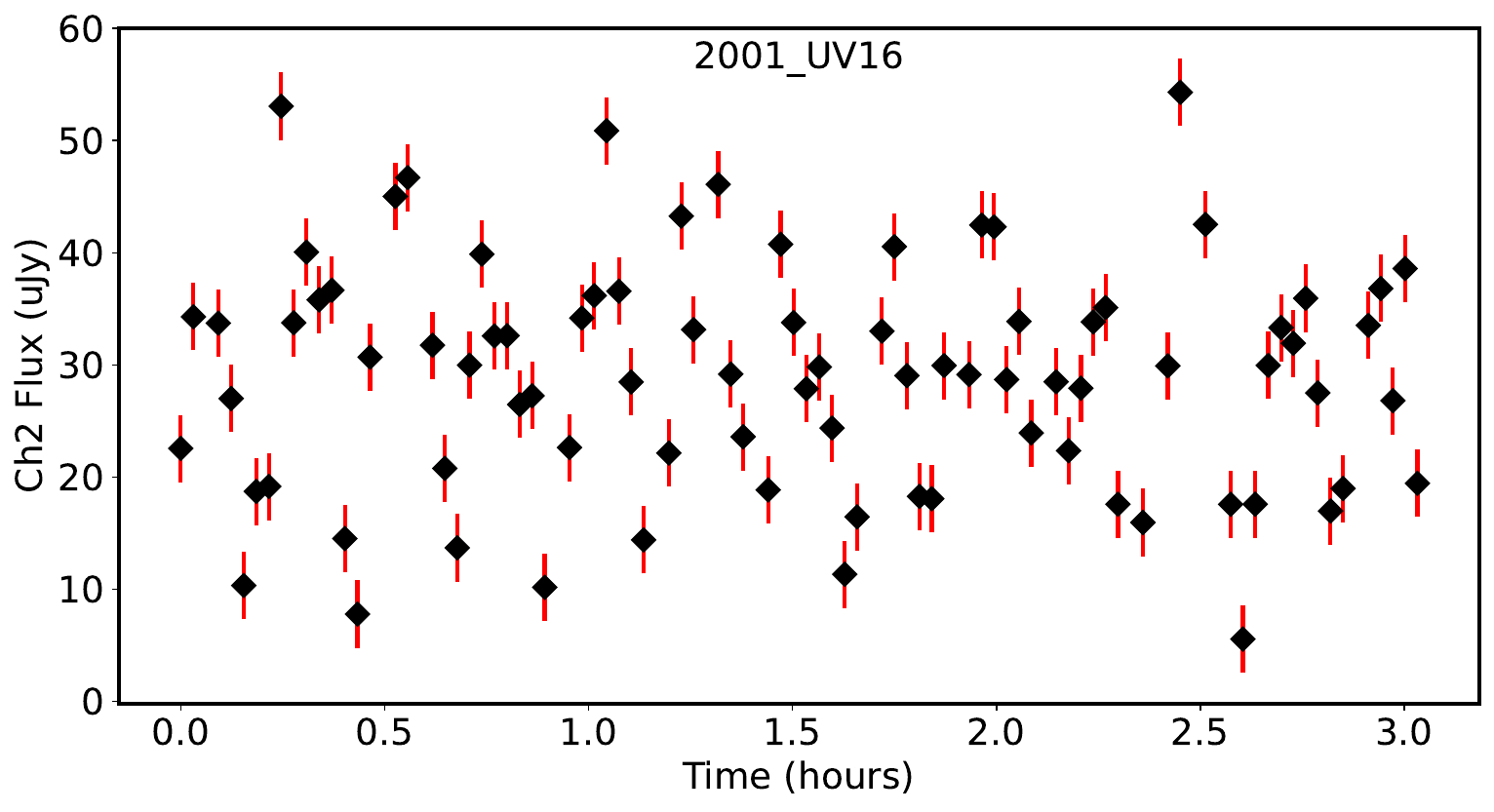}
    \includegraphics[width=0.495\linewidth]{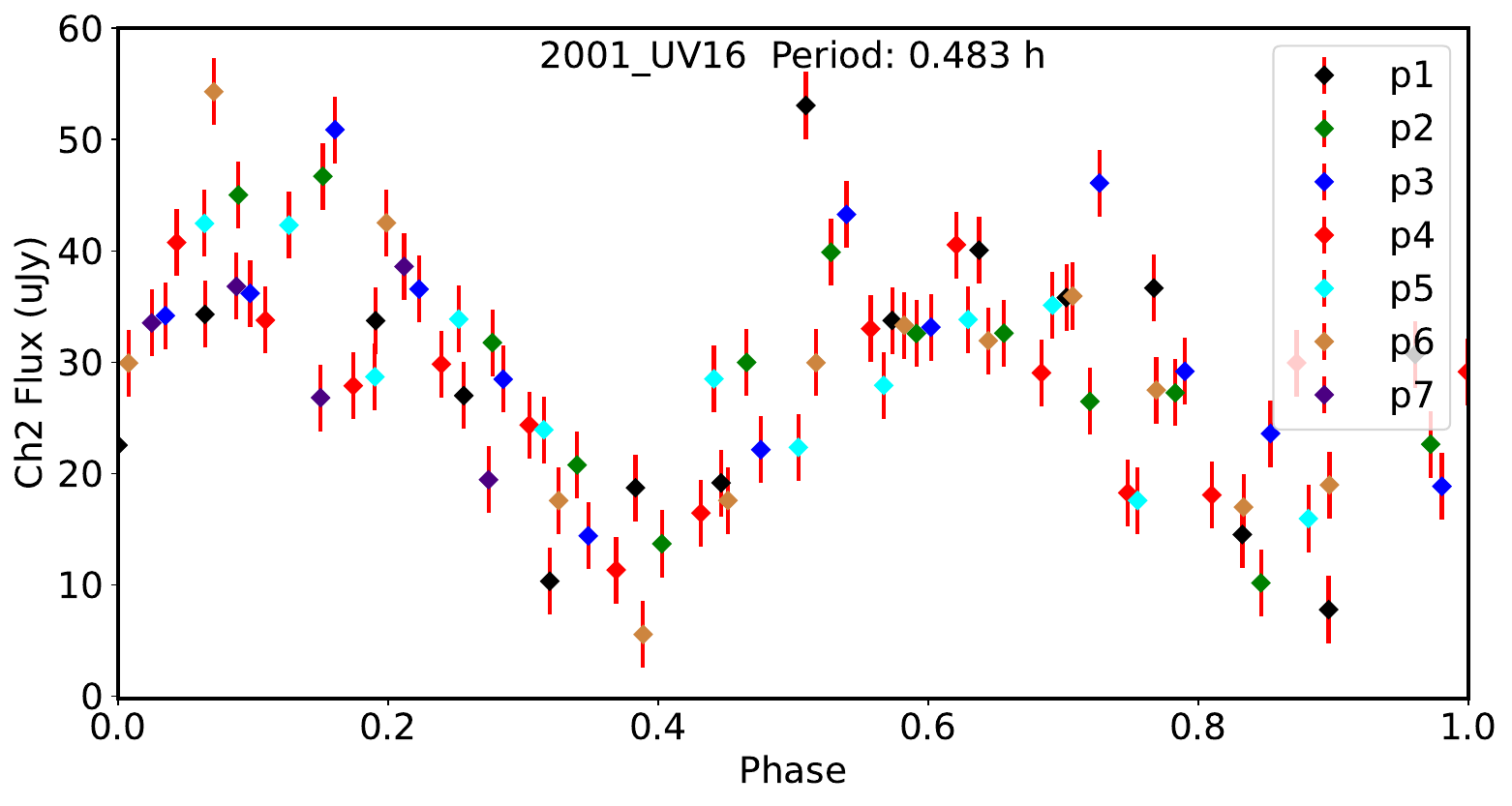}
    \includegraphics[width=0.495\linewidth]{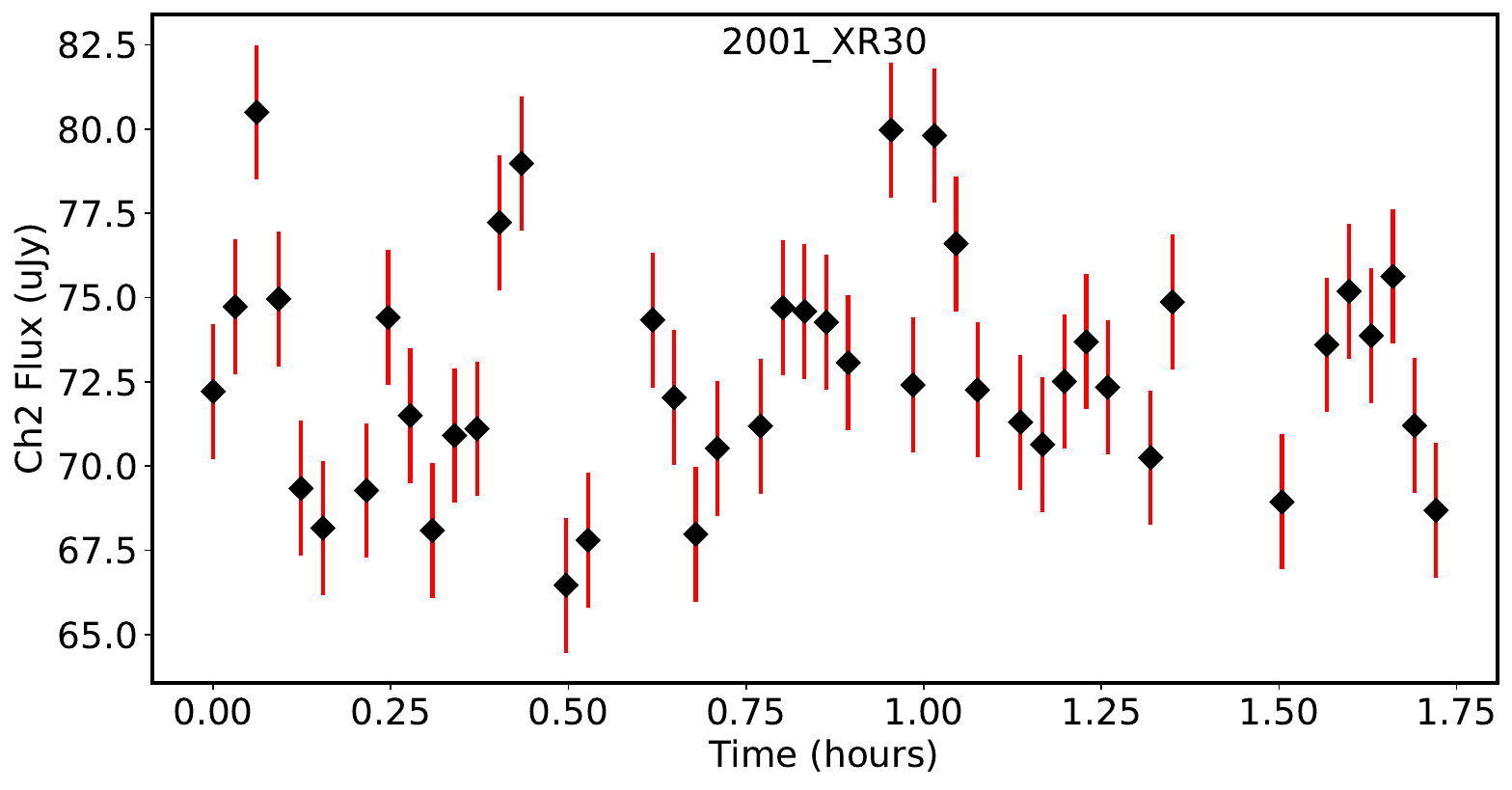}
    \includegraphics[width=0.495\linewidth]{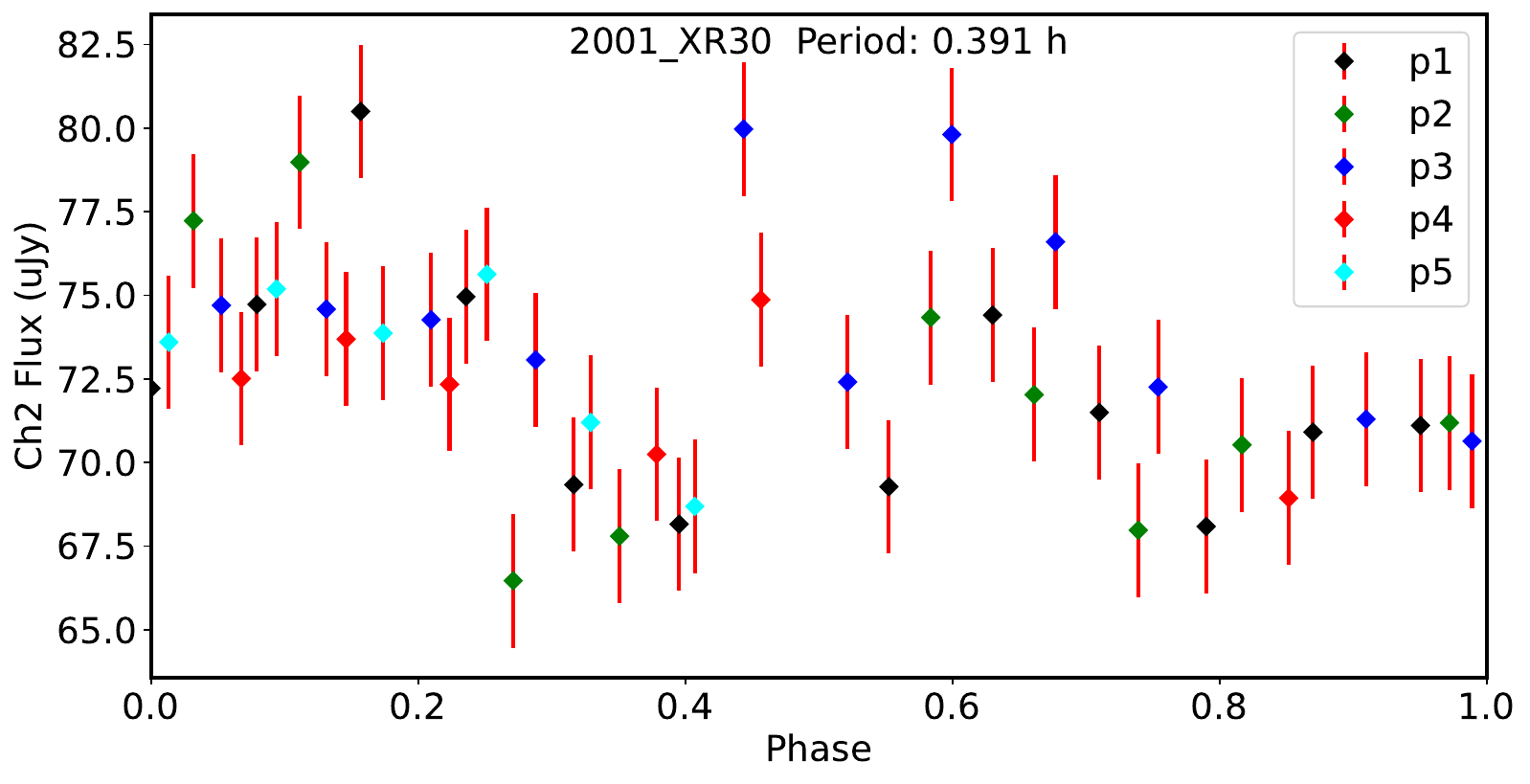}
    \includegraphics[width=0.495\linewidth]{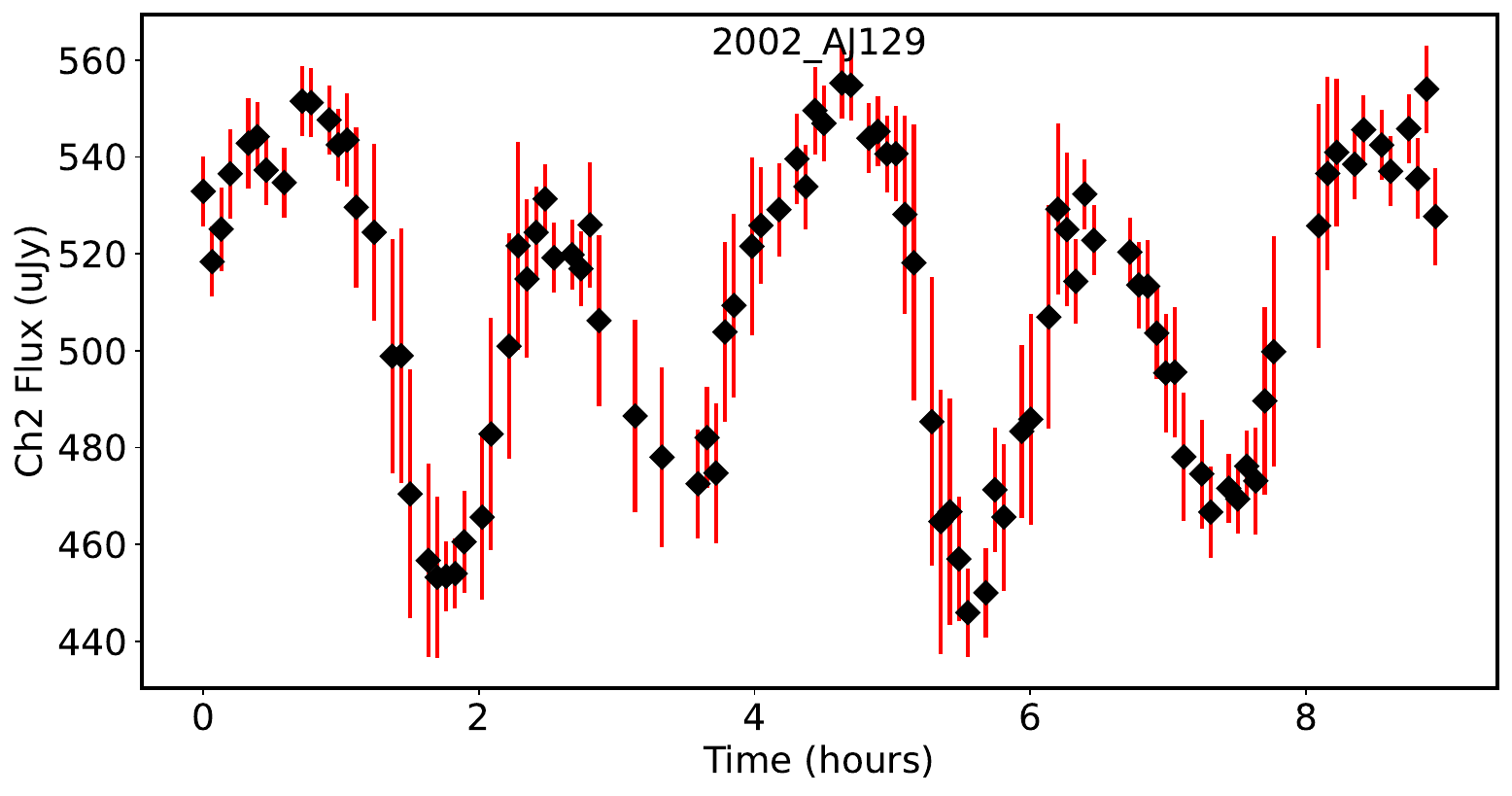}
    \includegraphics[width=0.495\linewidth]{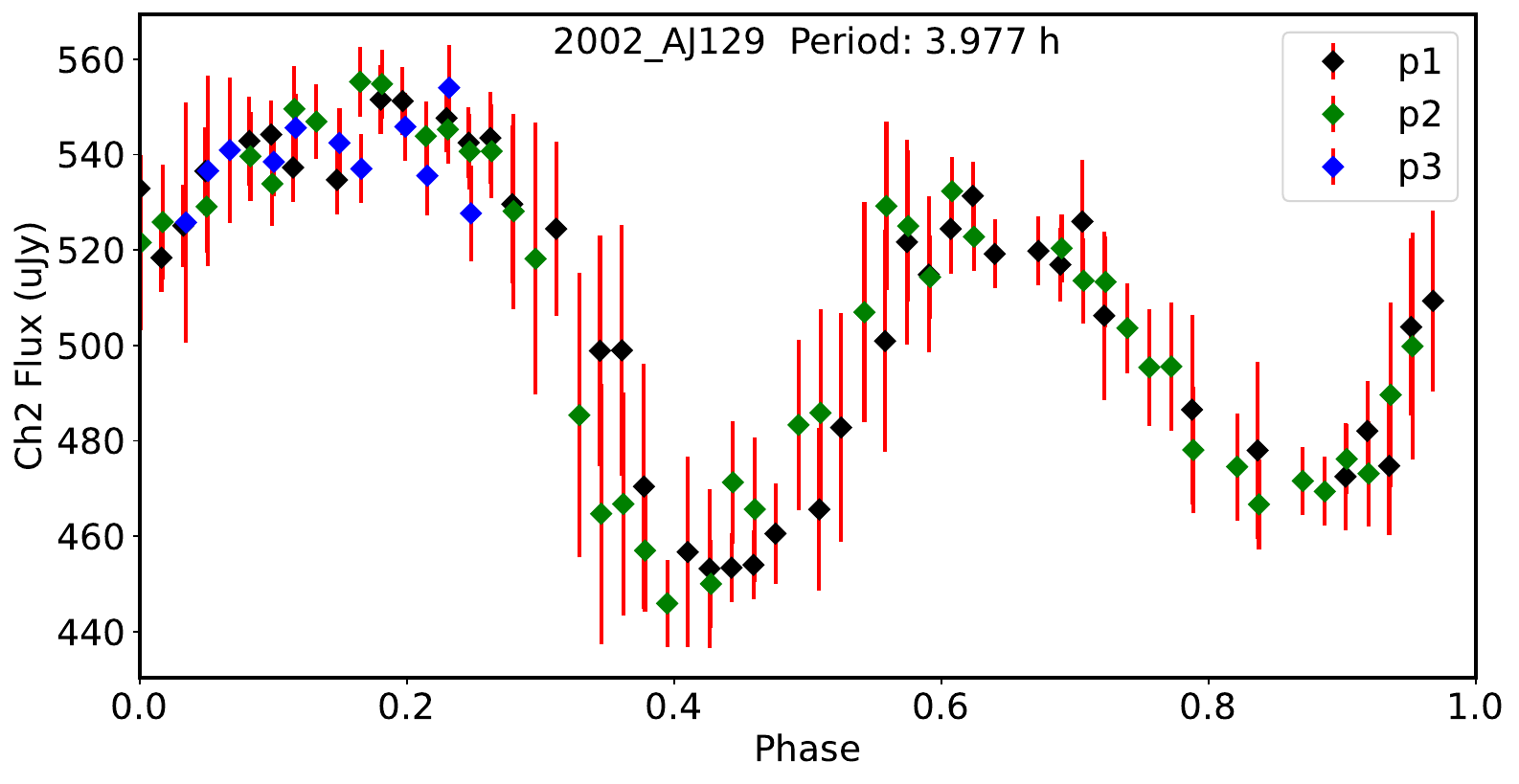}
 
    \caption{Lightcurves (left column) and phased lightcurves (right column) for sources with one or more periods sampled and periods determined. The periods are plotted with different colors in the plots on the right.}
    \label{fig:lc3}
\end{figure*}

\begin{figure*}
    \centering
   \includegraphics[width=0.495\linewidth]{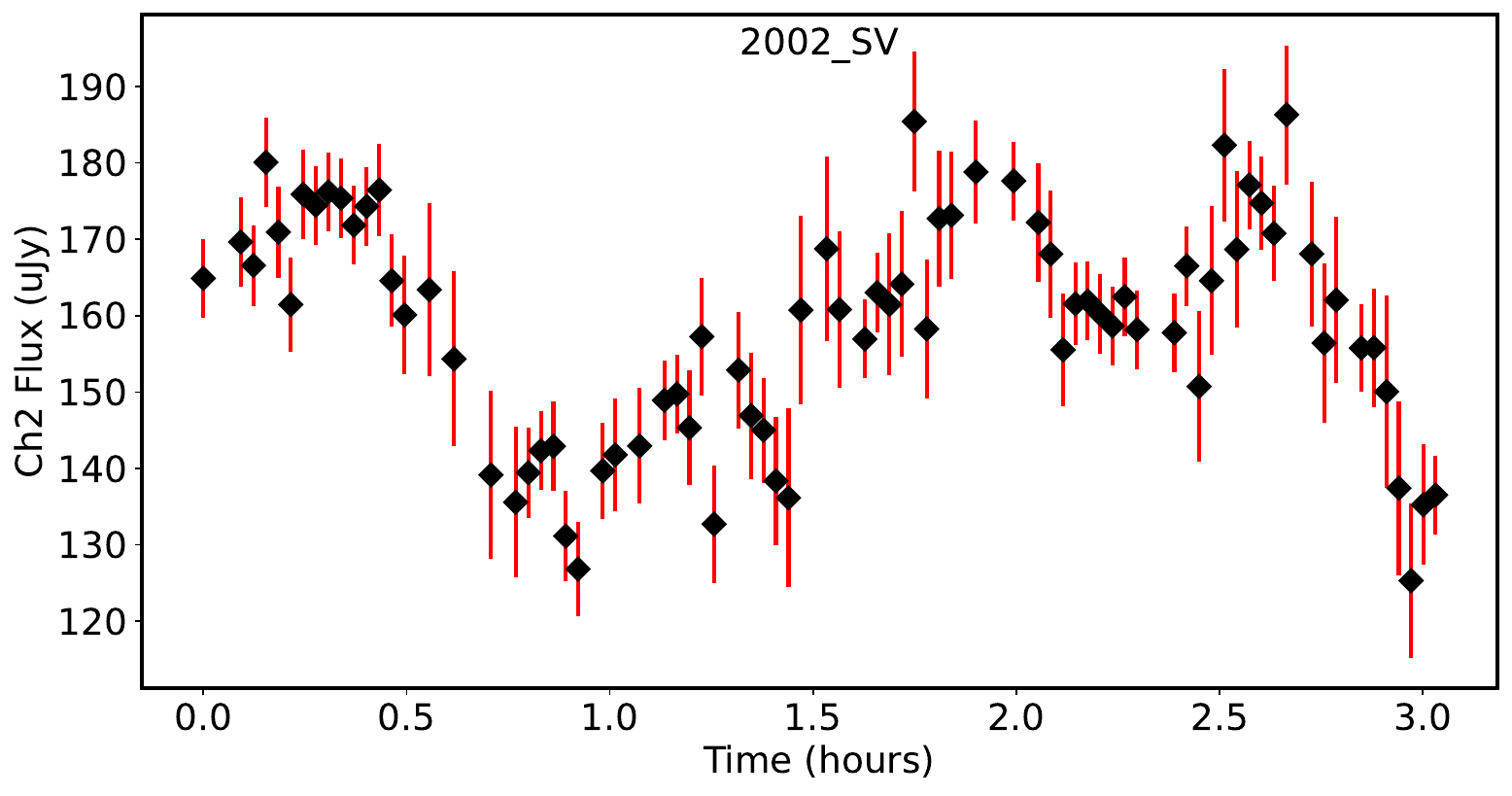}
    \includegraphics[width=0.495\linewidth]{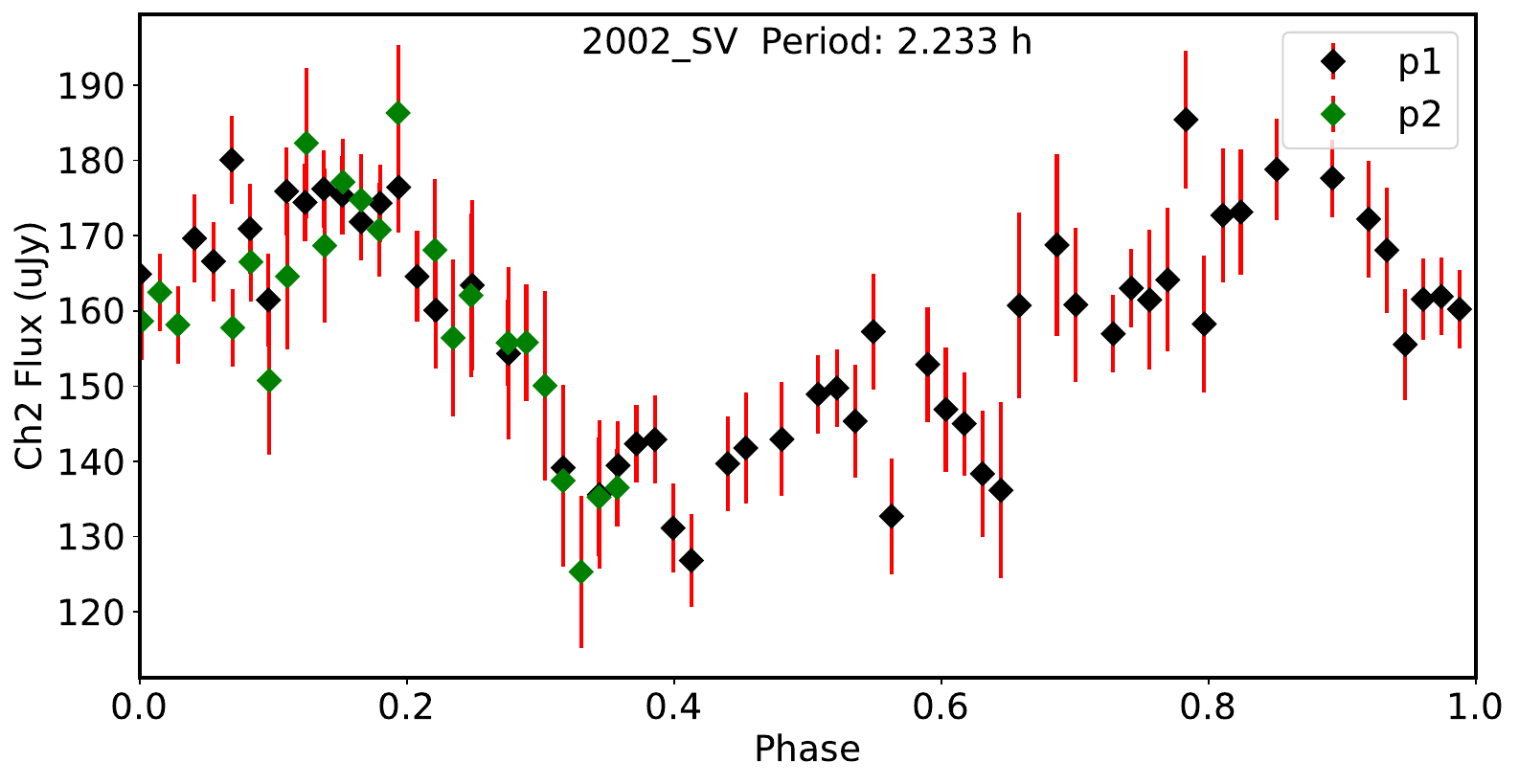}
    \includegraphics[width=0.495\linewidth]{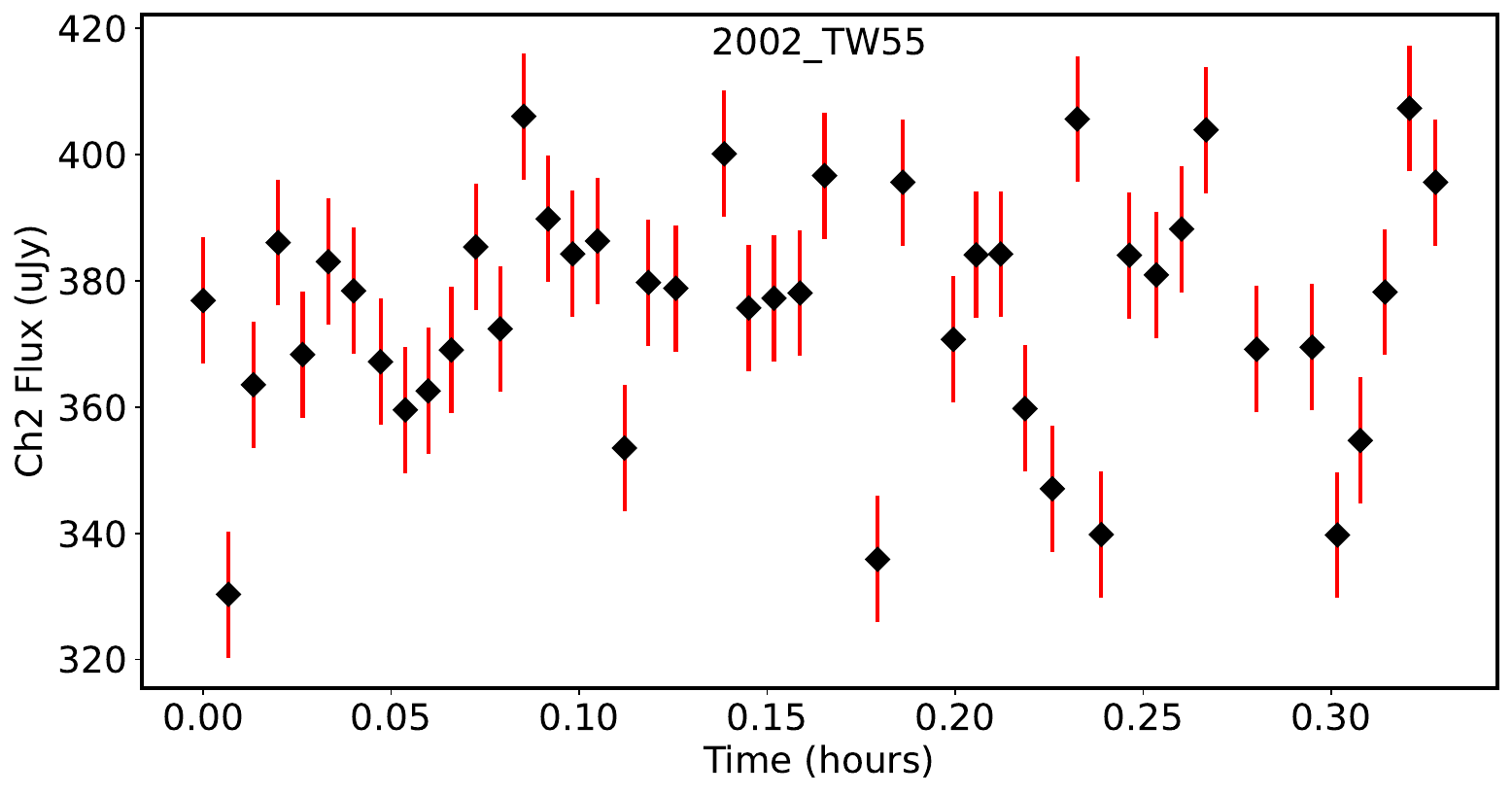}
    \includegraphics[width=0.495\linewidth]{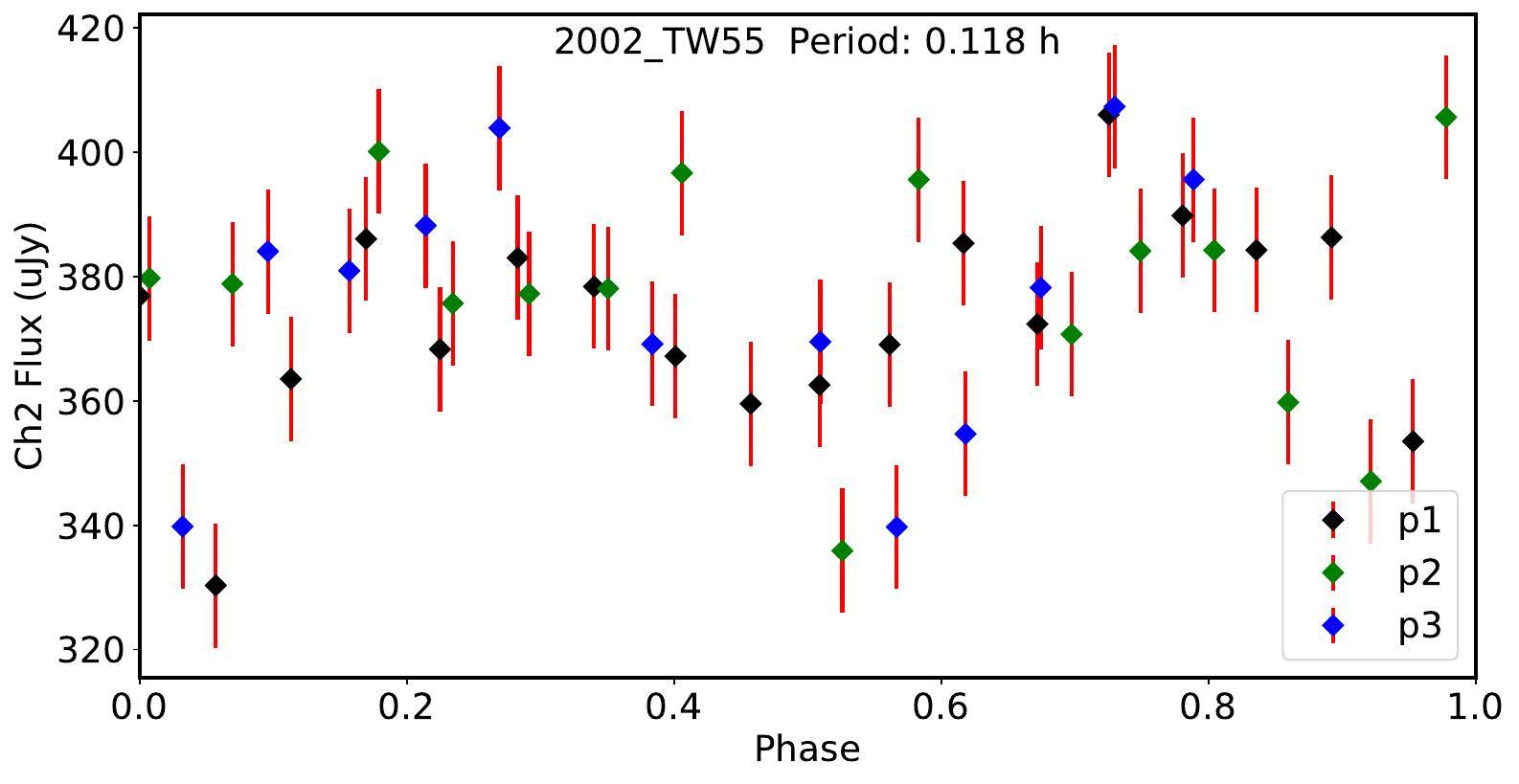}
    \includegraphics[width=0.495\linewidth]{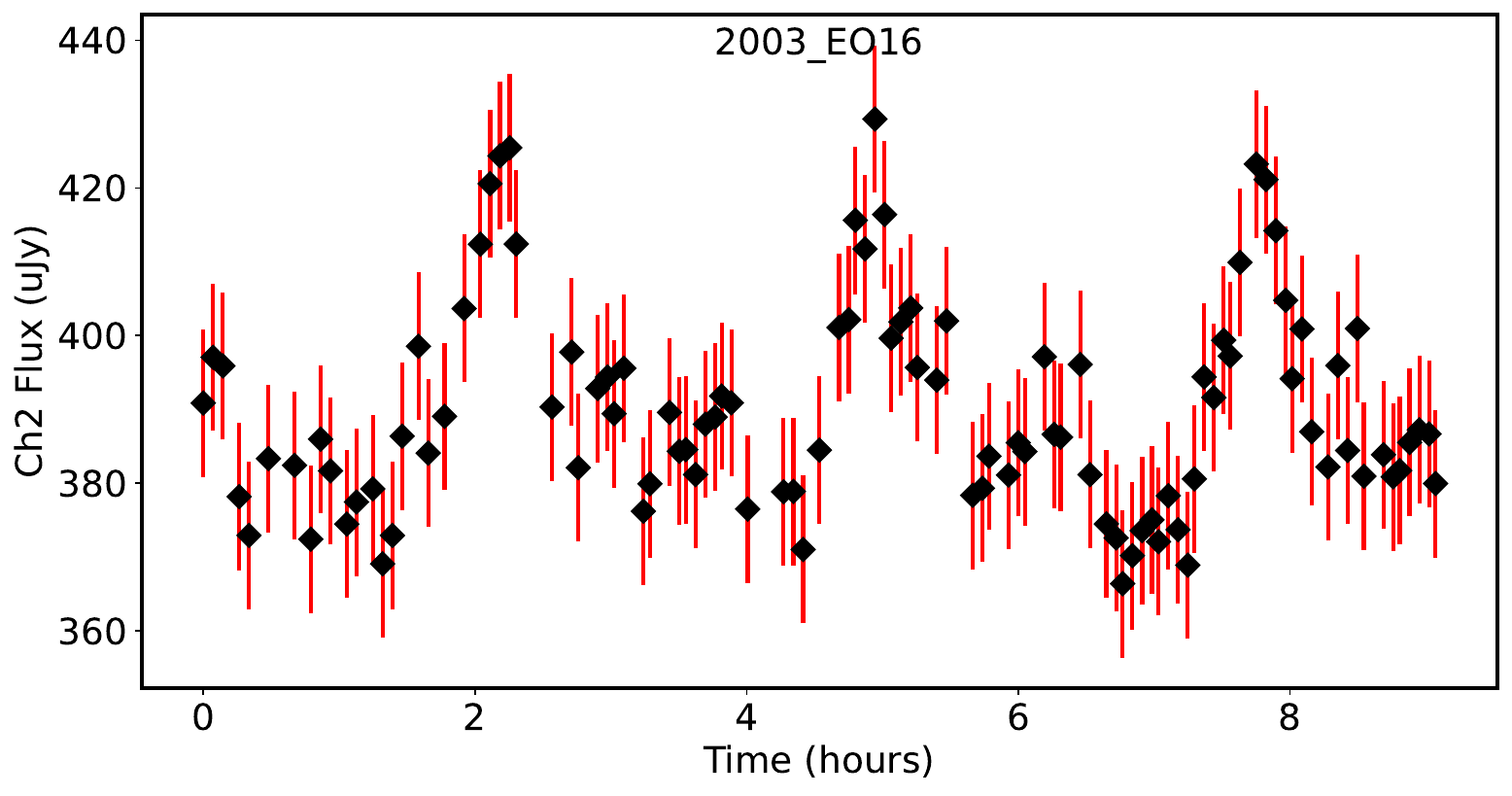}
    \includegraphics[width=0.495\linewidth]{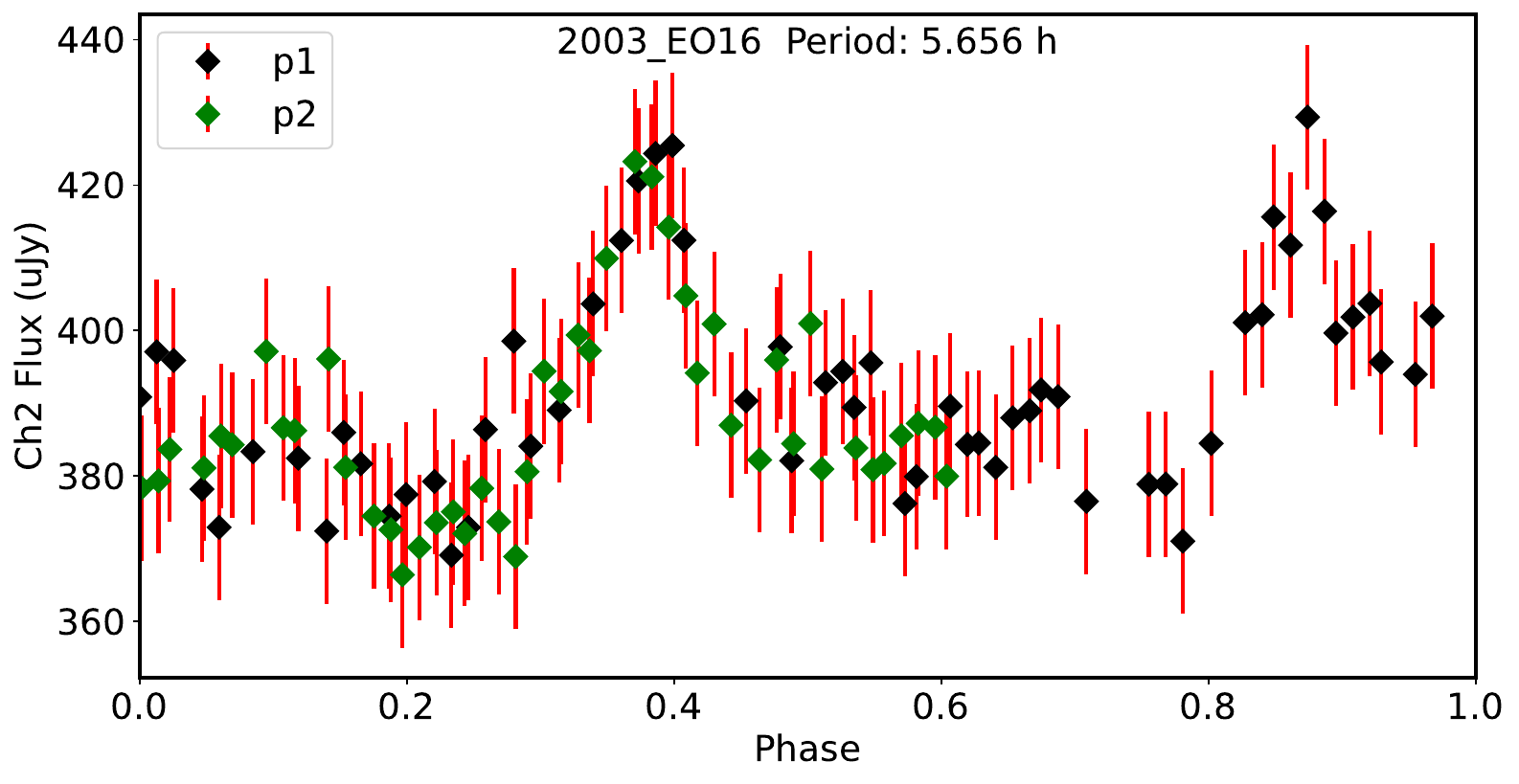}
    \includegraphics[width=0.495\linewidth]{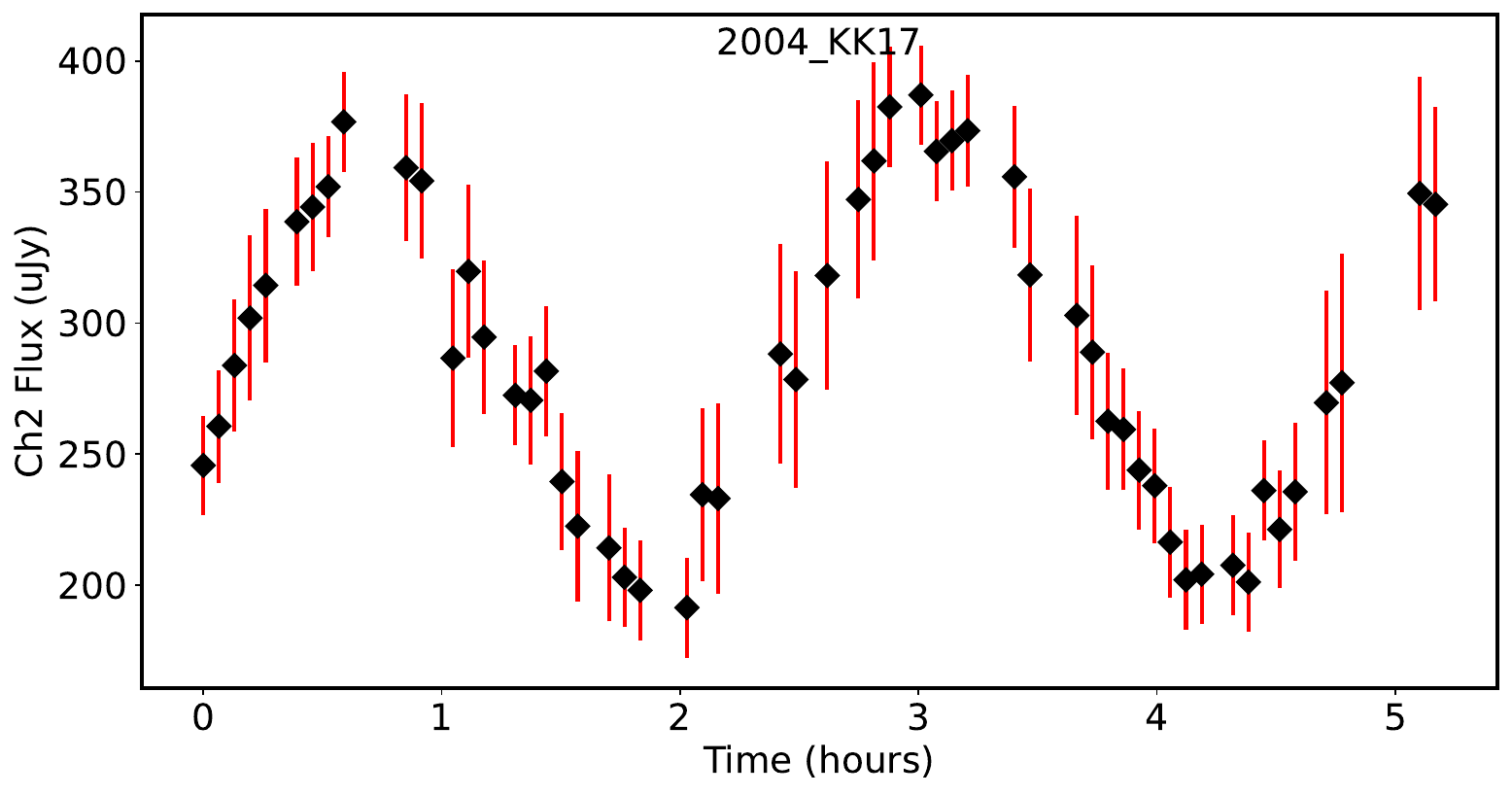}
    \includegraphics[width=0.495\linewidth]{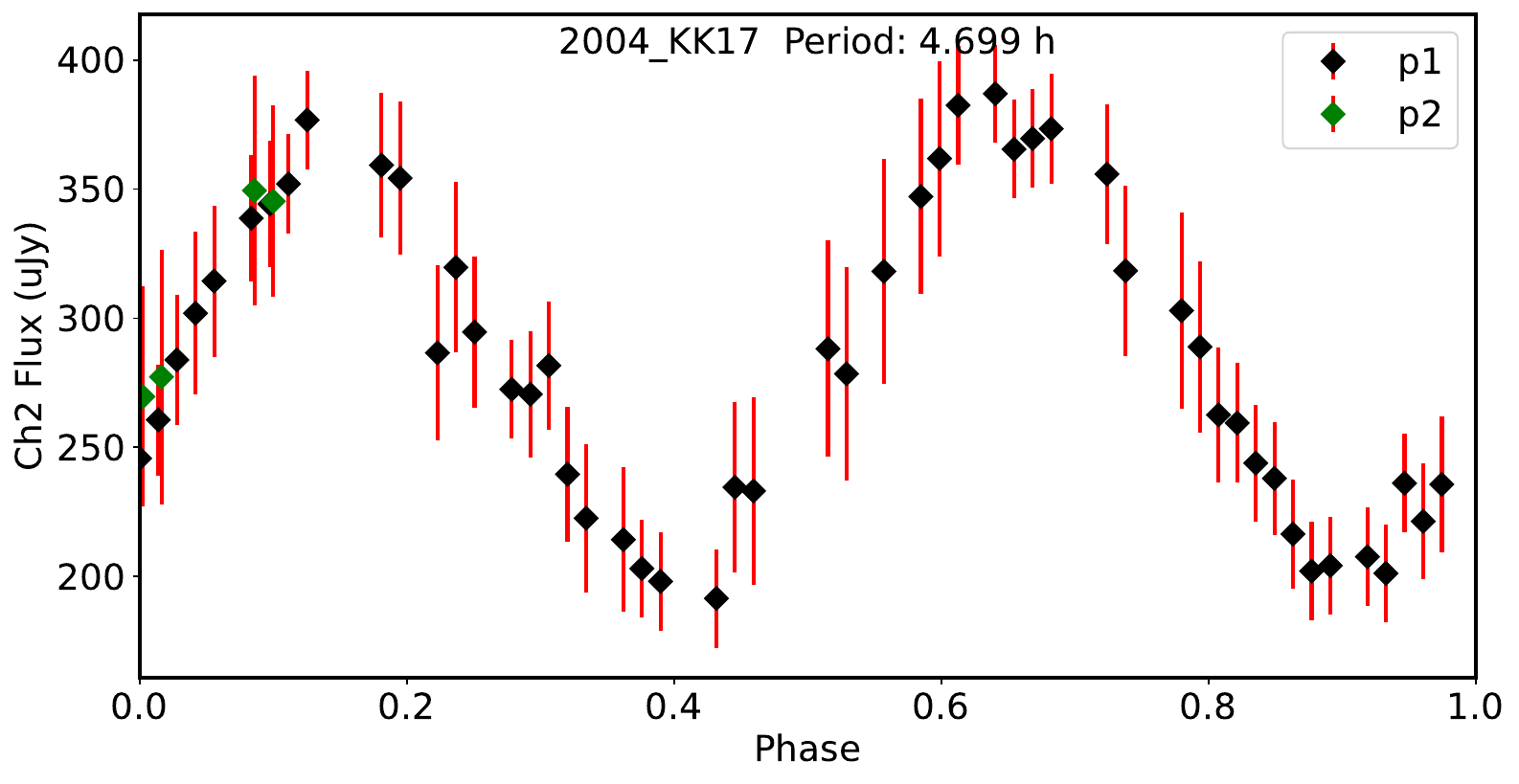}
    \includegraphics[width=0.495\linewidth]{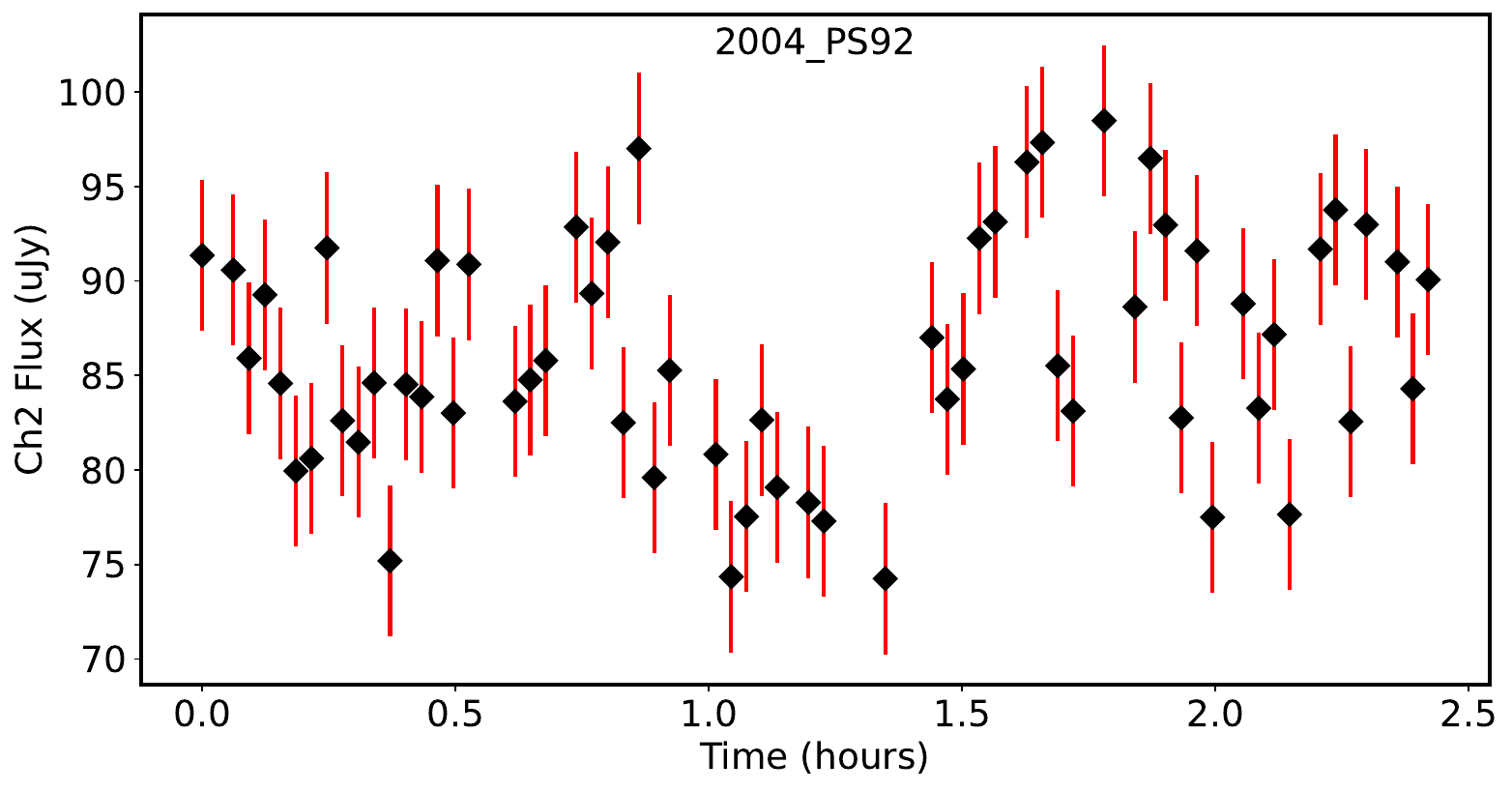}
    \includegraphics[width=0.495\linewidth]{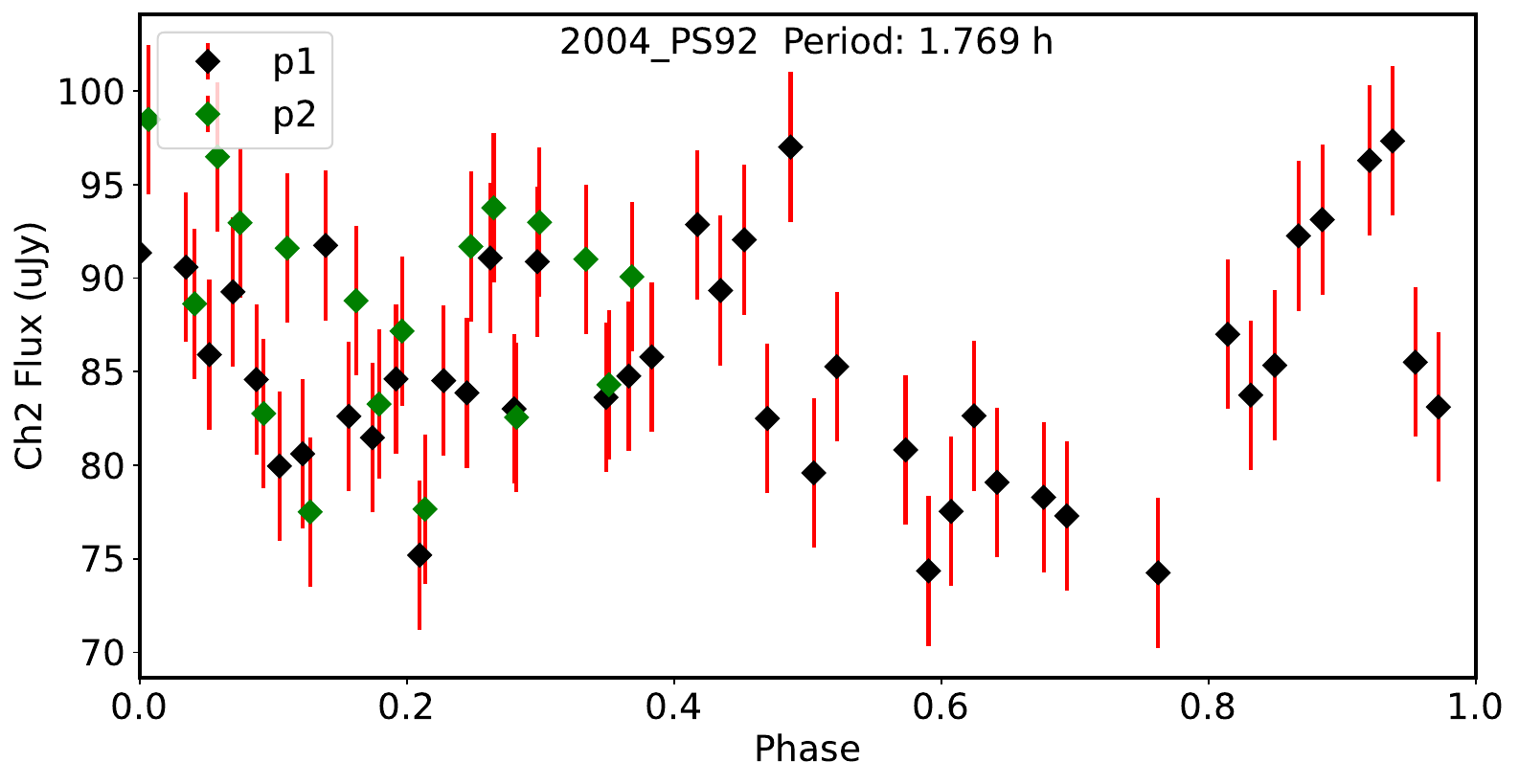}
    \caption{Lightcurves (left column) and phased lightcurves (right column) for sources with one or more periods sampled and periods determined. The periods are plotted with different colors in the plots on the right.}
    \label{fig:lc4}
\end{figure*}

\begin{figure*}
    \centering

    \includegraphics[width=0.495\linewidth]{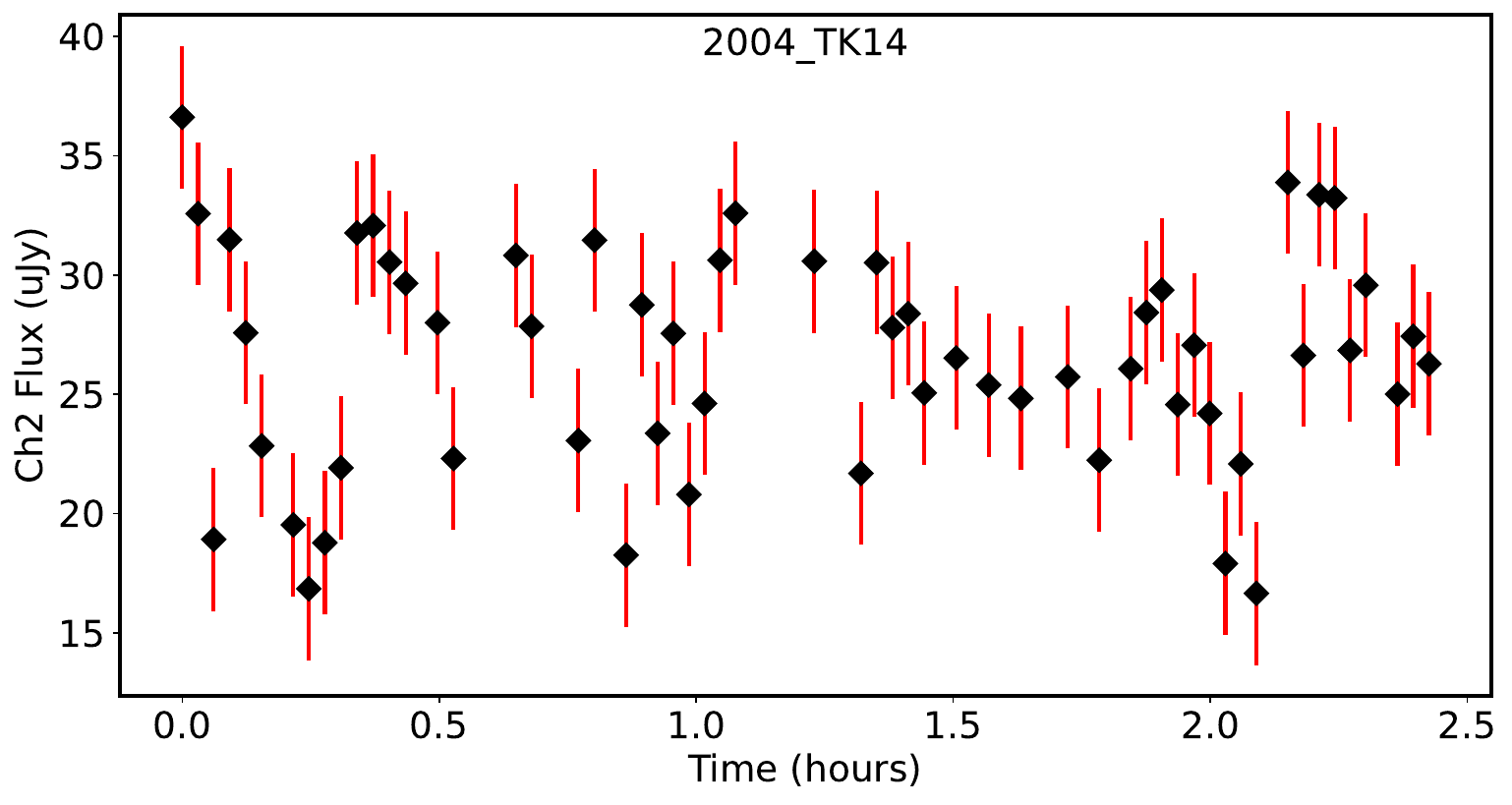}
    \includegraphics[width=0.495\linewidth]{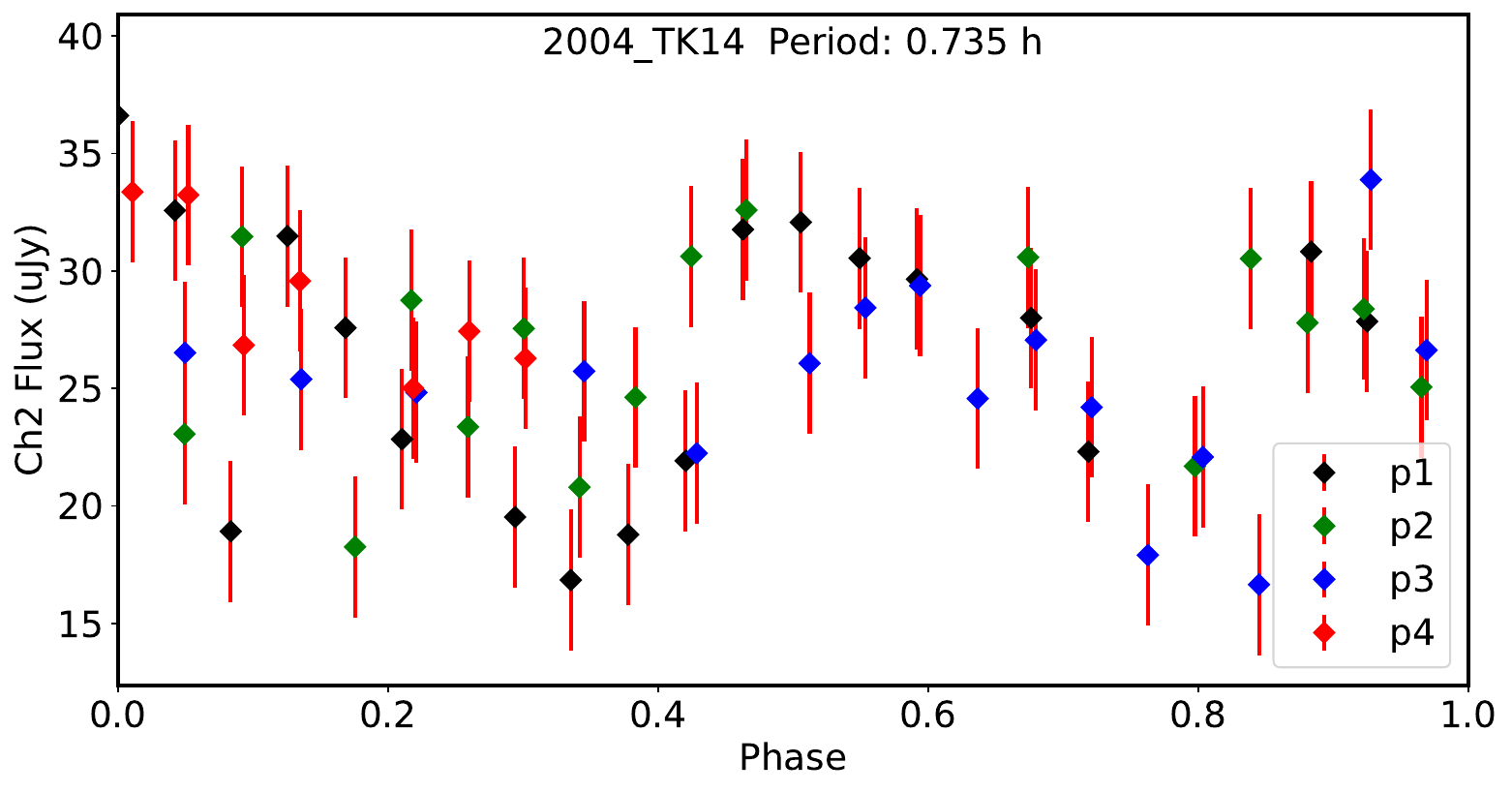}
    \includegraphics[width=0.495\linewidth]{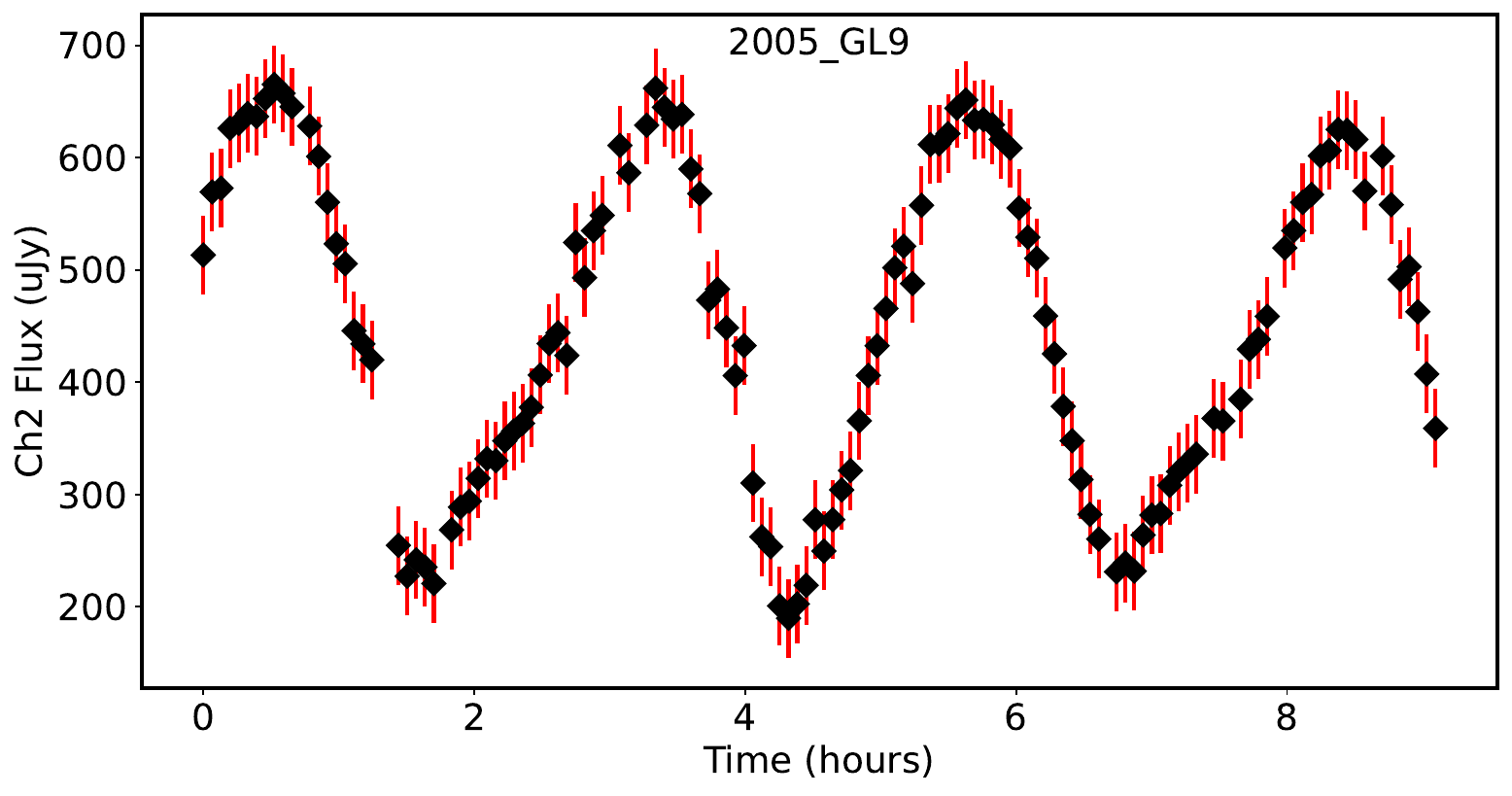}
    \includegraphics[width=0.495\linewidth]{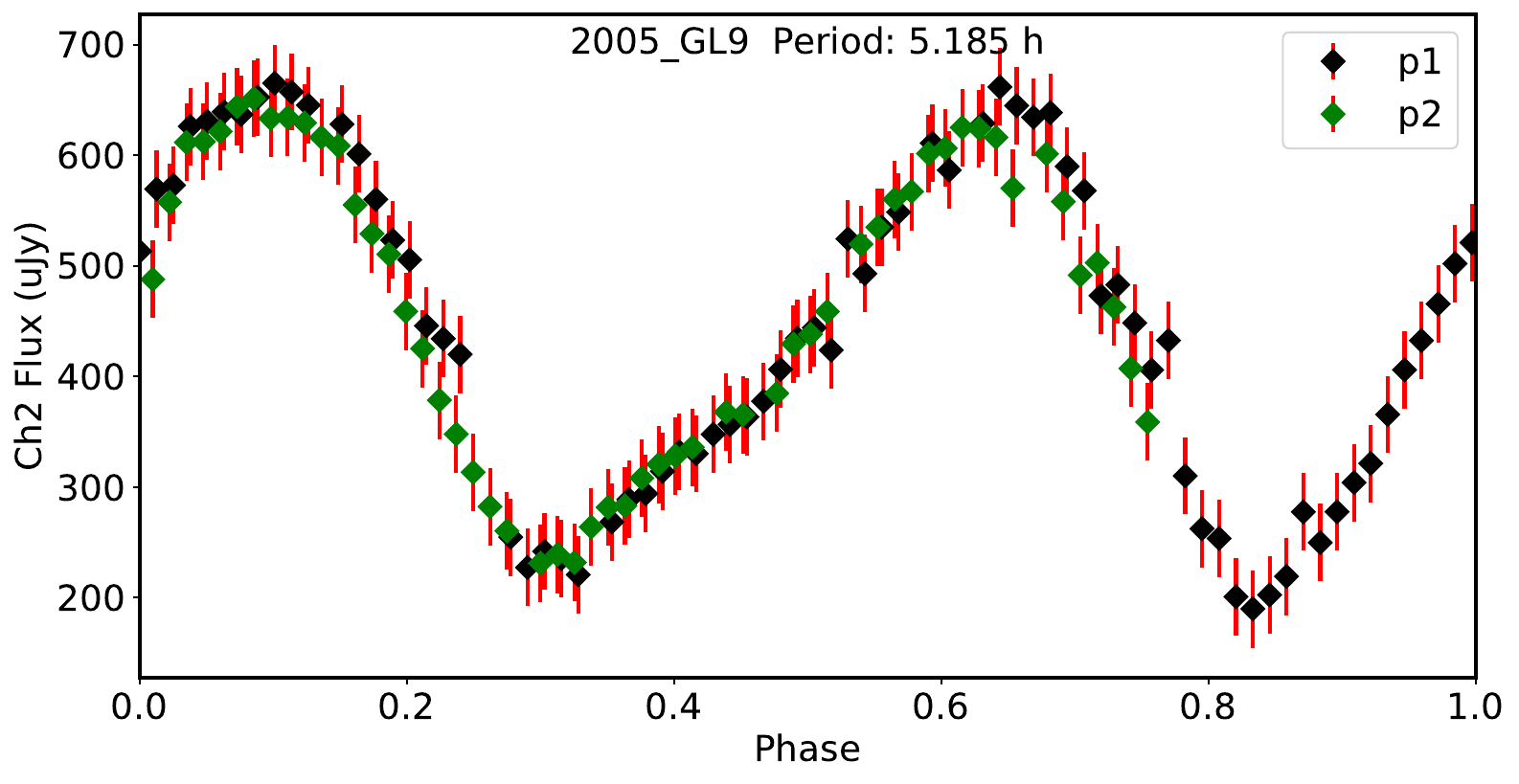}
    \includegraphics[width=0.495\linewidth]{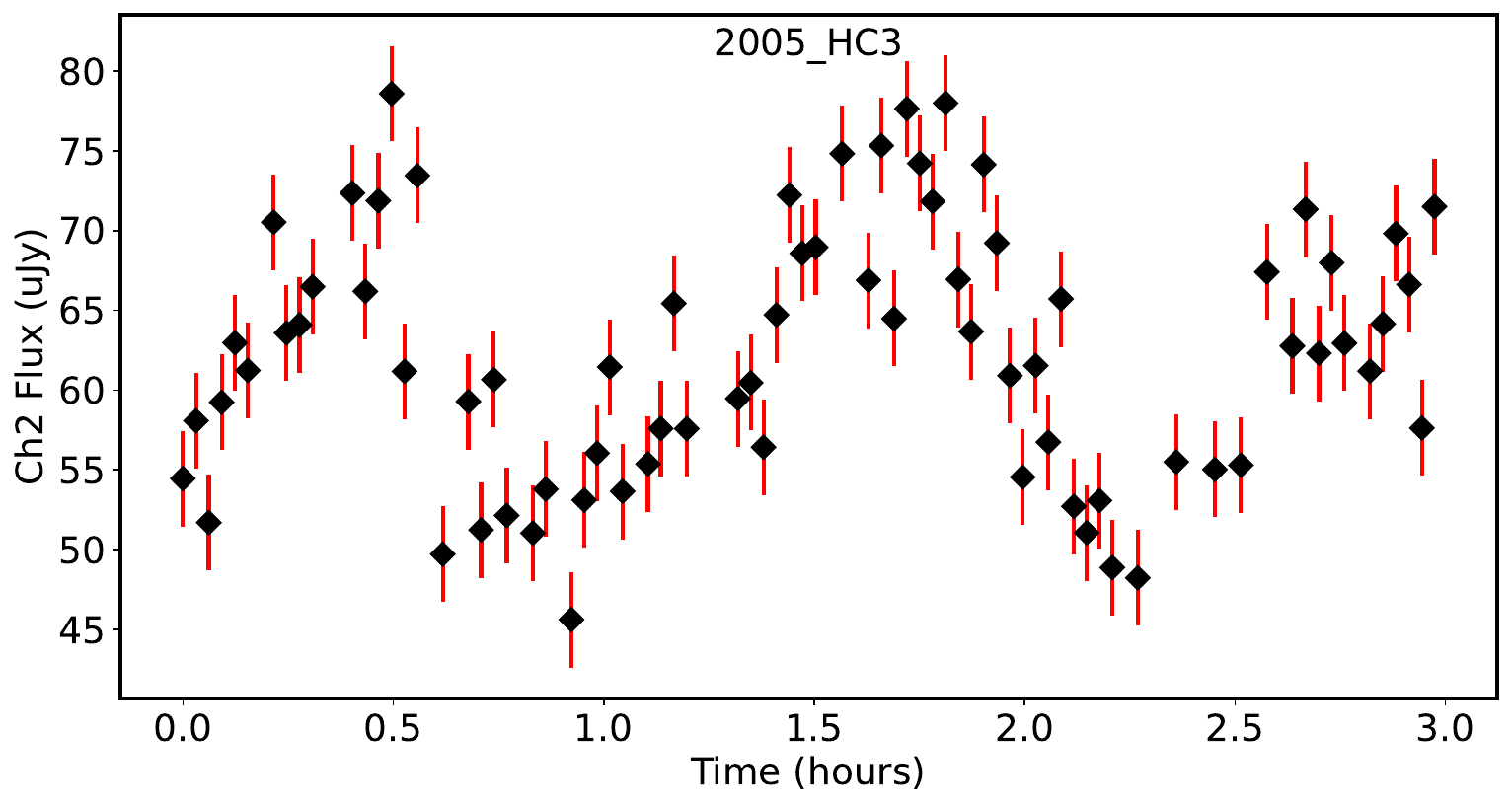}
    \includegraphics[width=0.495\linewidth]{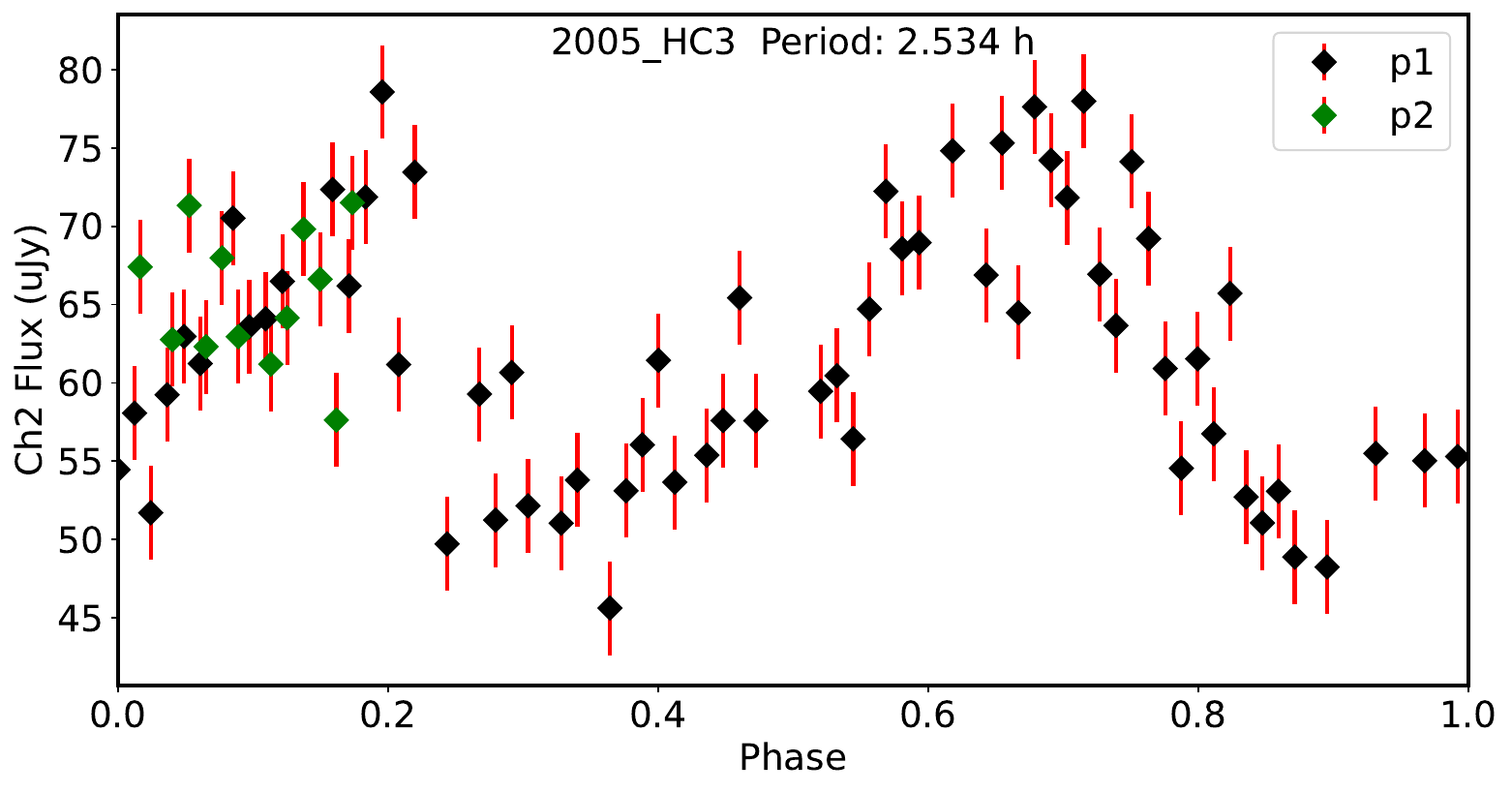}
    \includegraphics[width=0.495\linewidth]{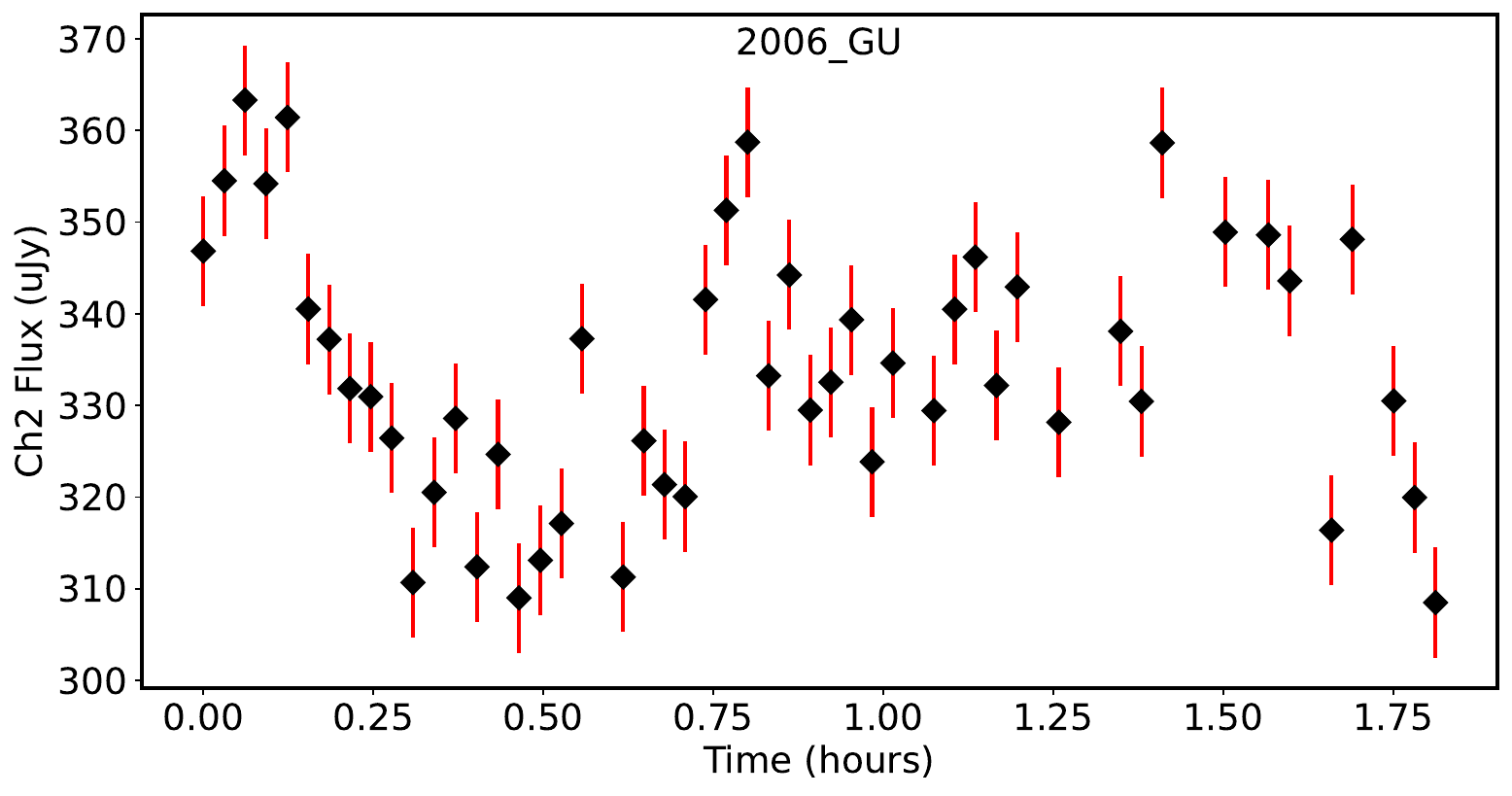}
    \includegraphics[width=0.495\linewidth]{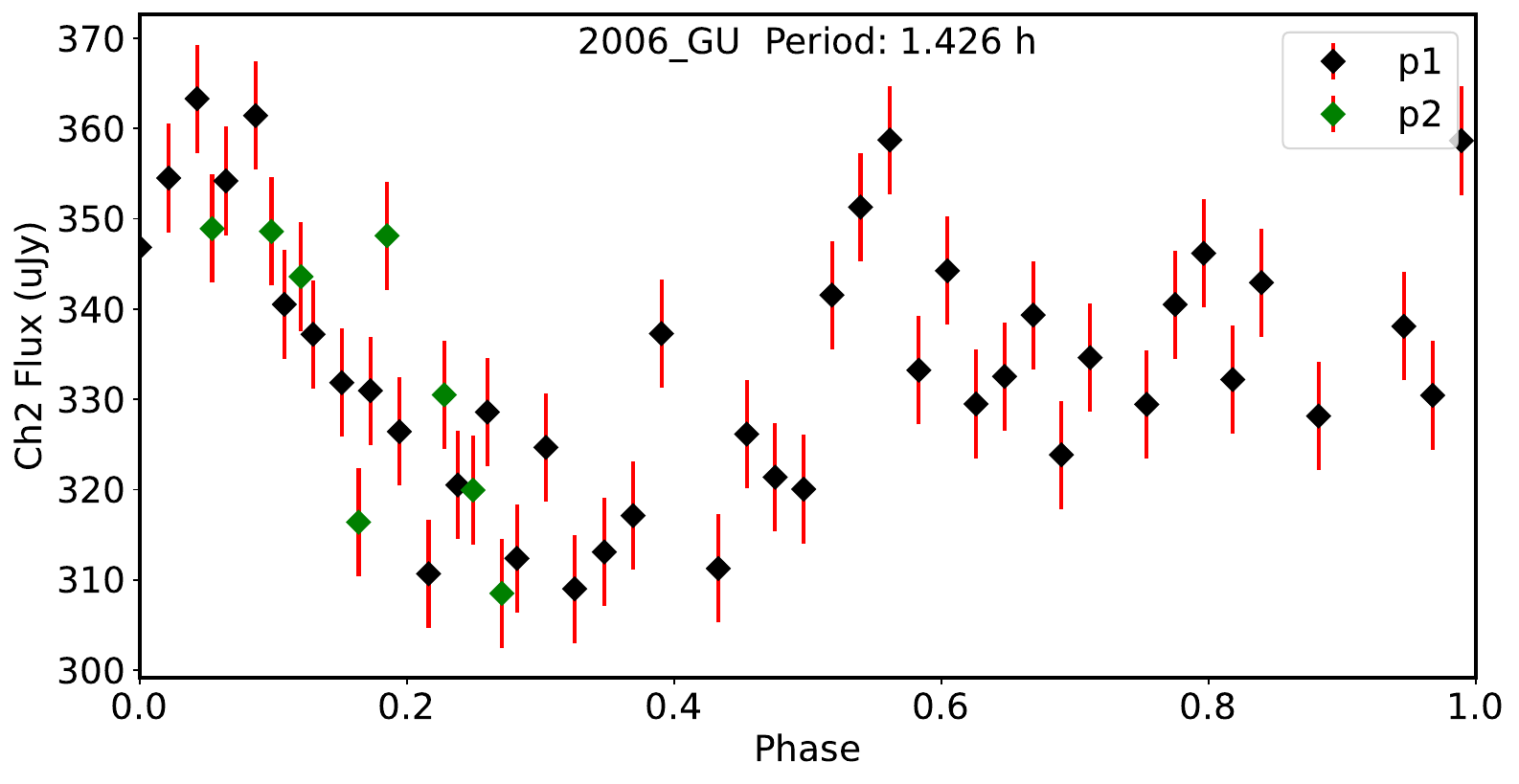}
    \includegraphics[width=0.495\linewidth]{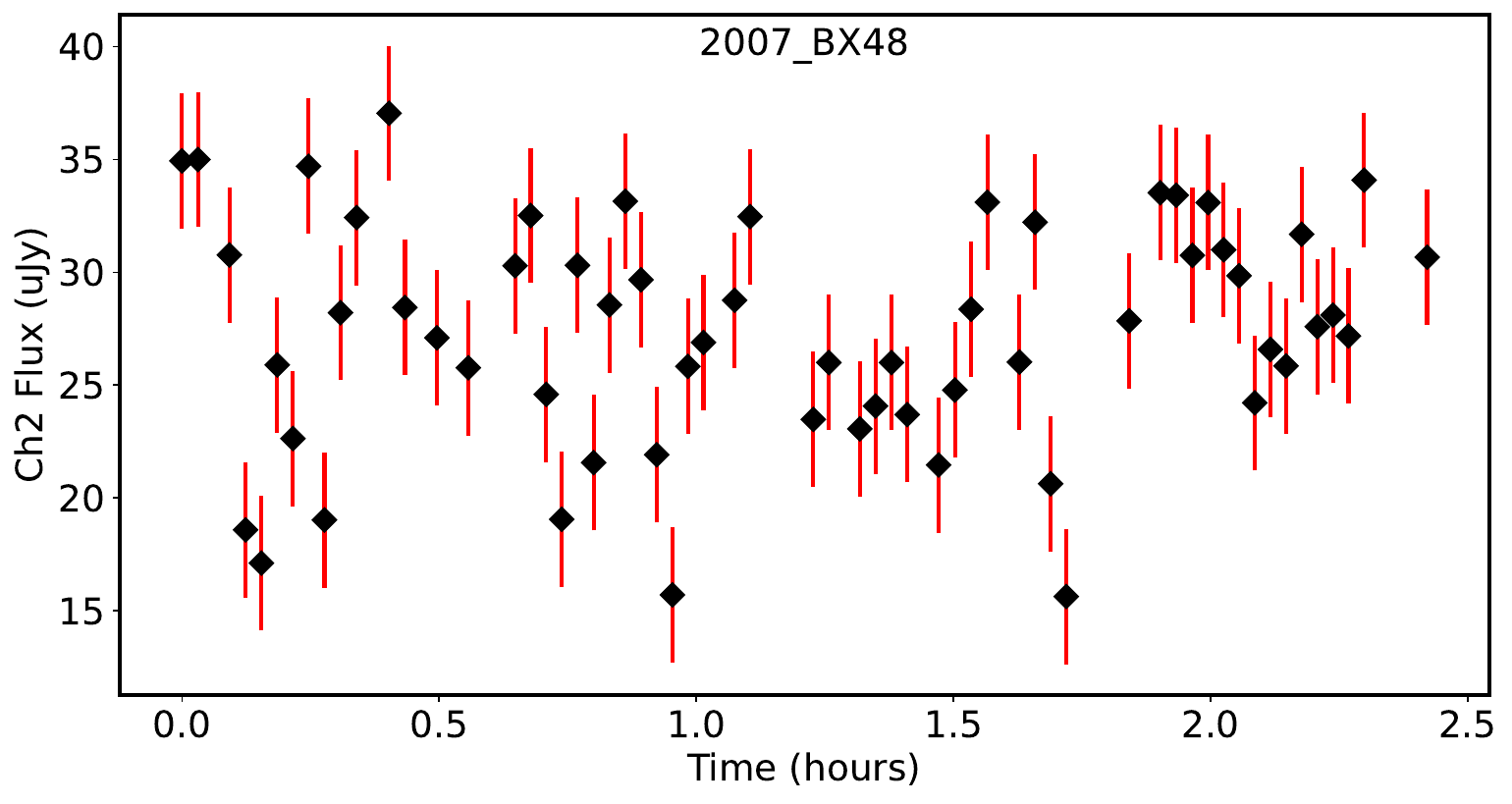}
    \includegraphics[width=0.495\linewidth]{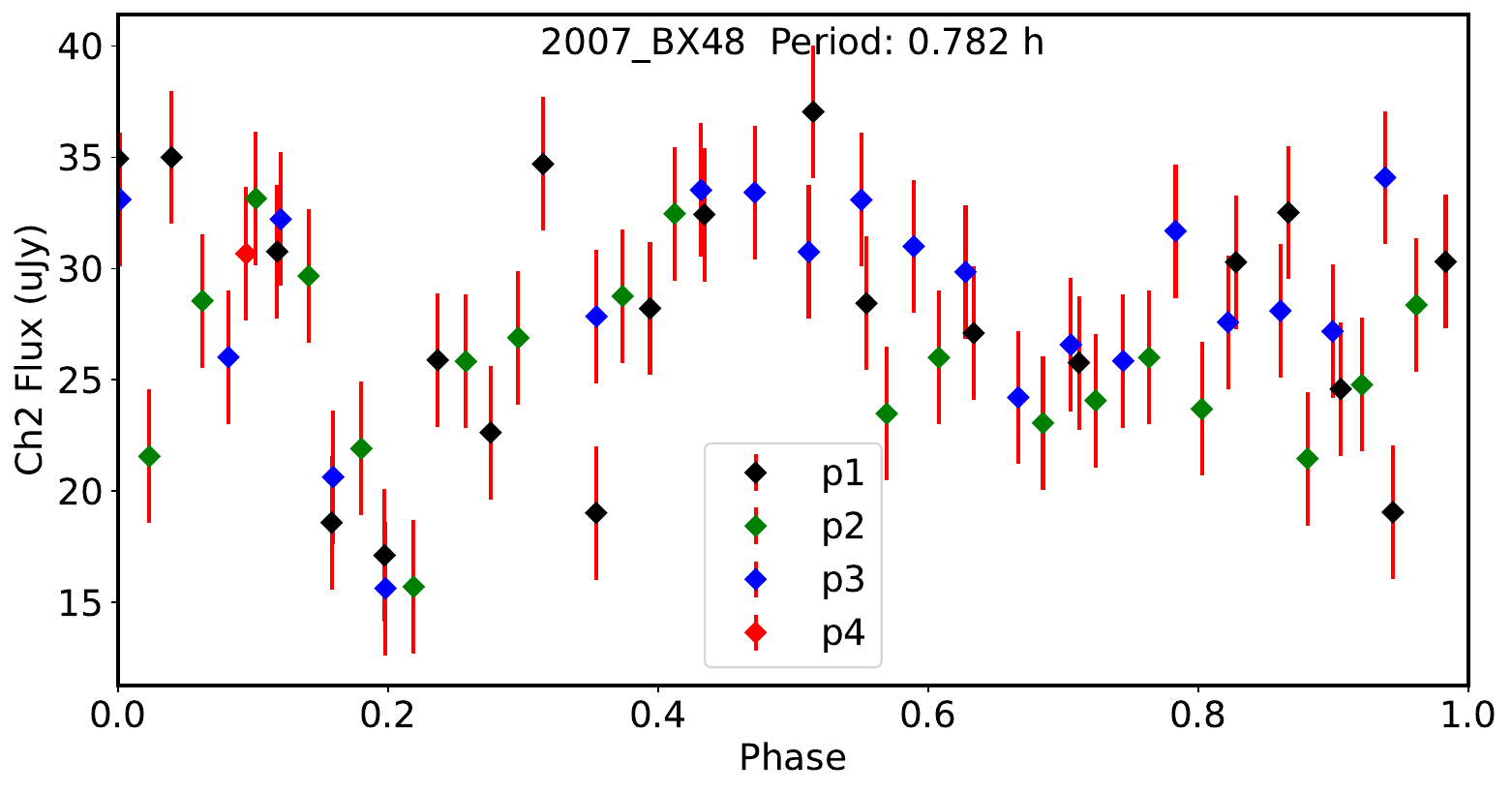}
 
    \caption{Lightcurves (left column) and phased lightcurves (right column) for sources with one or more periods sampled and periods determined. The periods are plotted with different colors in the plots on the right.}
    \label{fig:lc5}
\end{figure*}

\begin{figure*}
    \centering
   \includegraphics[width=0.495\linewidth]{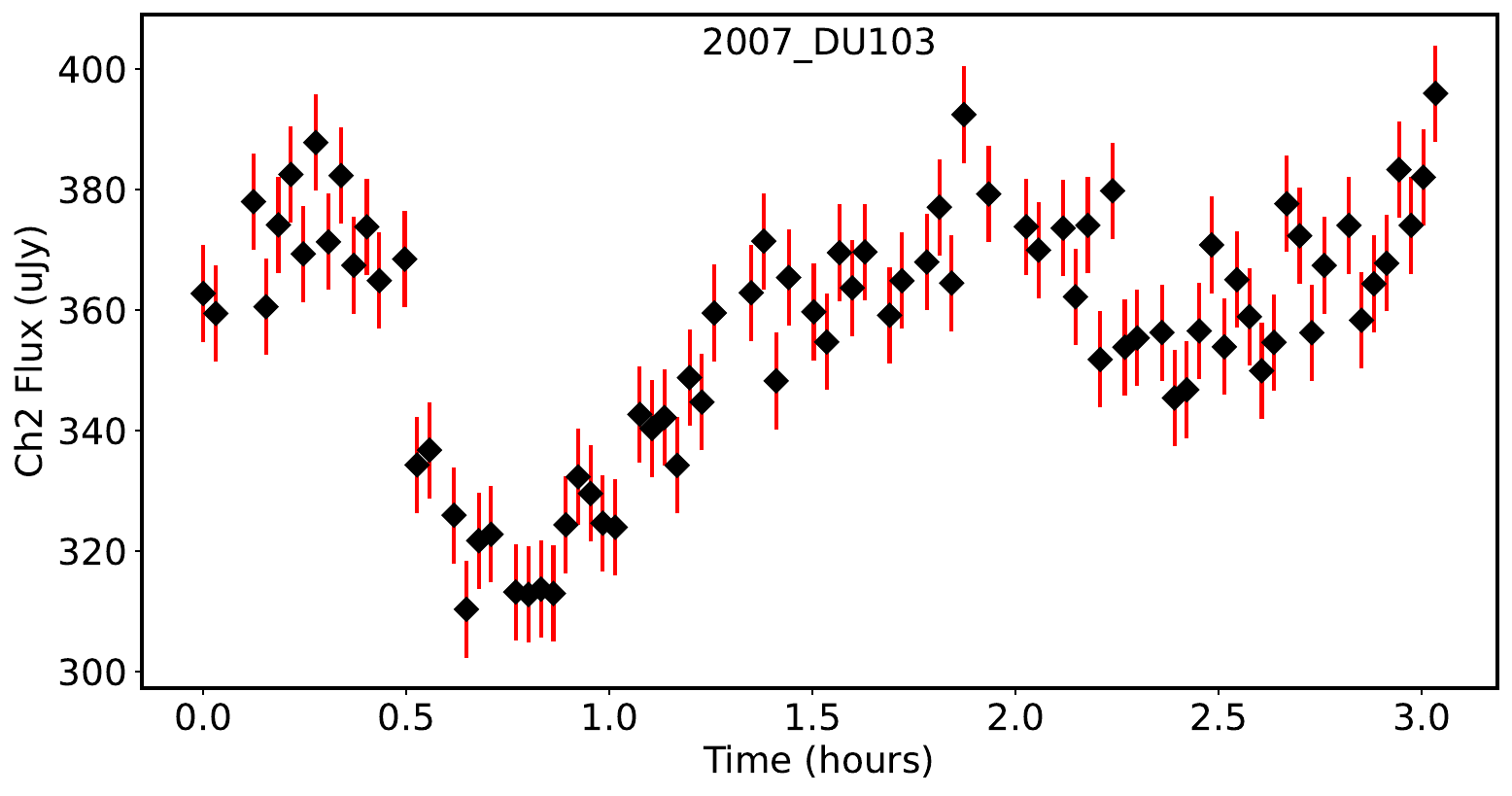}
    \includegraphics[width=0.495\linewidth]{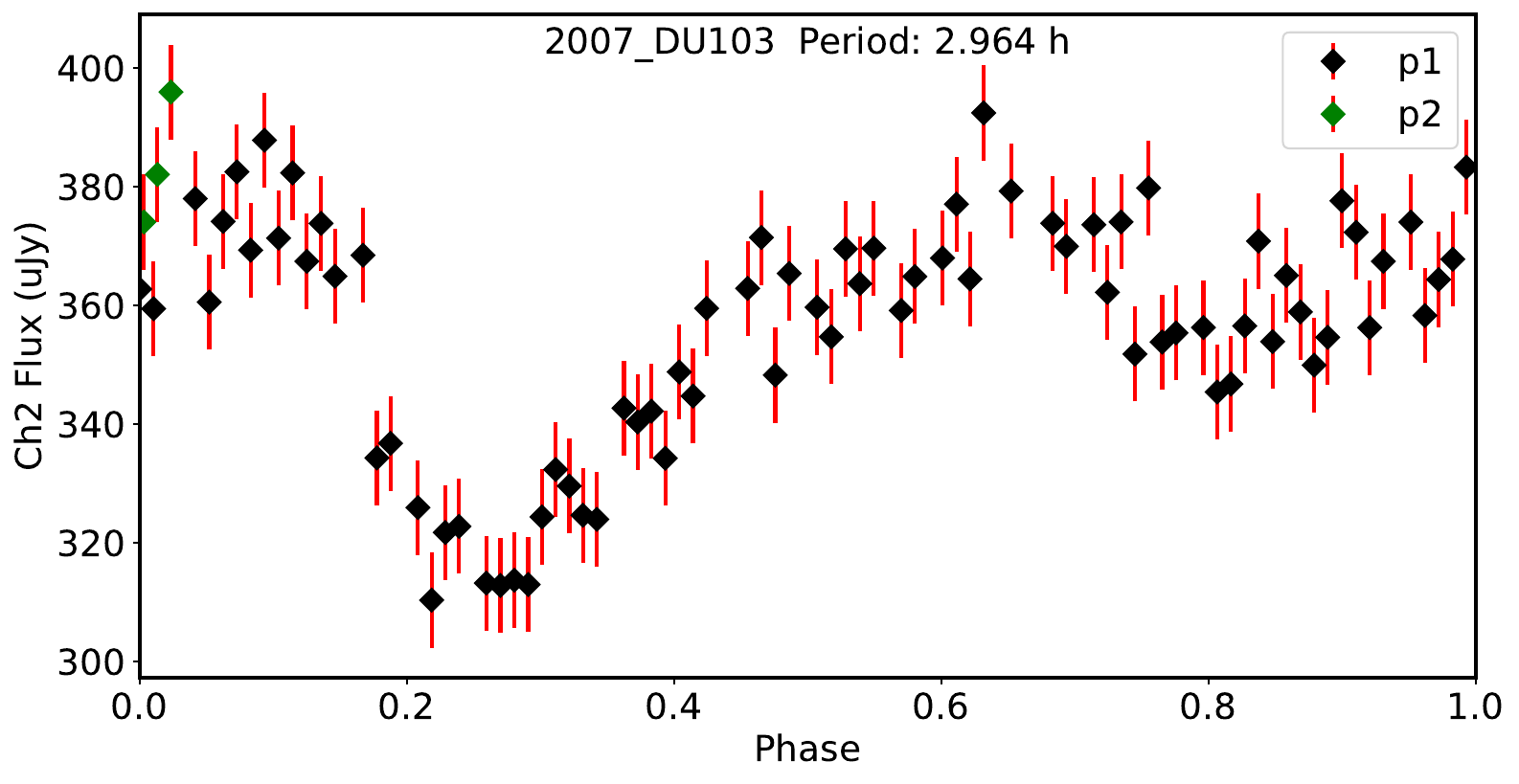}
    \includegraphics[width=0.495\linewidth]{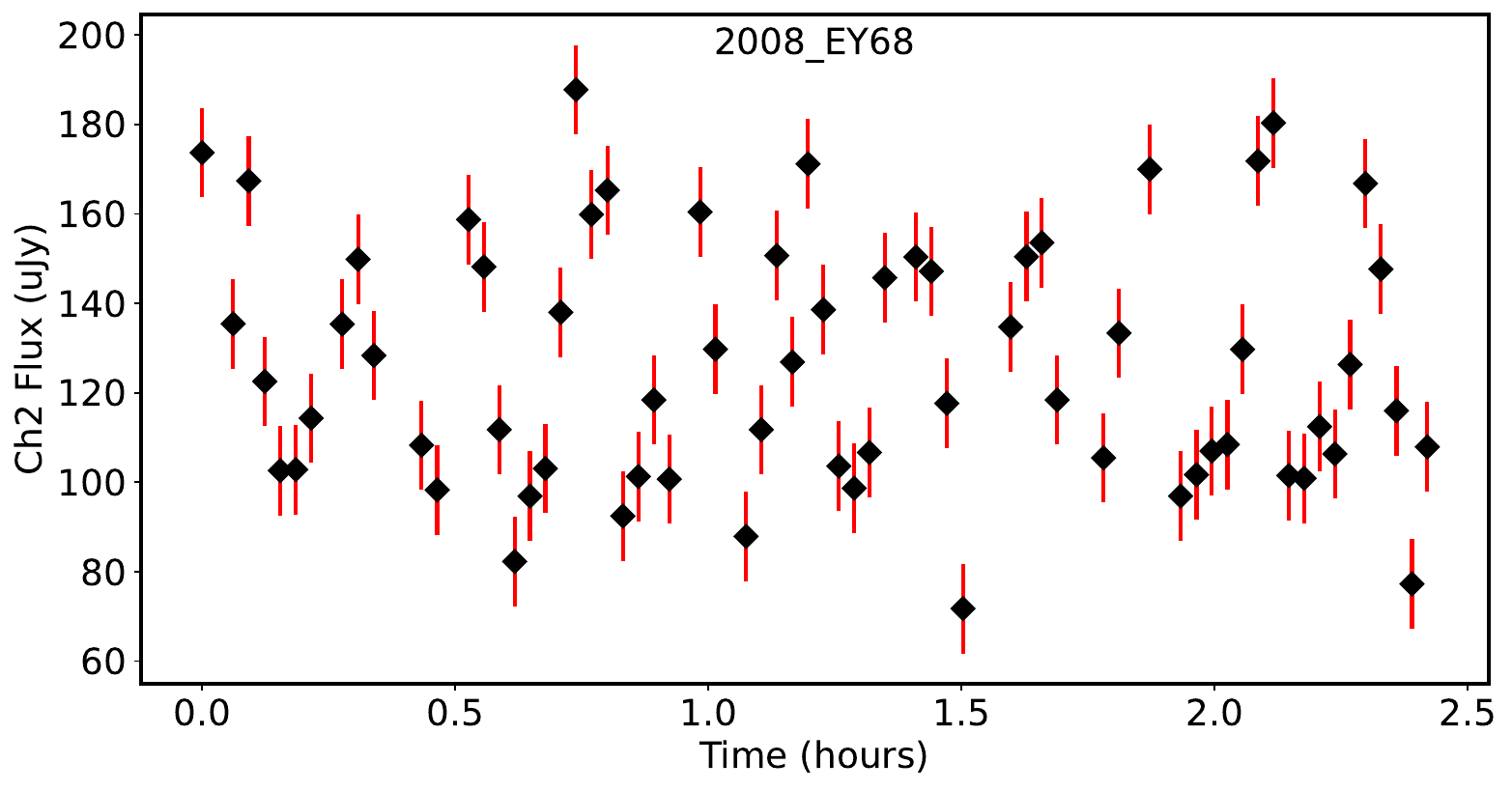}
    \includegraphics[width=0.495\linewidth]{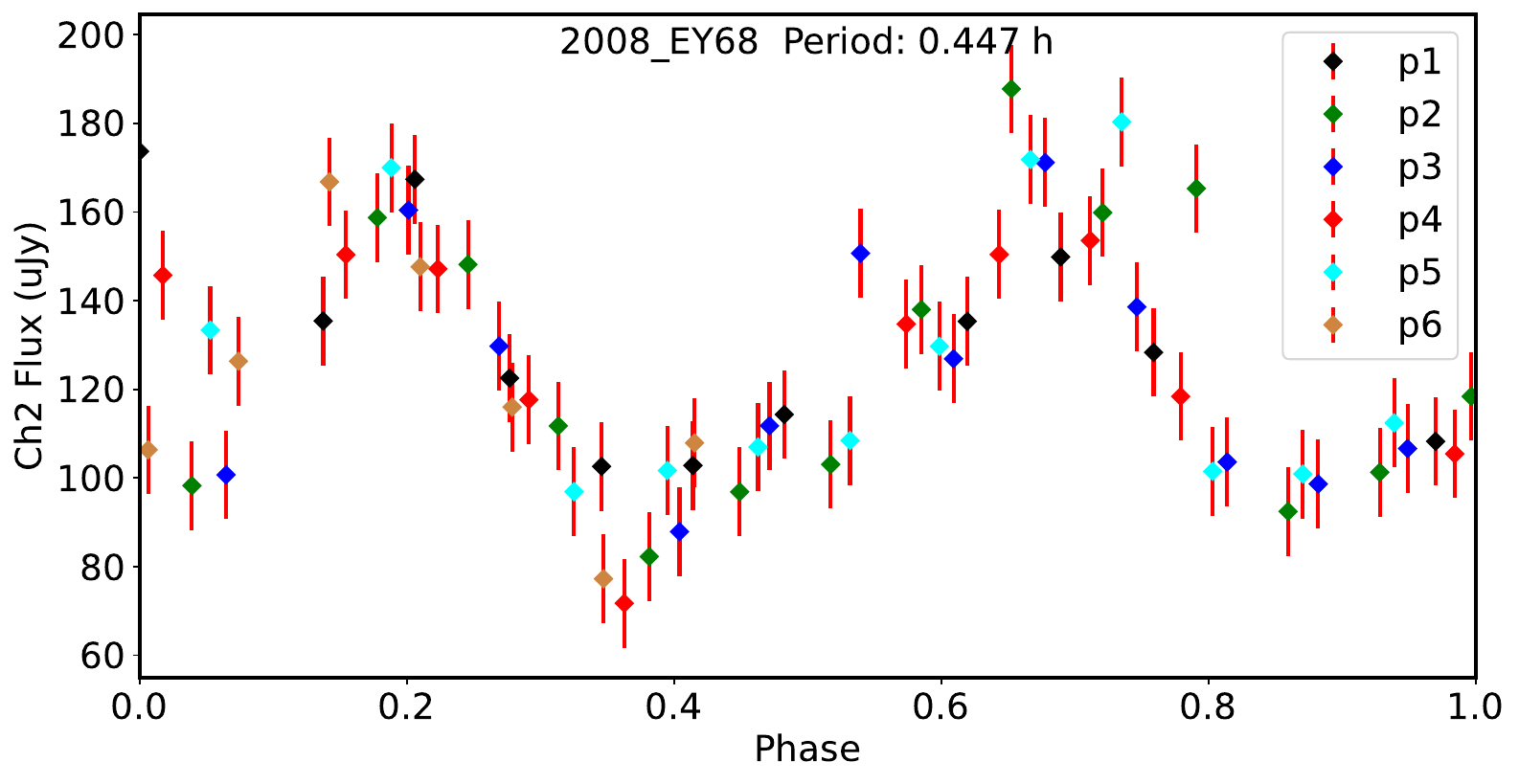}
        \includegraphics[width=0.495\linewidth]{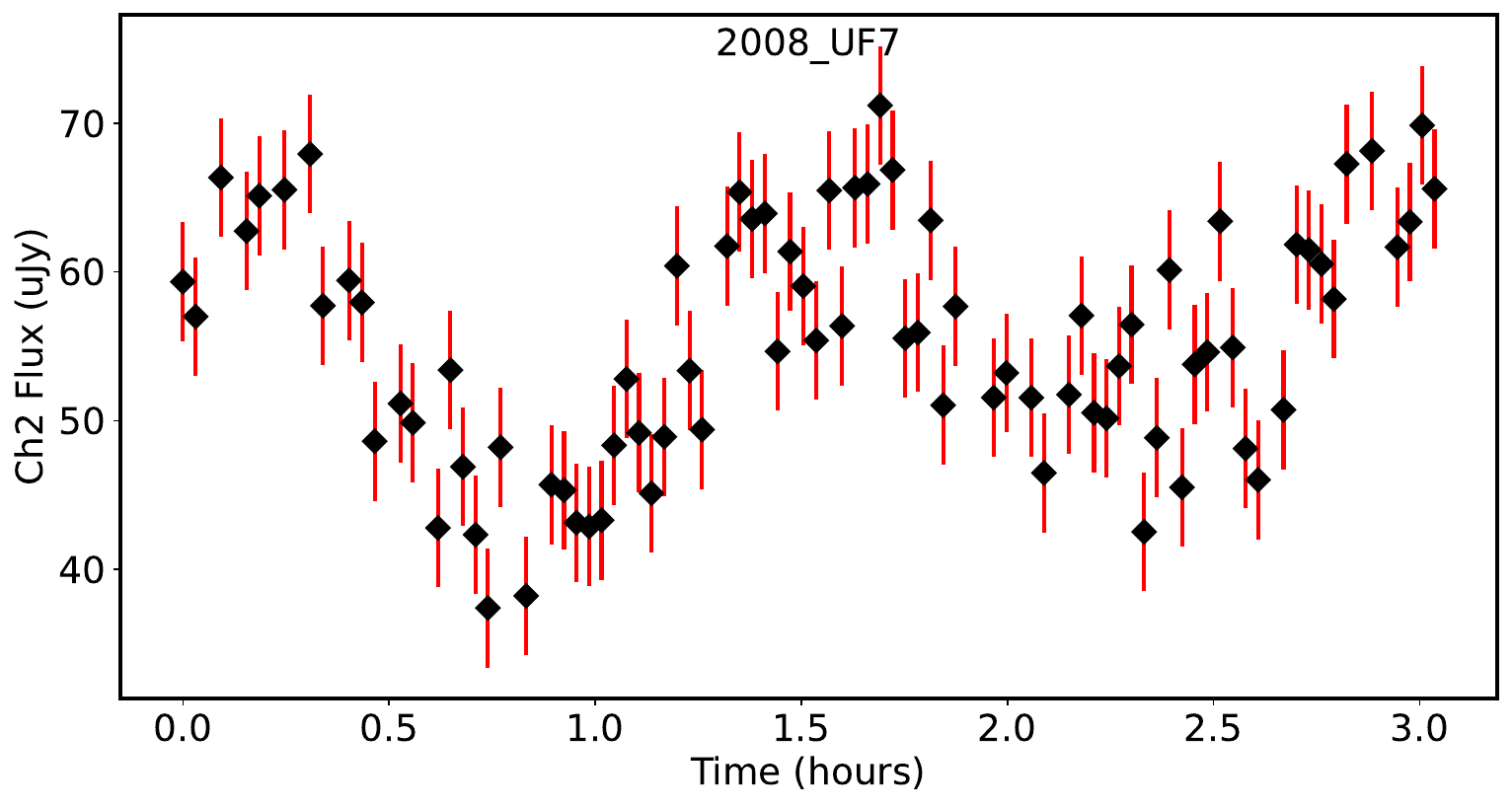}
    \includegraphics[width=0.495\linewidth]{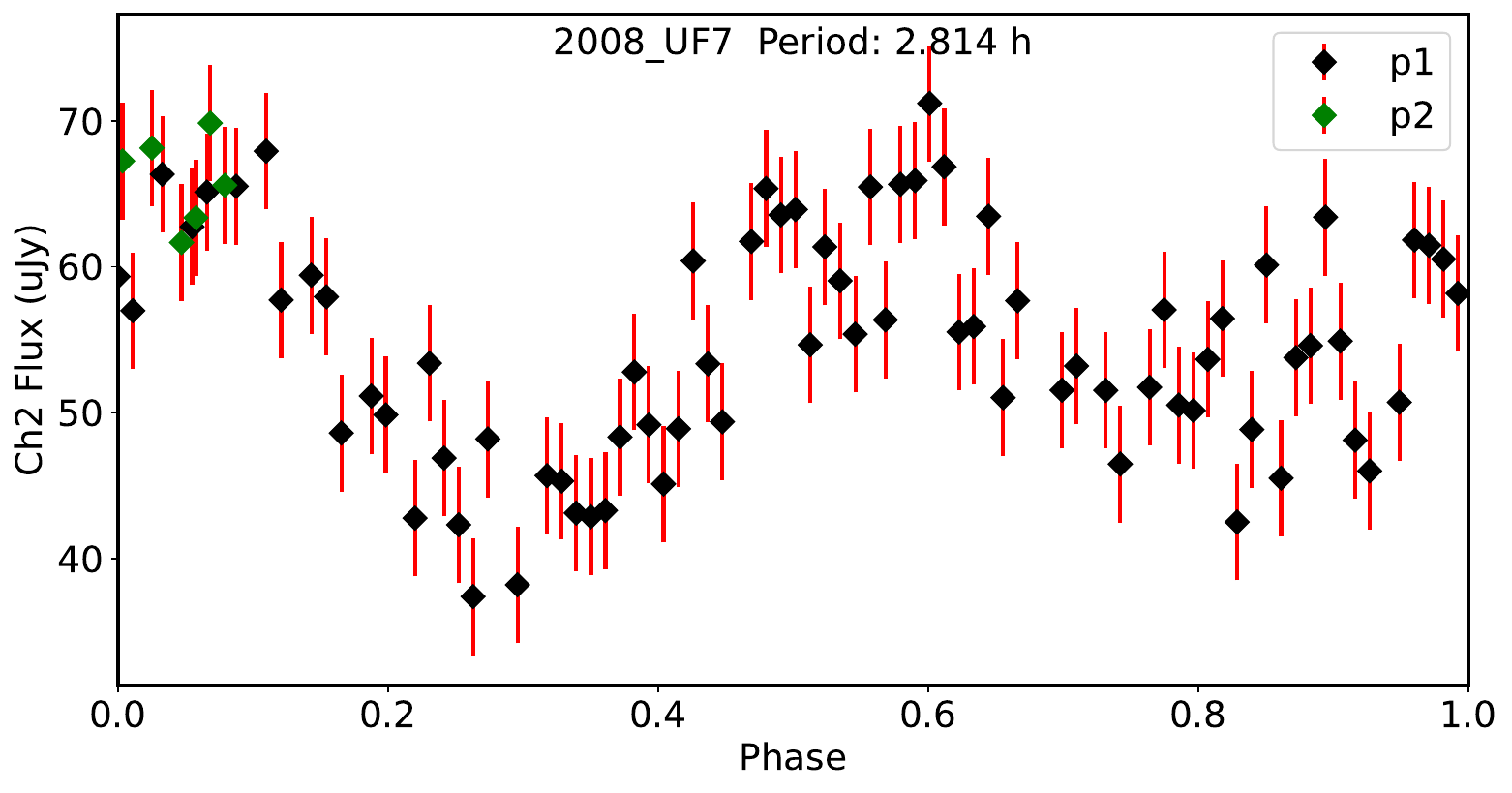}
    \includegraphics[width=0.495\linewidth]{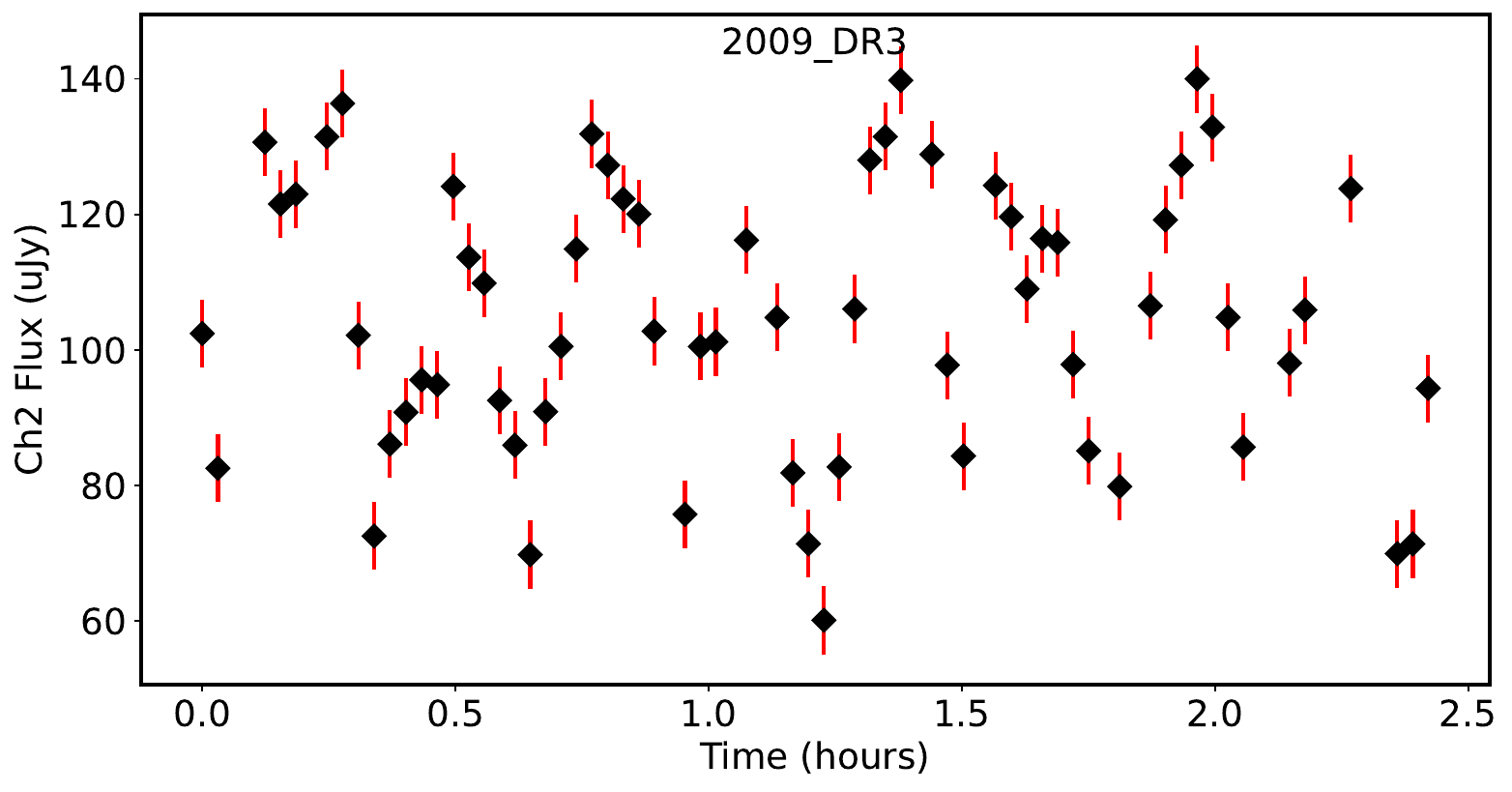}
    \includegraphics[width=0.495\linewidth]{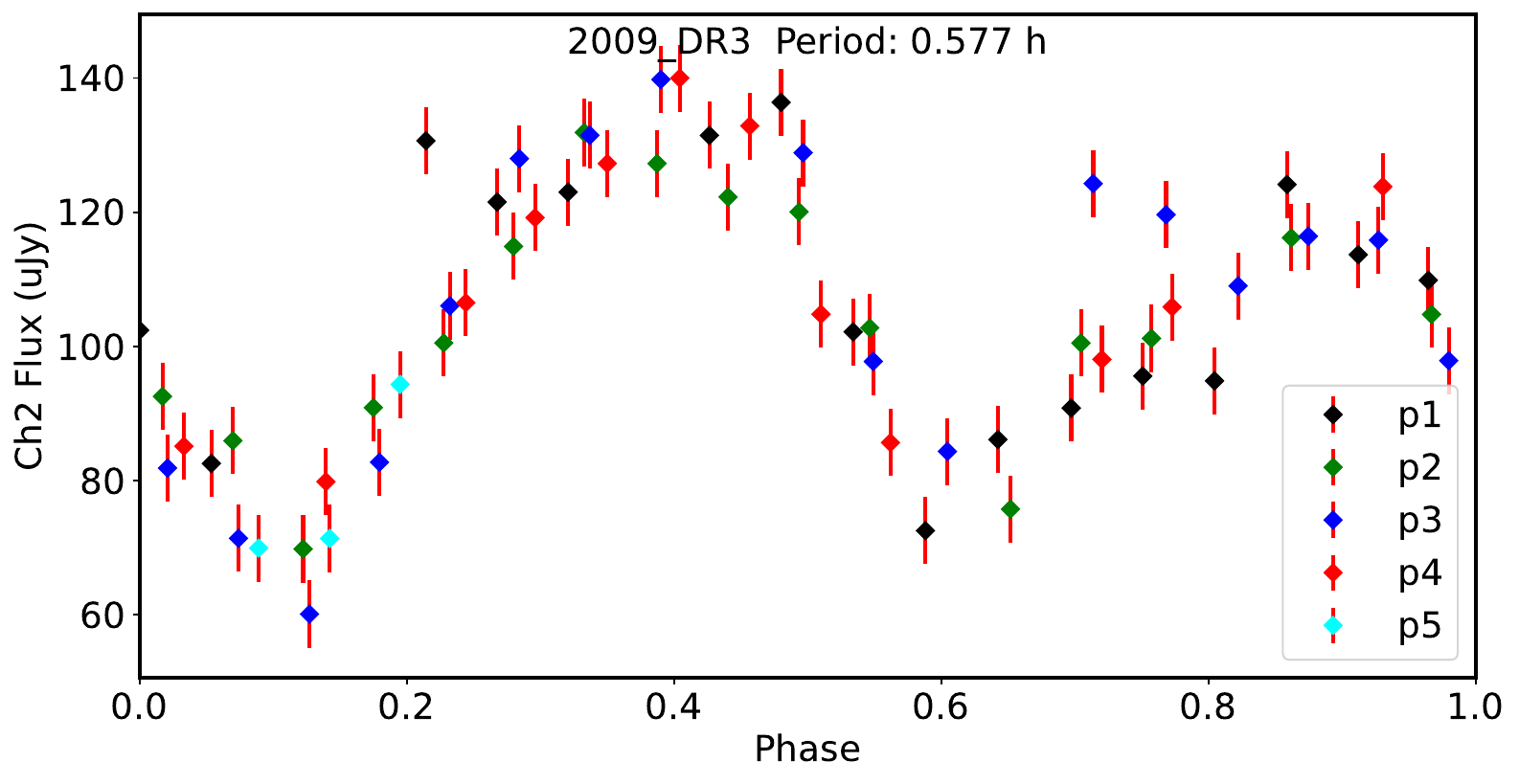}
    \includegraphics[width=0.495\linewidth]{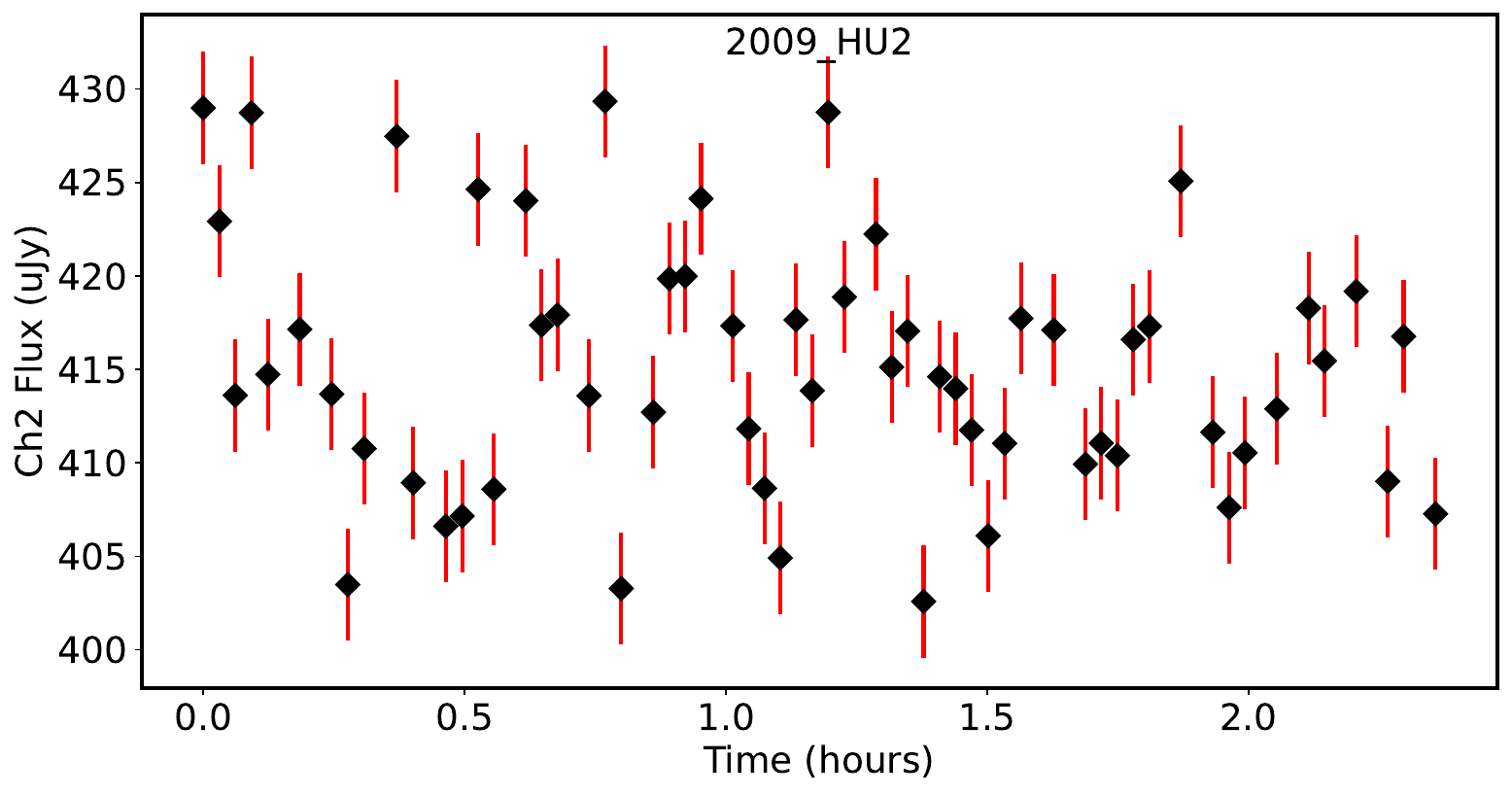}
    \includegraphics[width=0.495\linewidth]{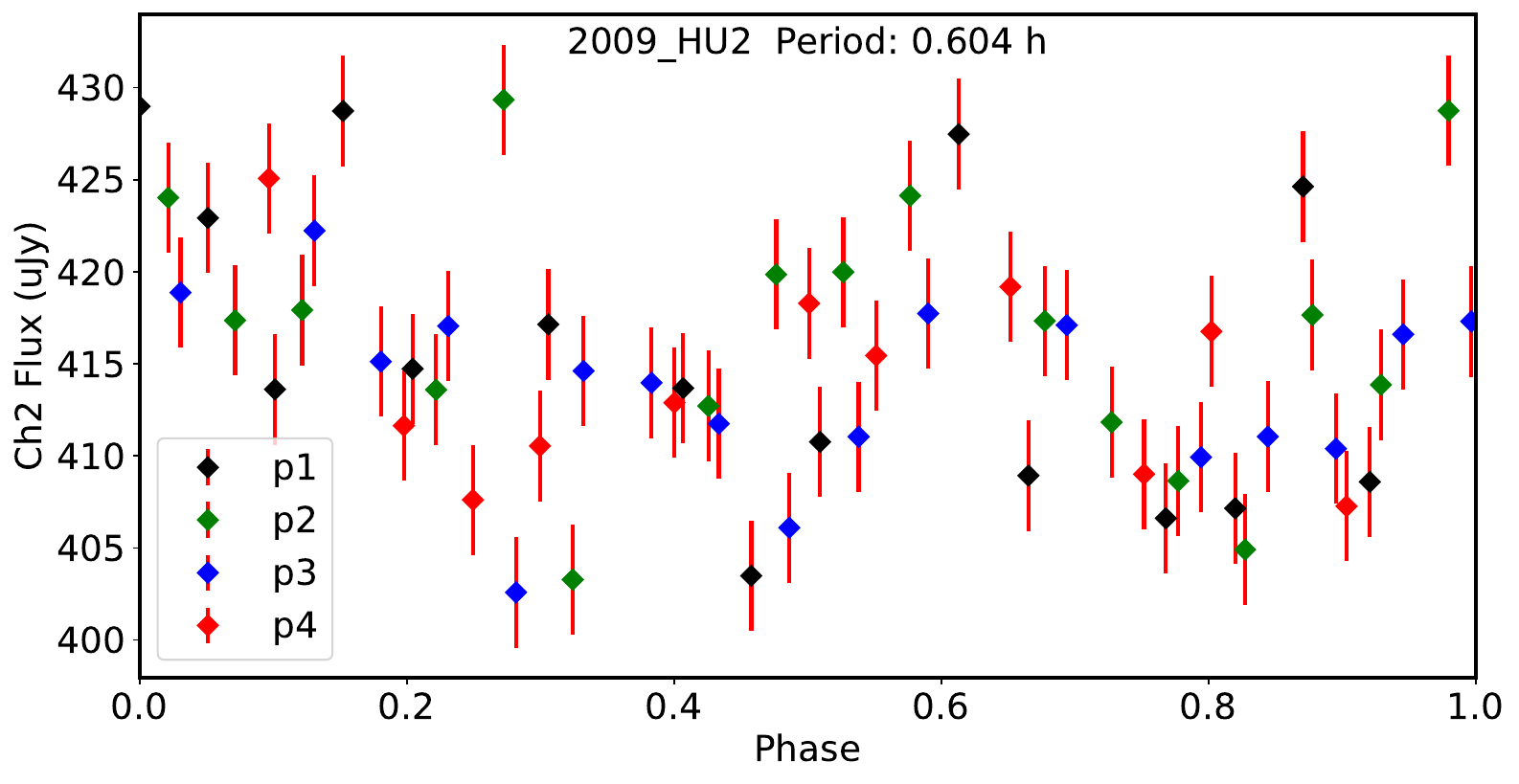}
    \caption{Lightcurves (left column) and phased lightcurves (right column) for sources with one or more periods sampled and periods determined. The periods are plotted with different colors in the plots on the right.}
    \label{fig:lc6}
\end{figure*}

\begin{figure*}
    \centering
    \includegraphics[width=0.495\linewidth]{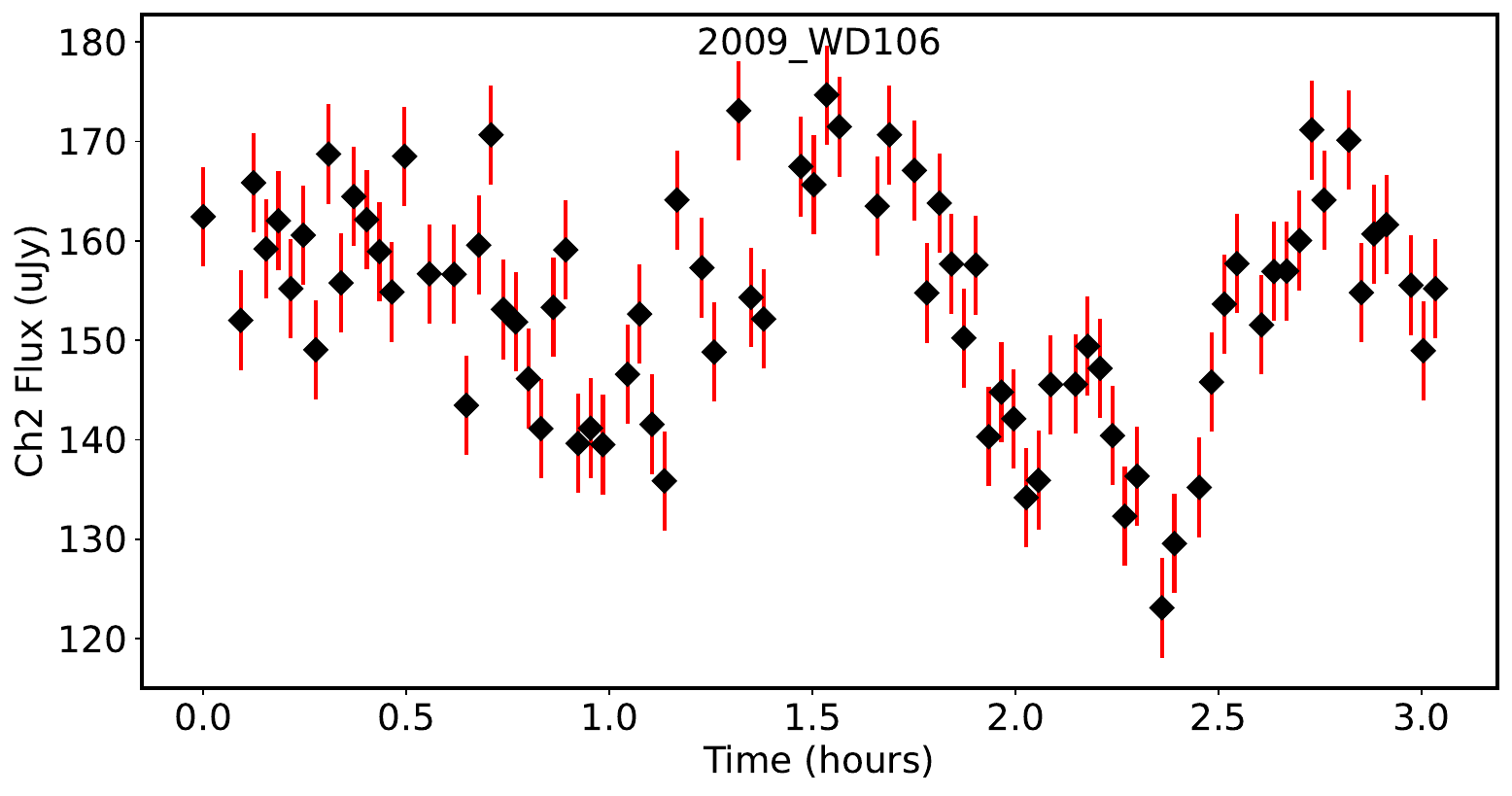}
    \includegraphics[width=0.495\linewidth]{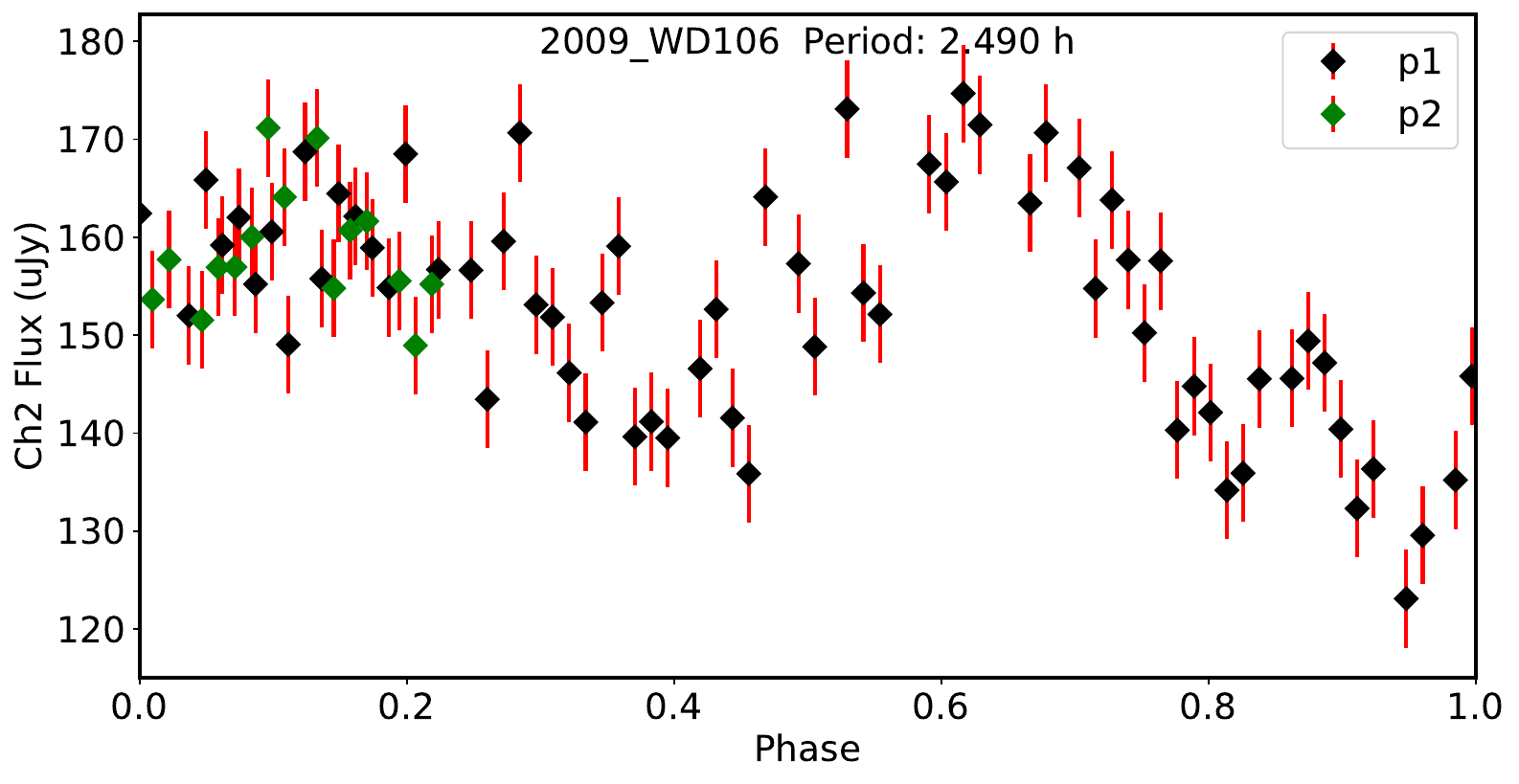}
    \includegraphics[width=0.495\linewidth]{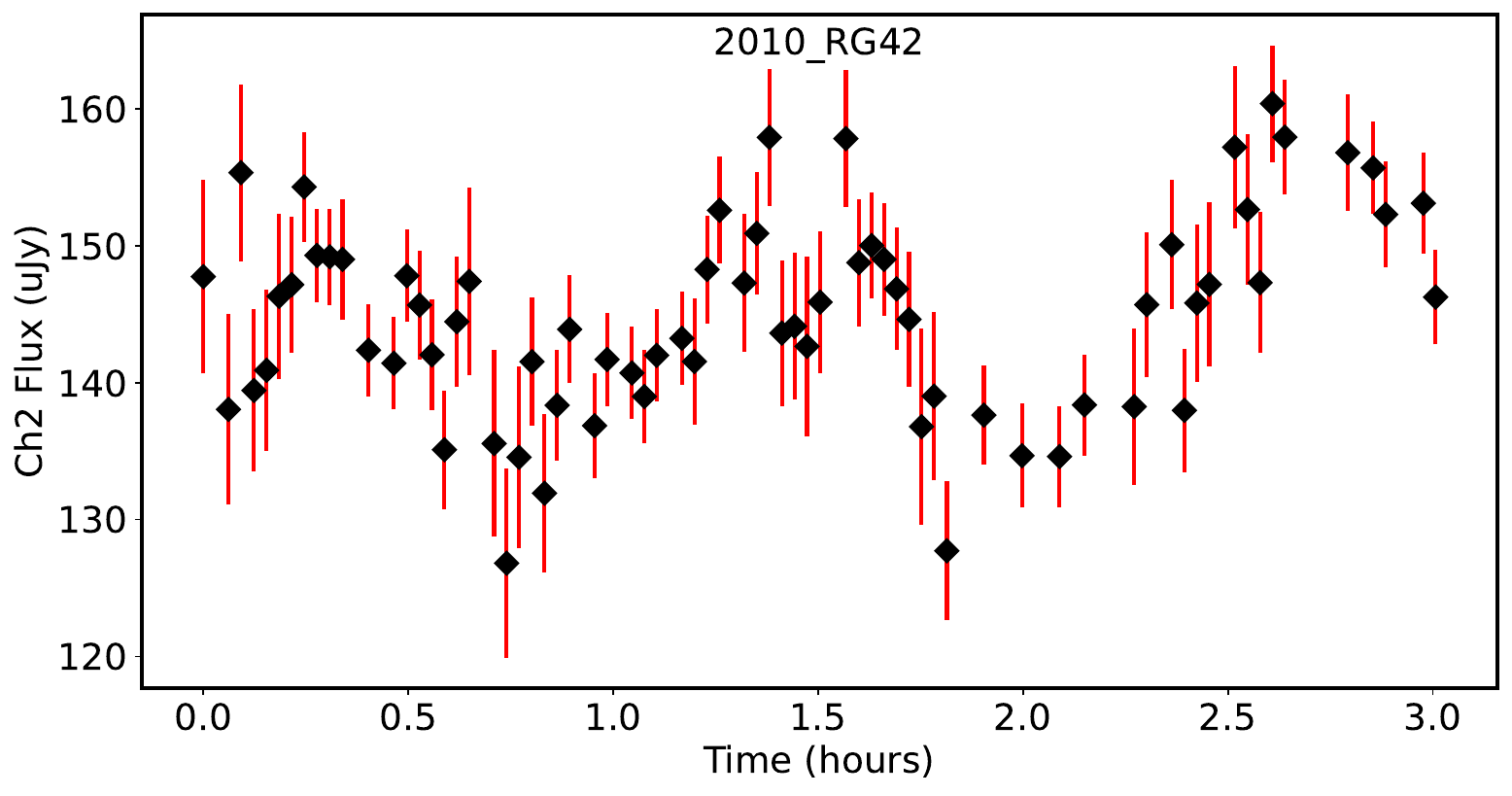}
    \includegraphics[width=0.495\linewidth]{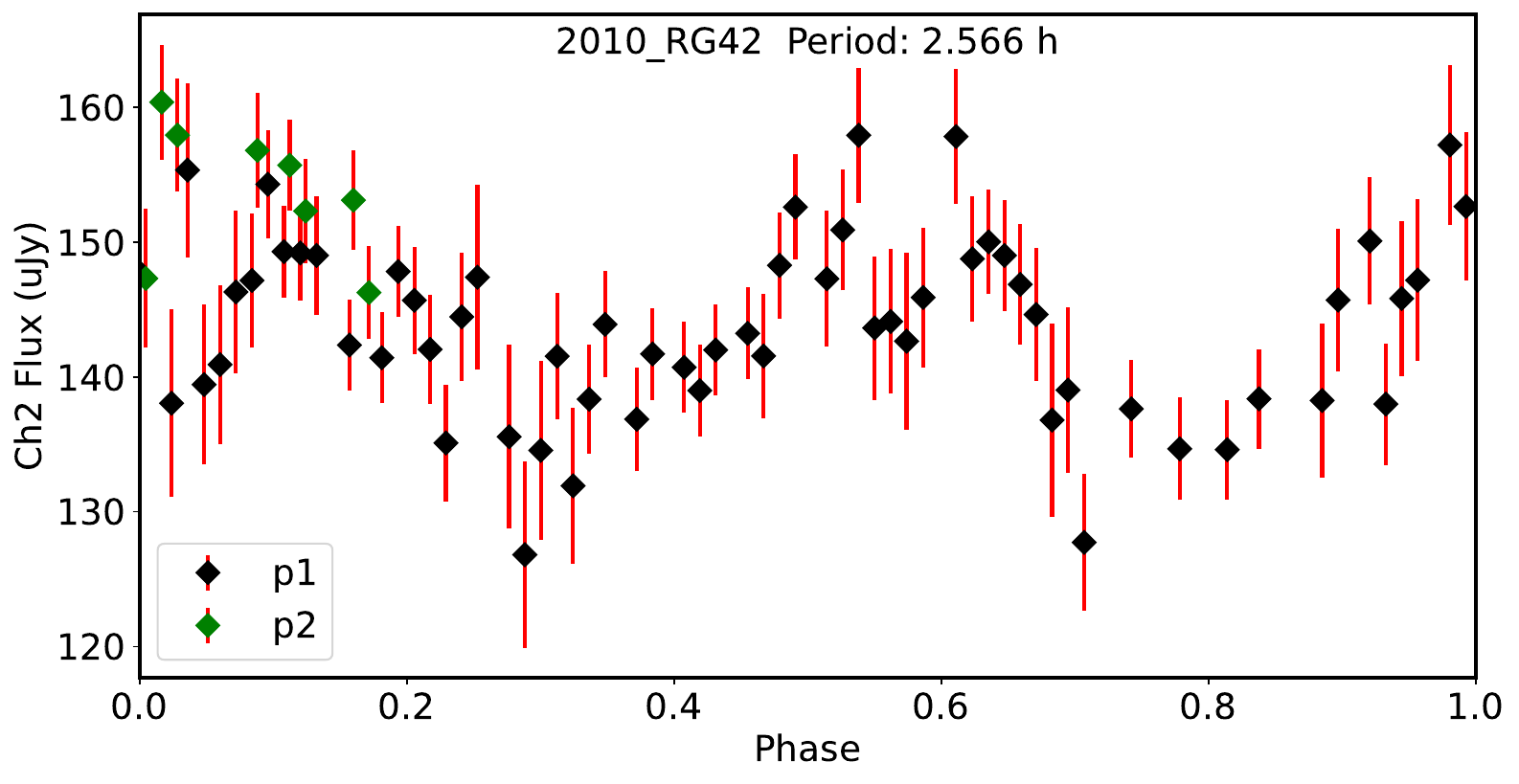}
    \includegraphics[width=0.495\linewidth]{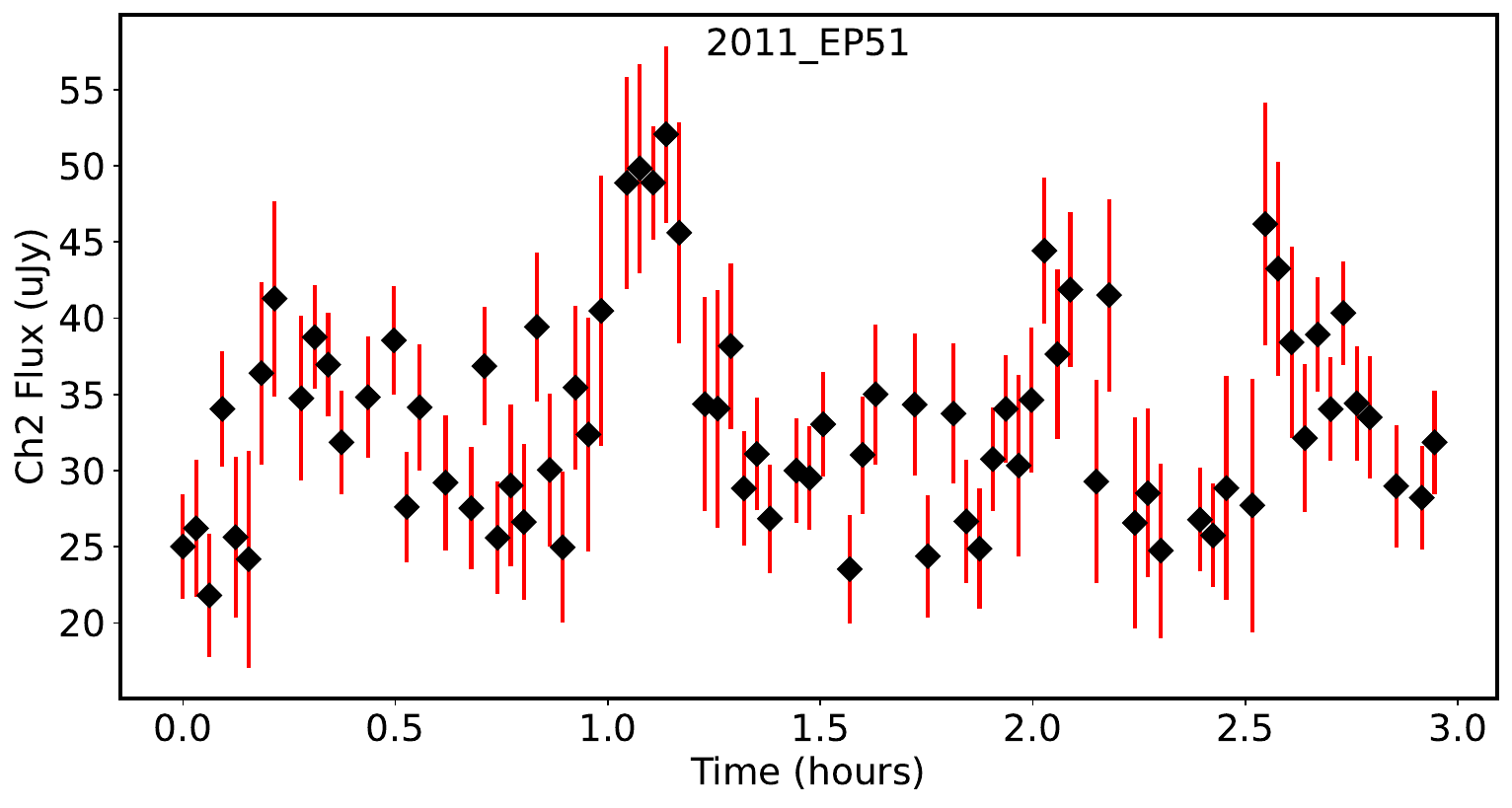}
    \includegraphics[width=0.495\linewidth]{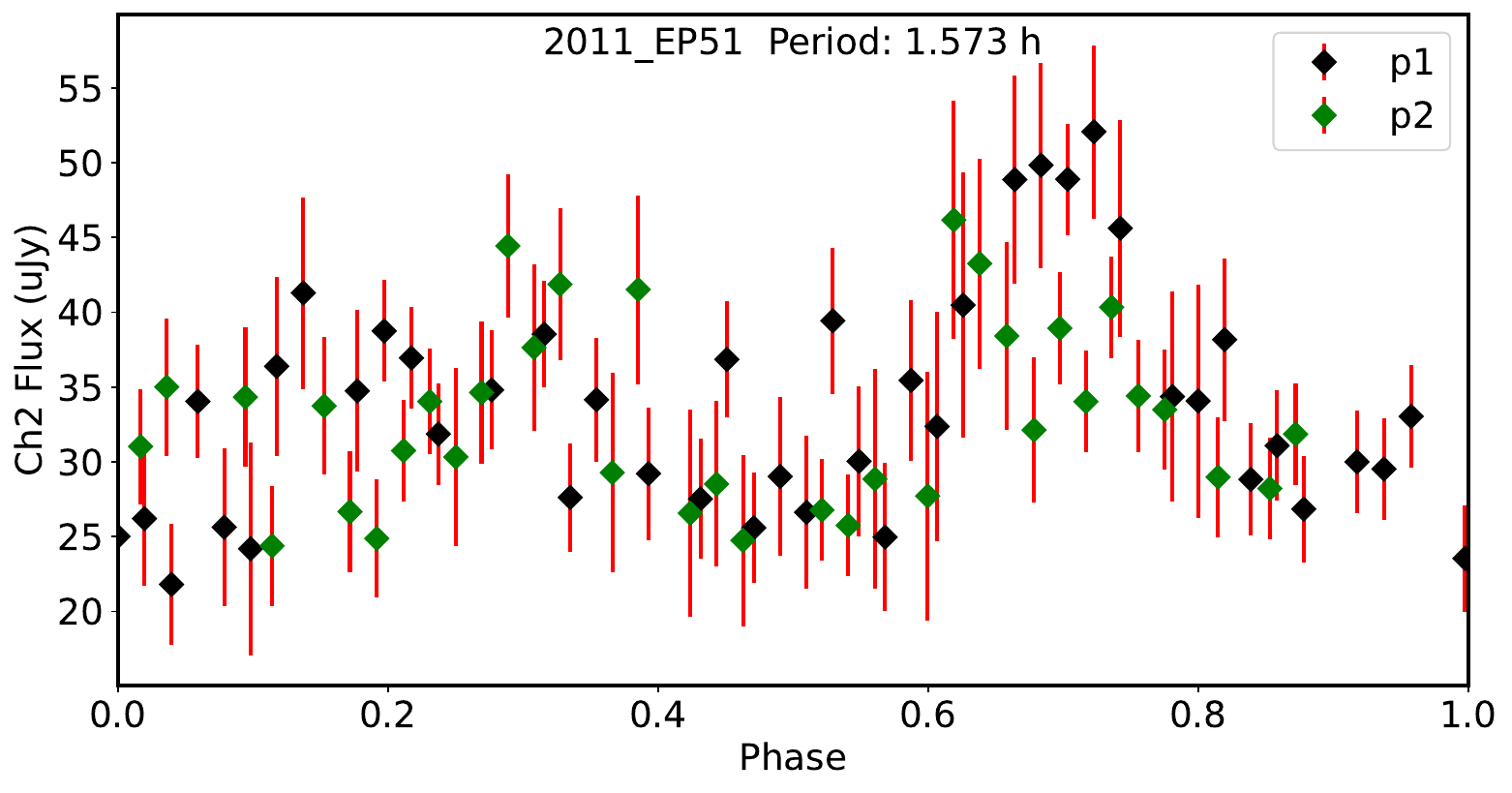}
    \includegraphics[width=0.495\linewidth]{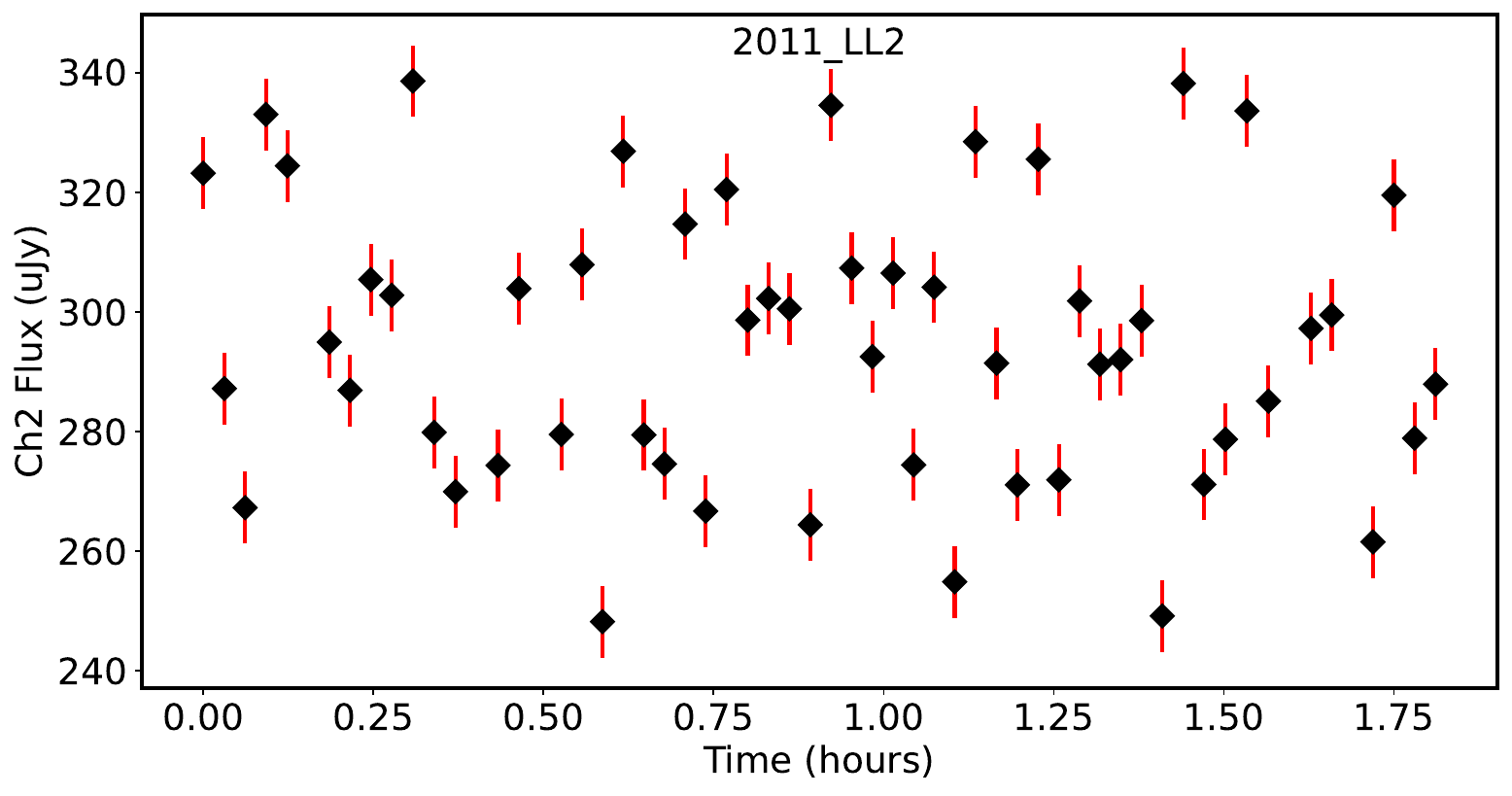}
    \includegraphics[width=0.495\linewidth]{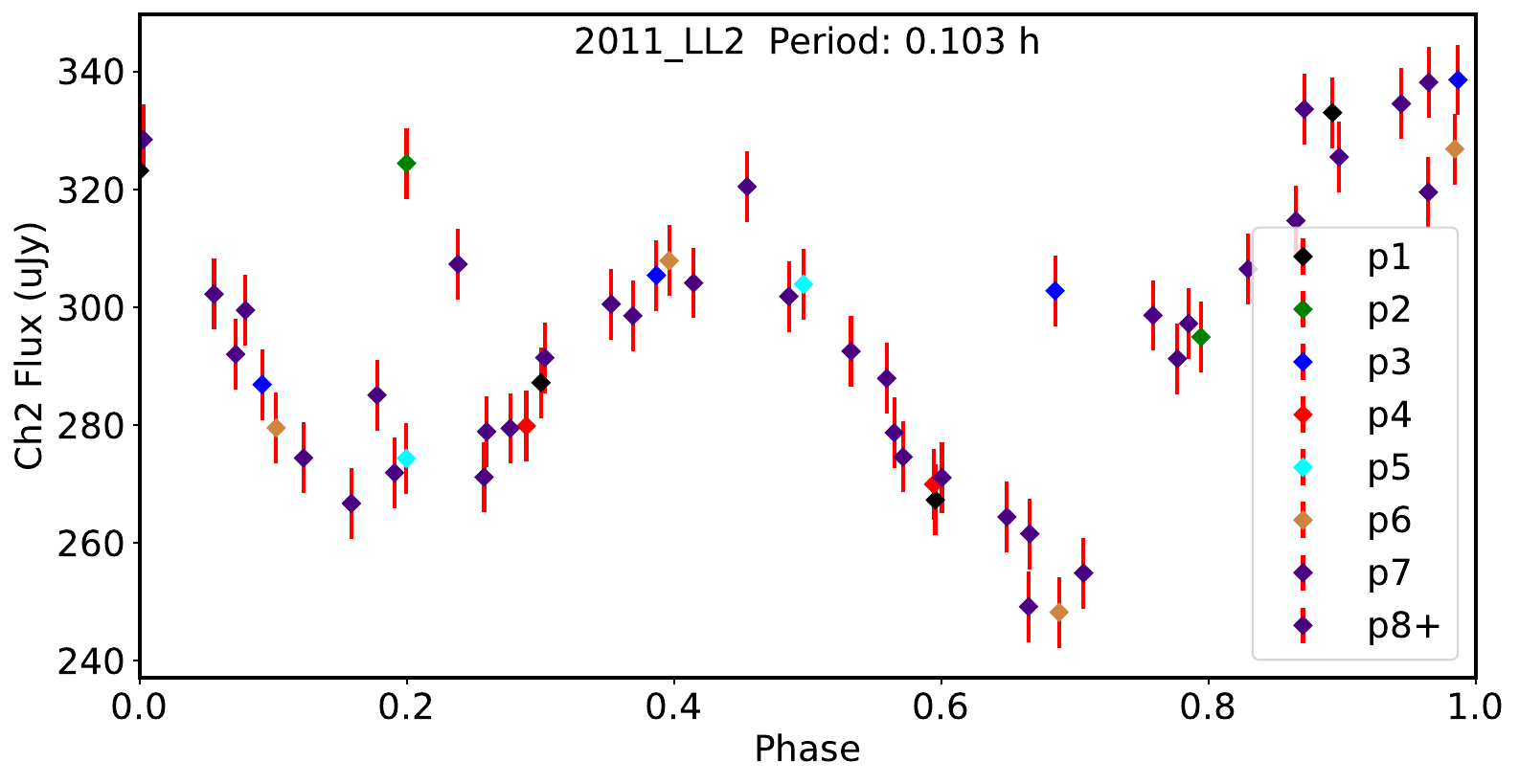}
    \includegraphics[width=0.495\linewidth]{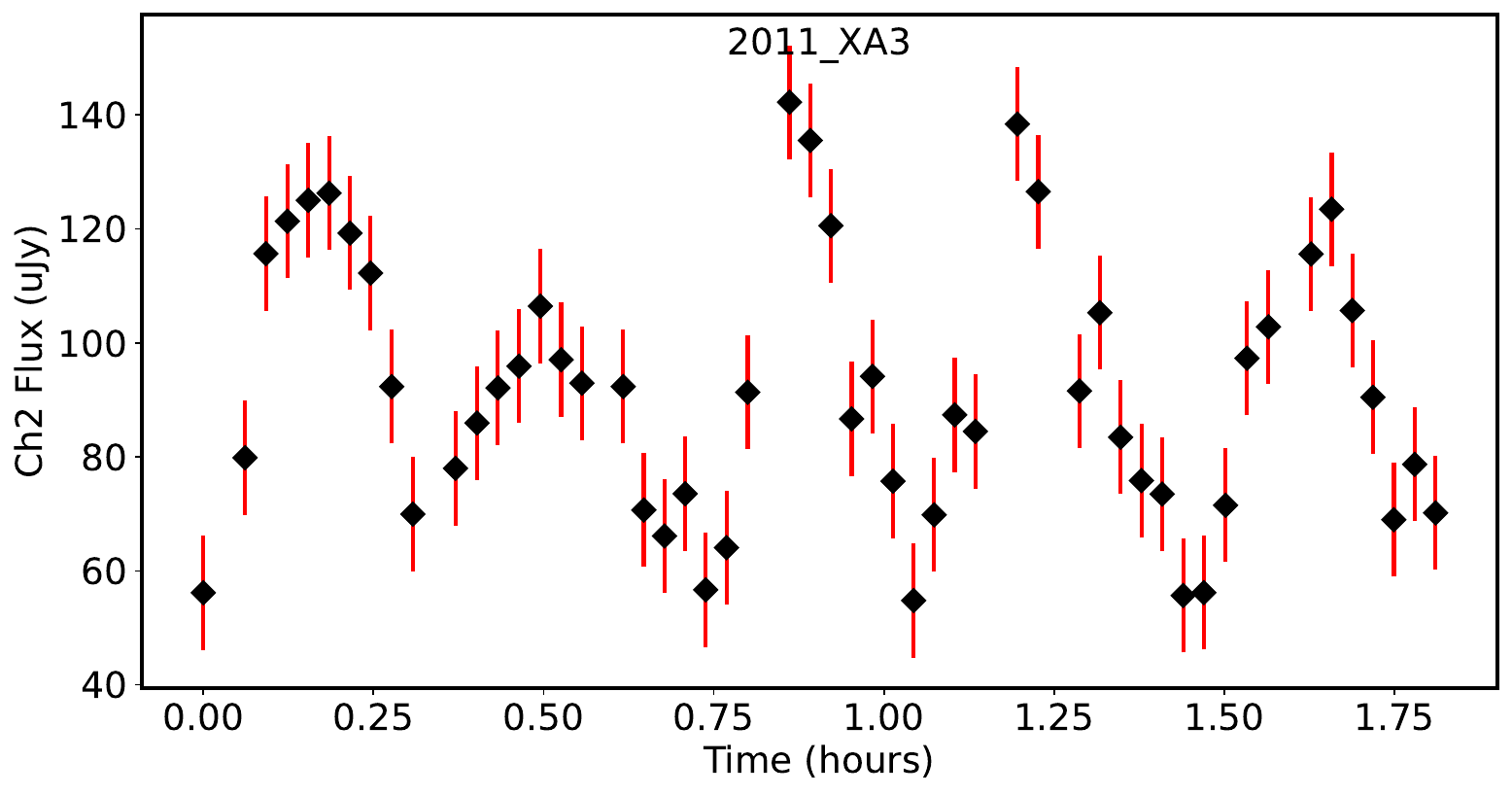}
    \includegraphics[width=0.495\linewidth]{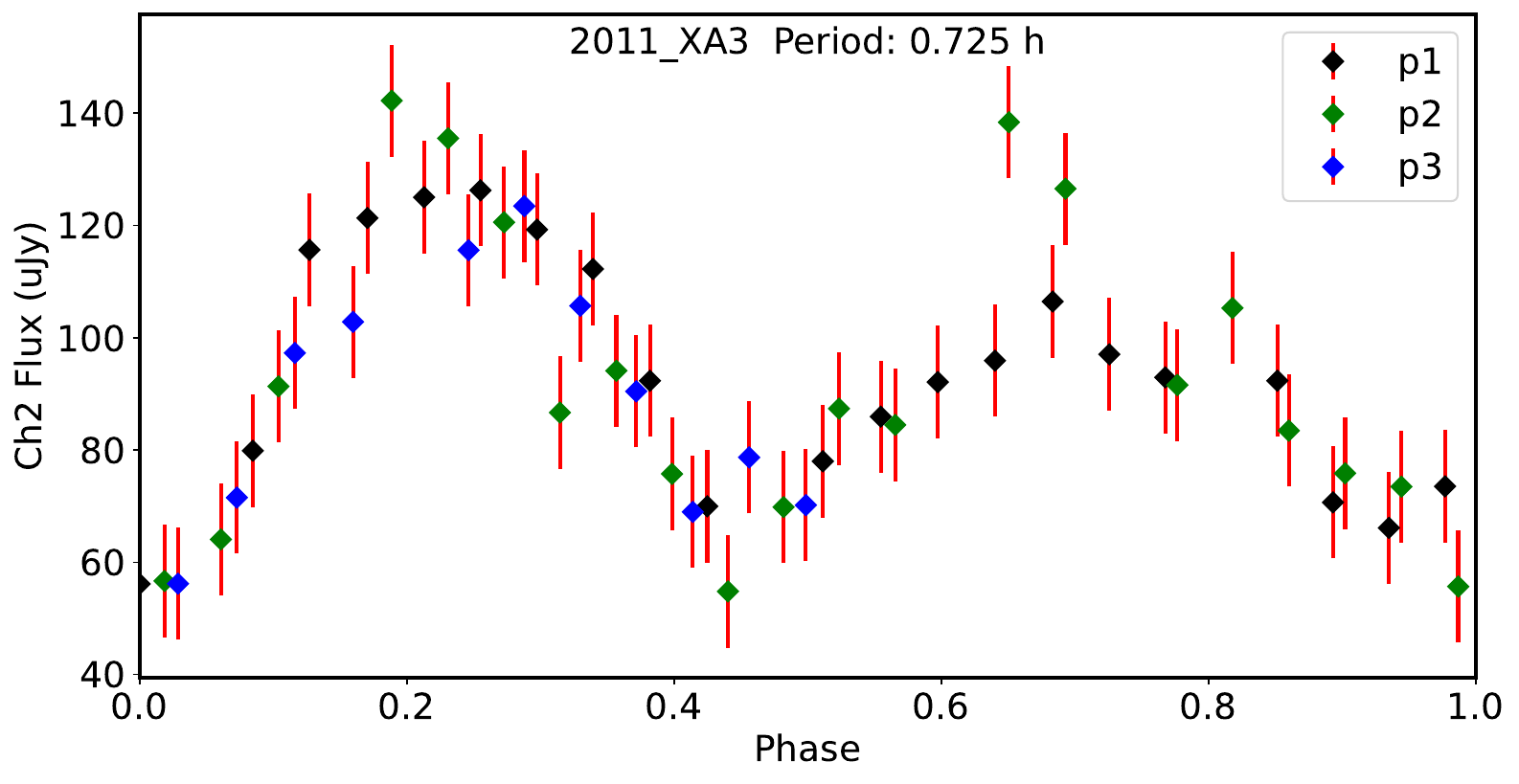}
\caption{Lightcurves (left column) and phased lightcurves (right column) for sources with one or more periods sampled and periods determined. The periods are plotted with different colors in the plots on the right.}
    \label{fig:lc7}
\end{figure*}

\begin{figure*}
    \centering
    \includegraphics[width=0.495\linewidth]{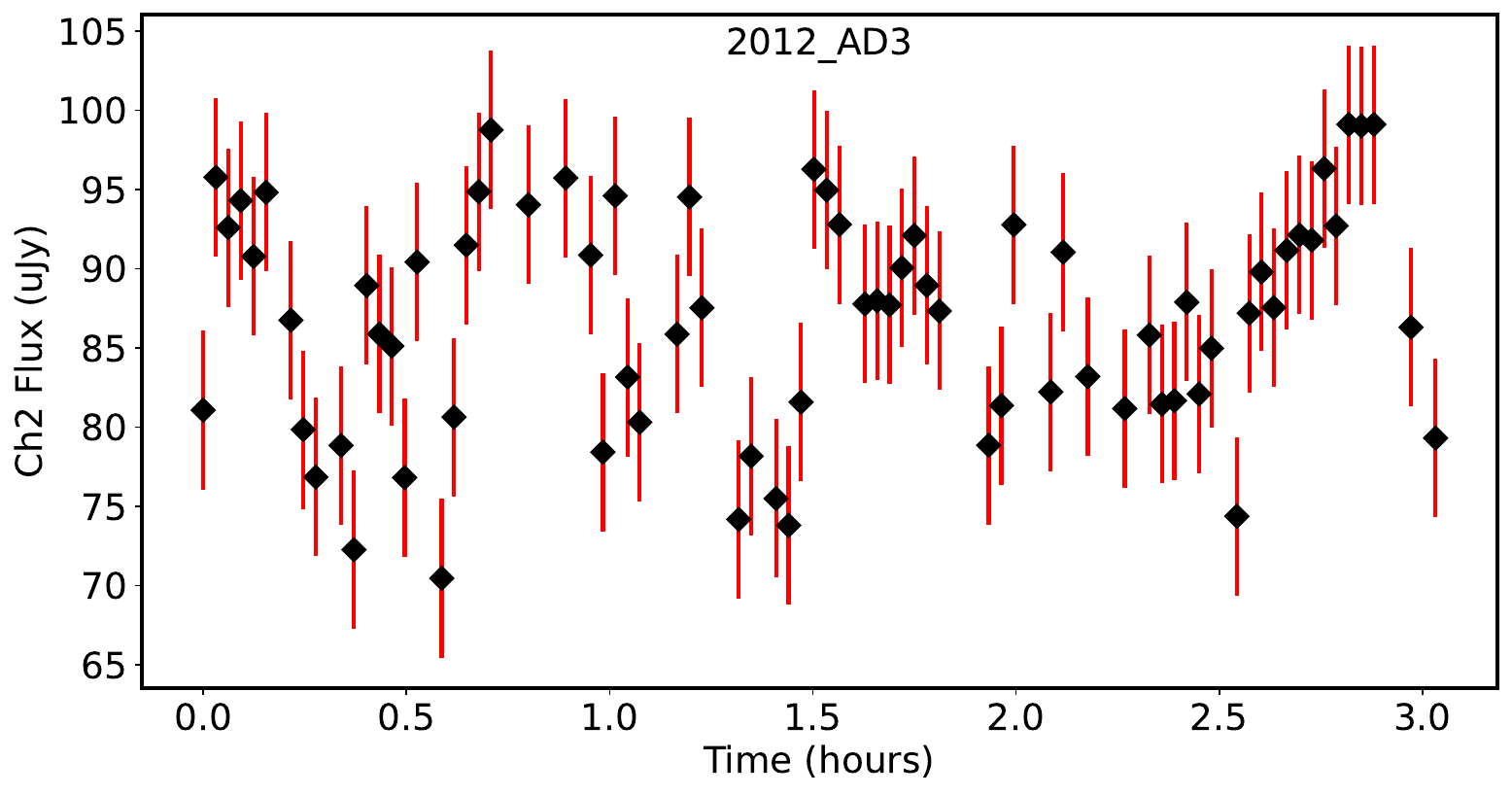}
    \includegraphics[width=0.495\linewidth]{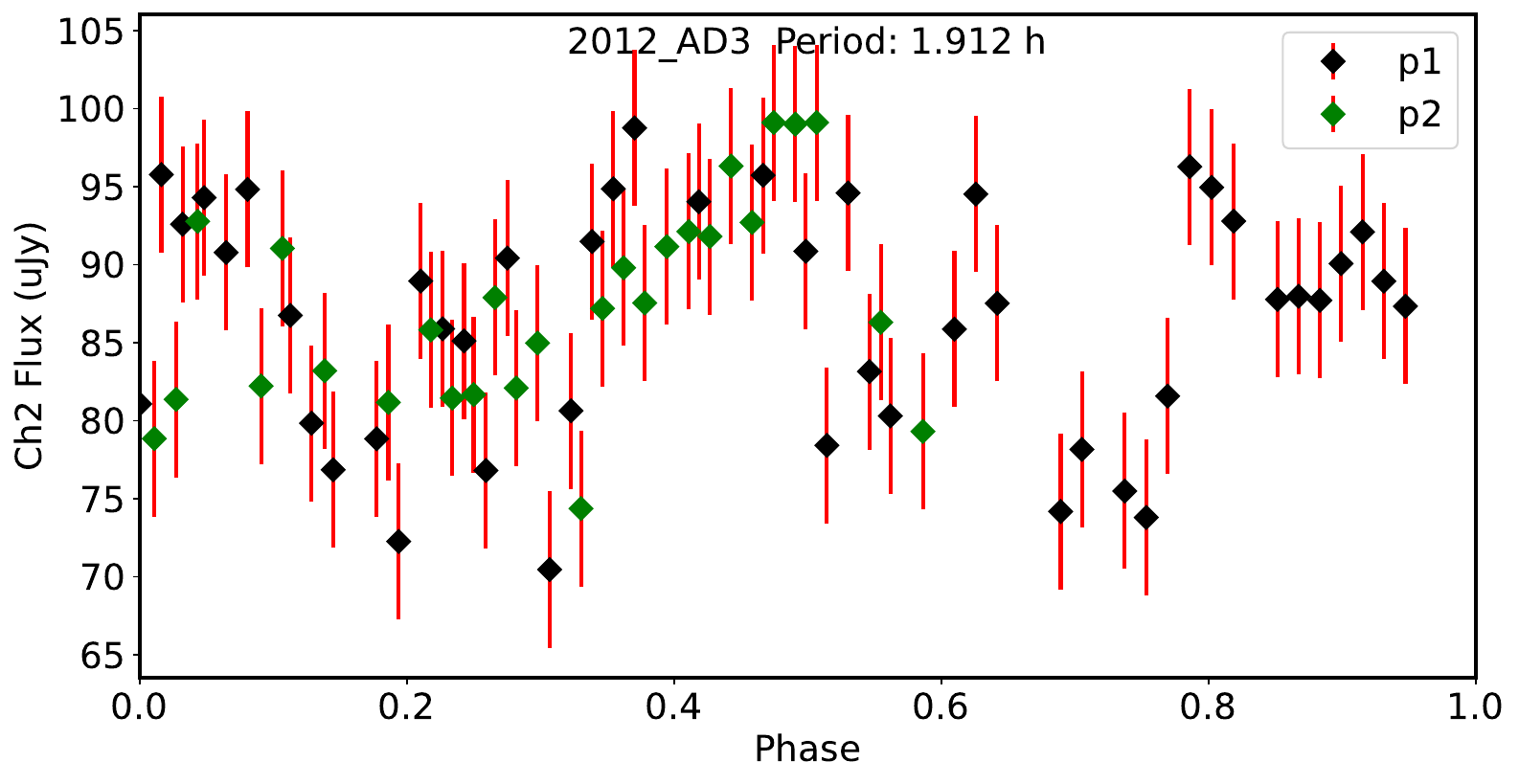}       \includegraphics[width=0.495\linewidth]{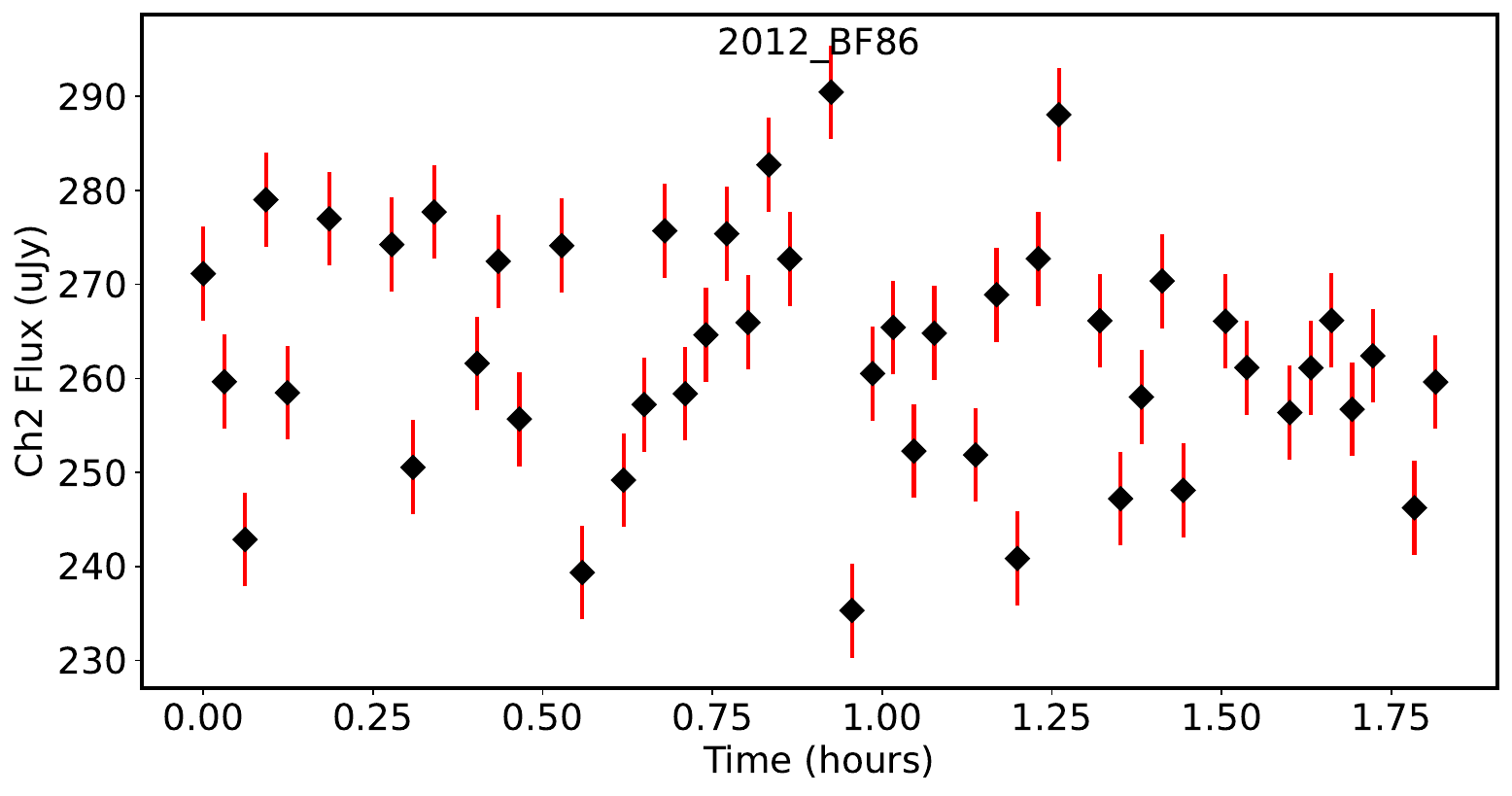}
    \includegraphics[width=0.495\linewidth]{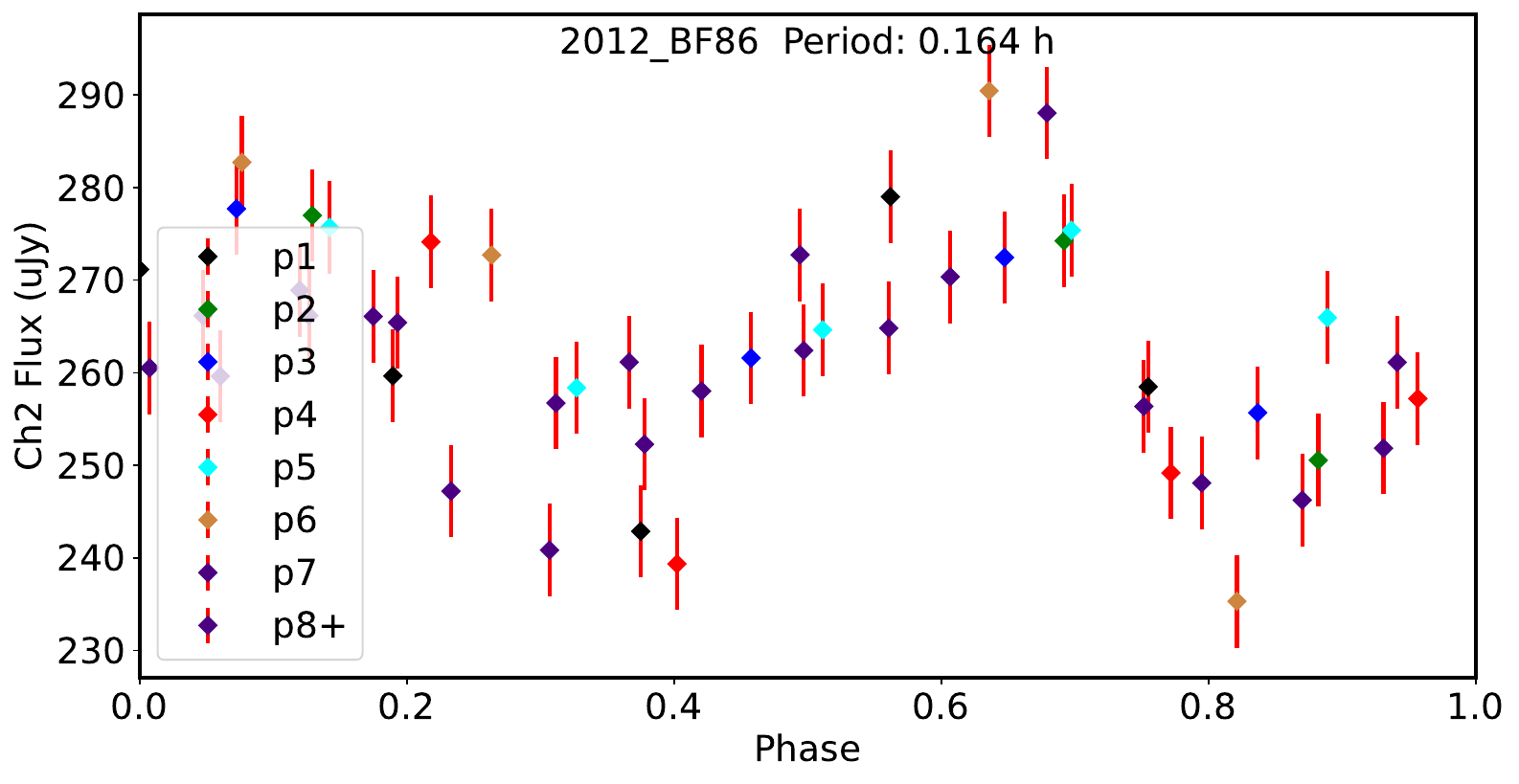}
    \includegraphics[width=0.495\linewidth]{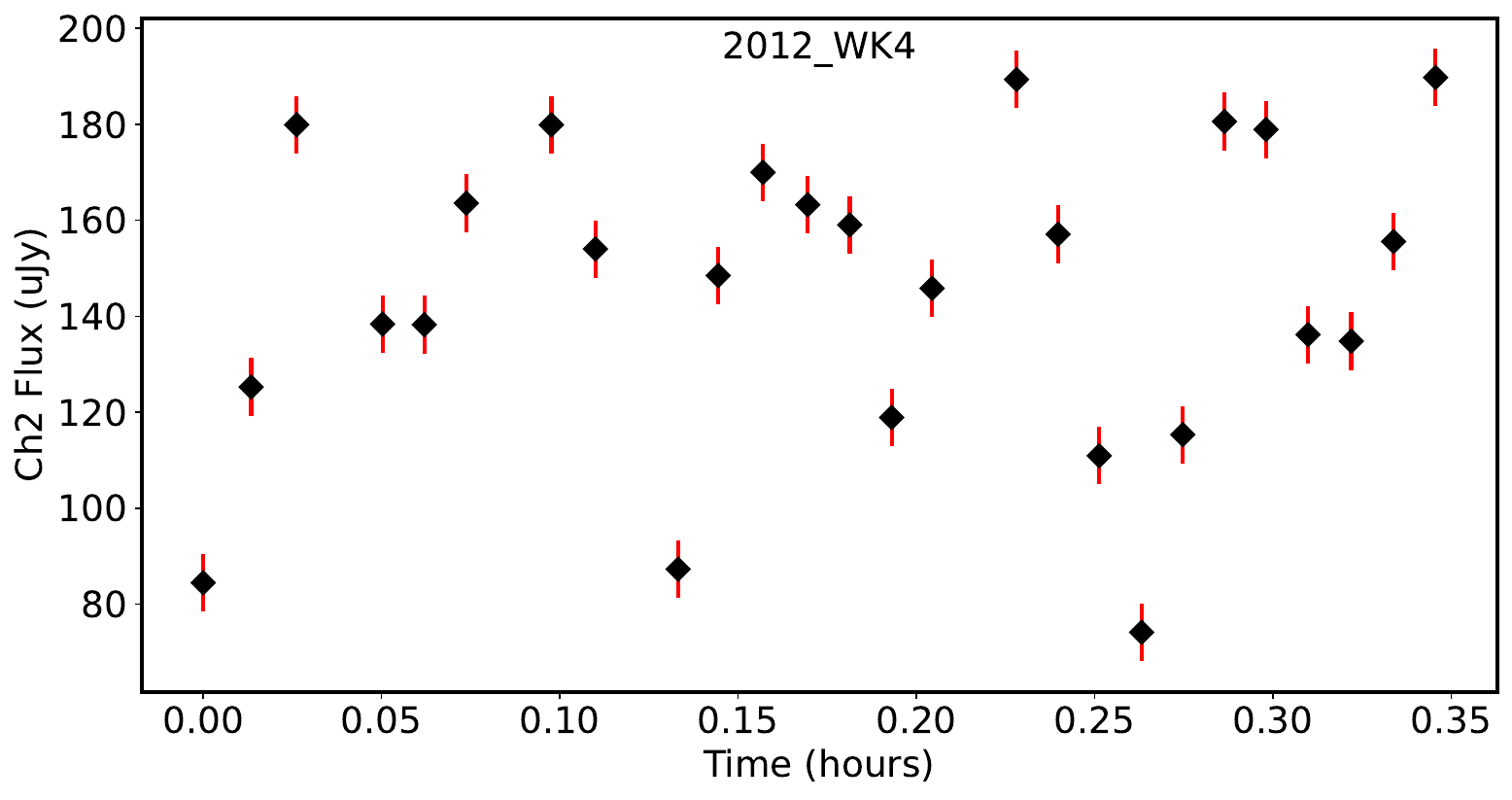}
    \includegraphics[width=0.495\linewidth]{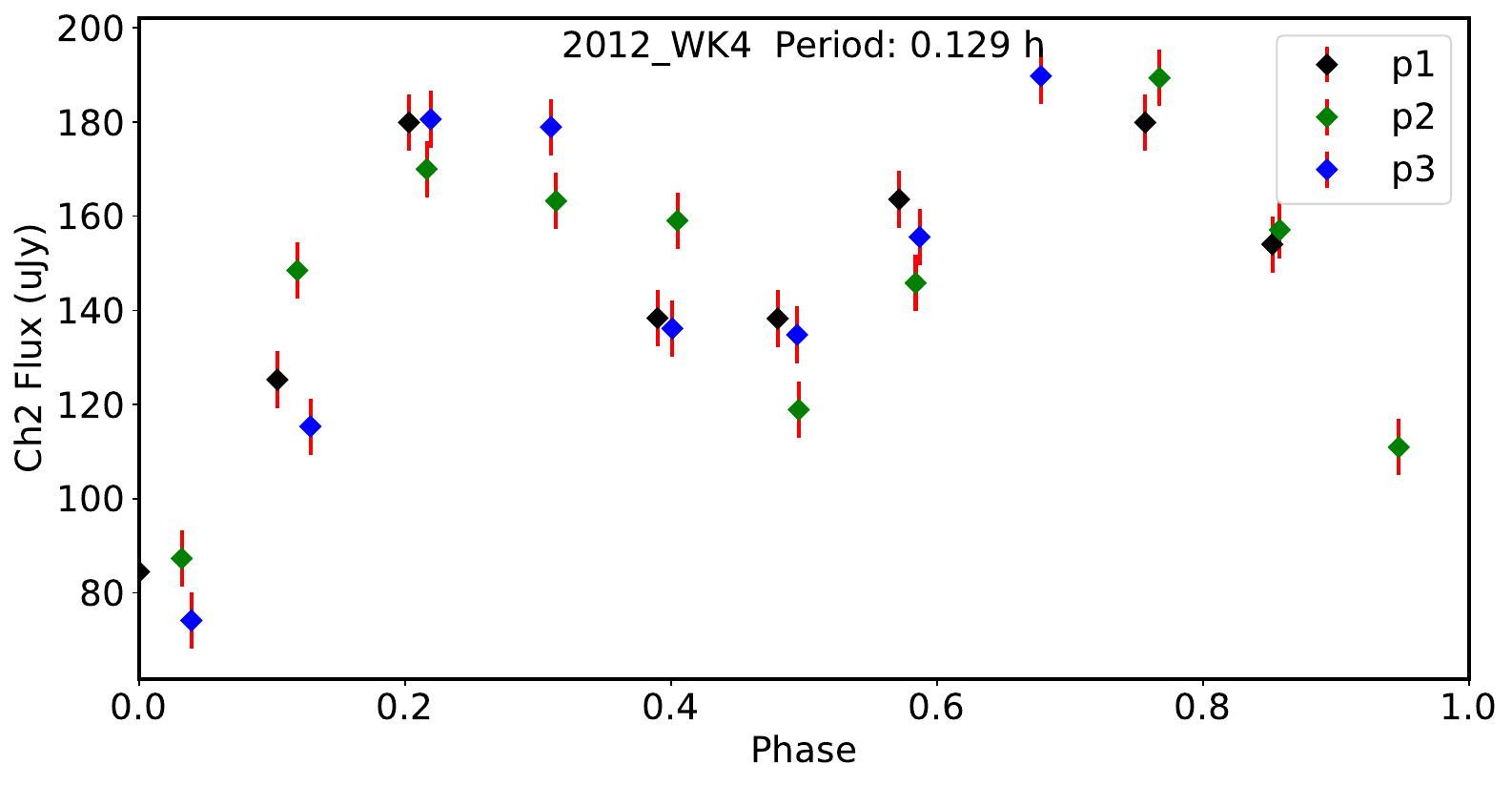}    
    \includegraphics[width=0.495\linewidth]{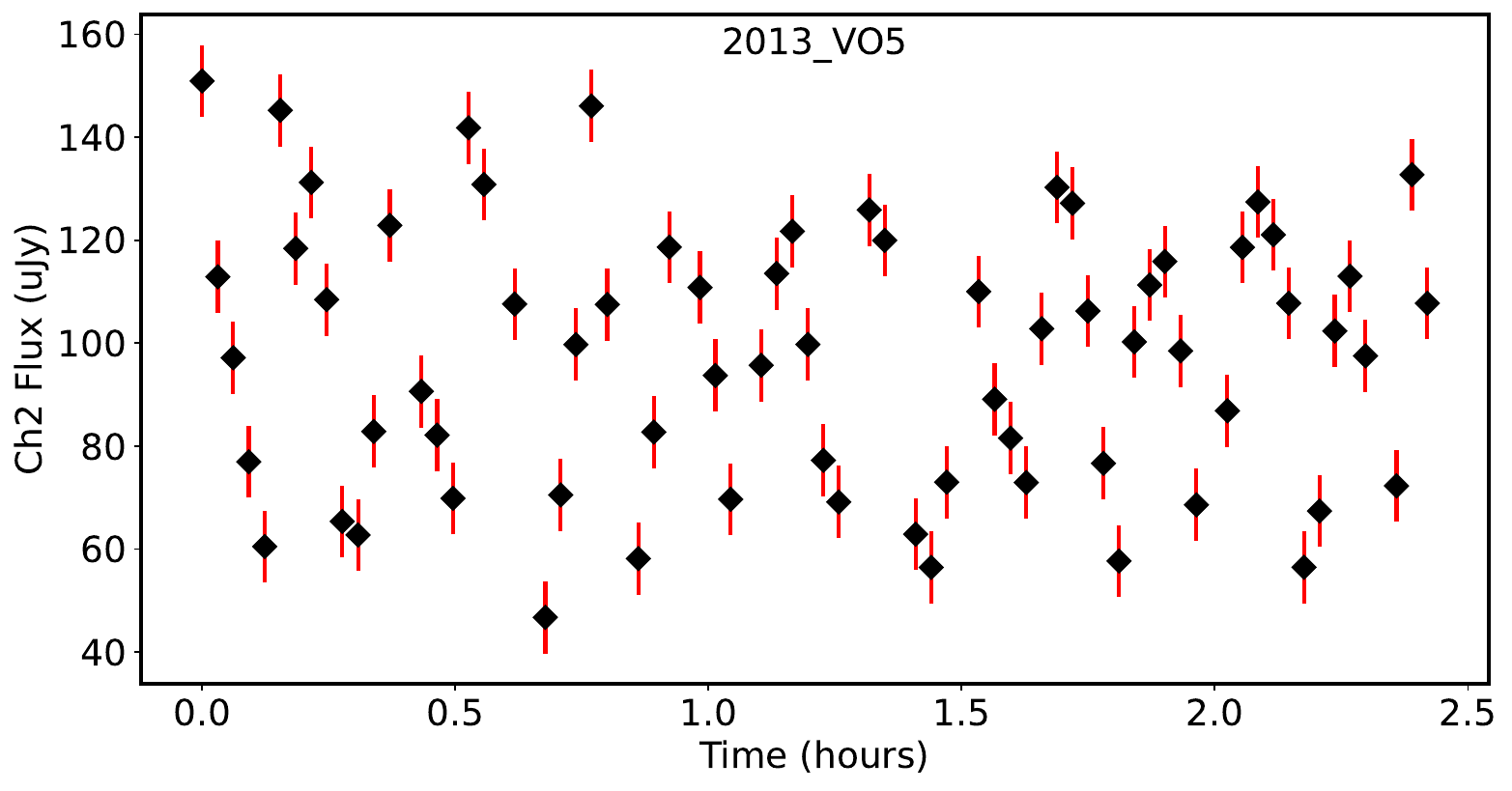}
    \includegraphics[width=0.495\linewidth]{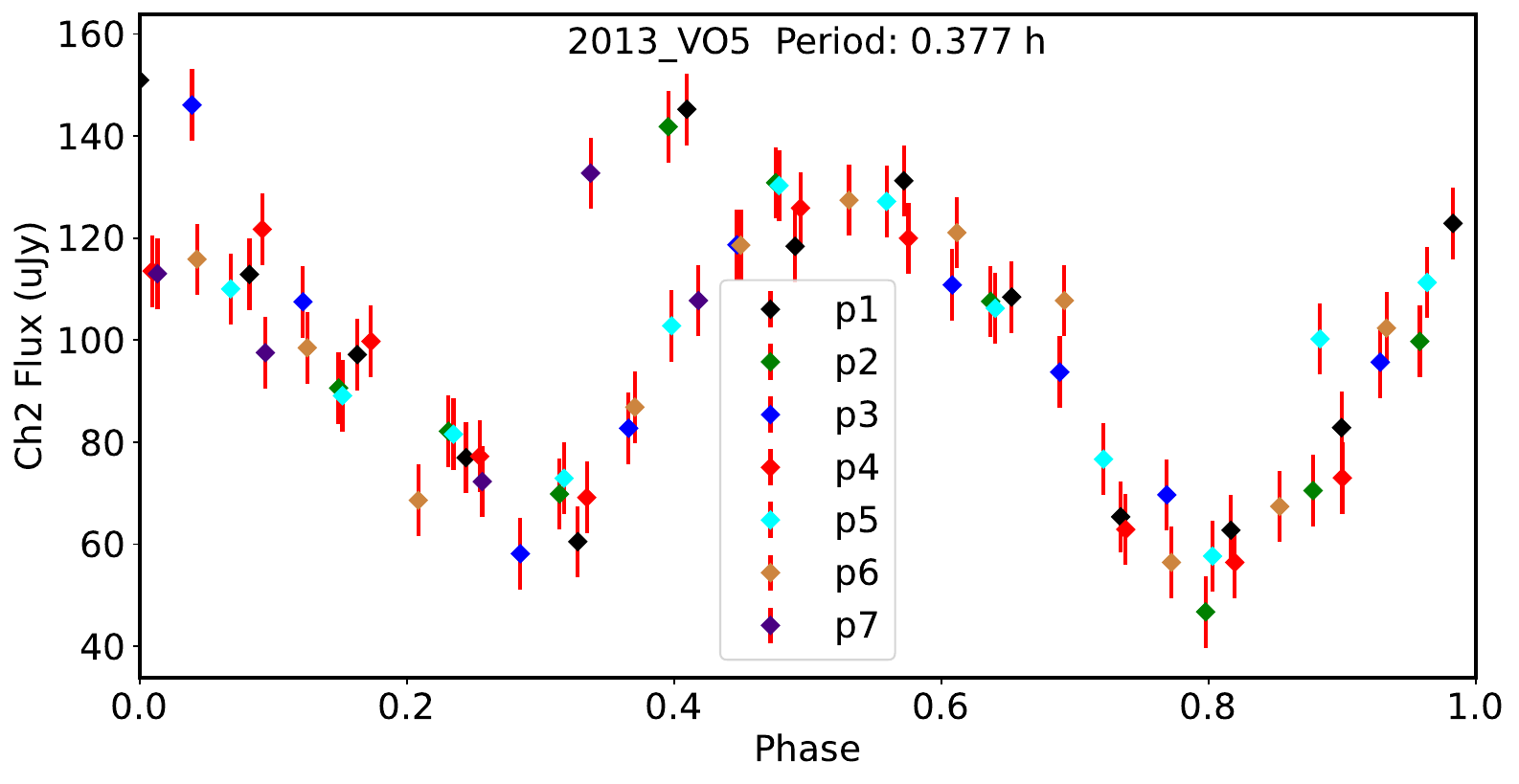}
    \includegraphics[width=0.495\linewidth]{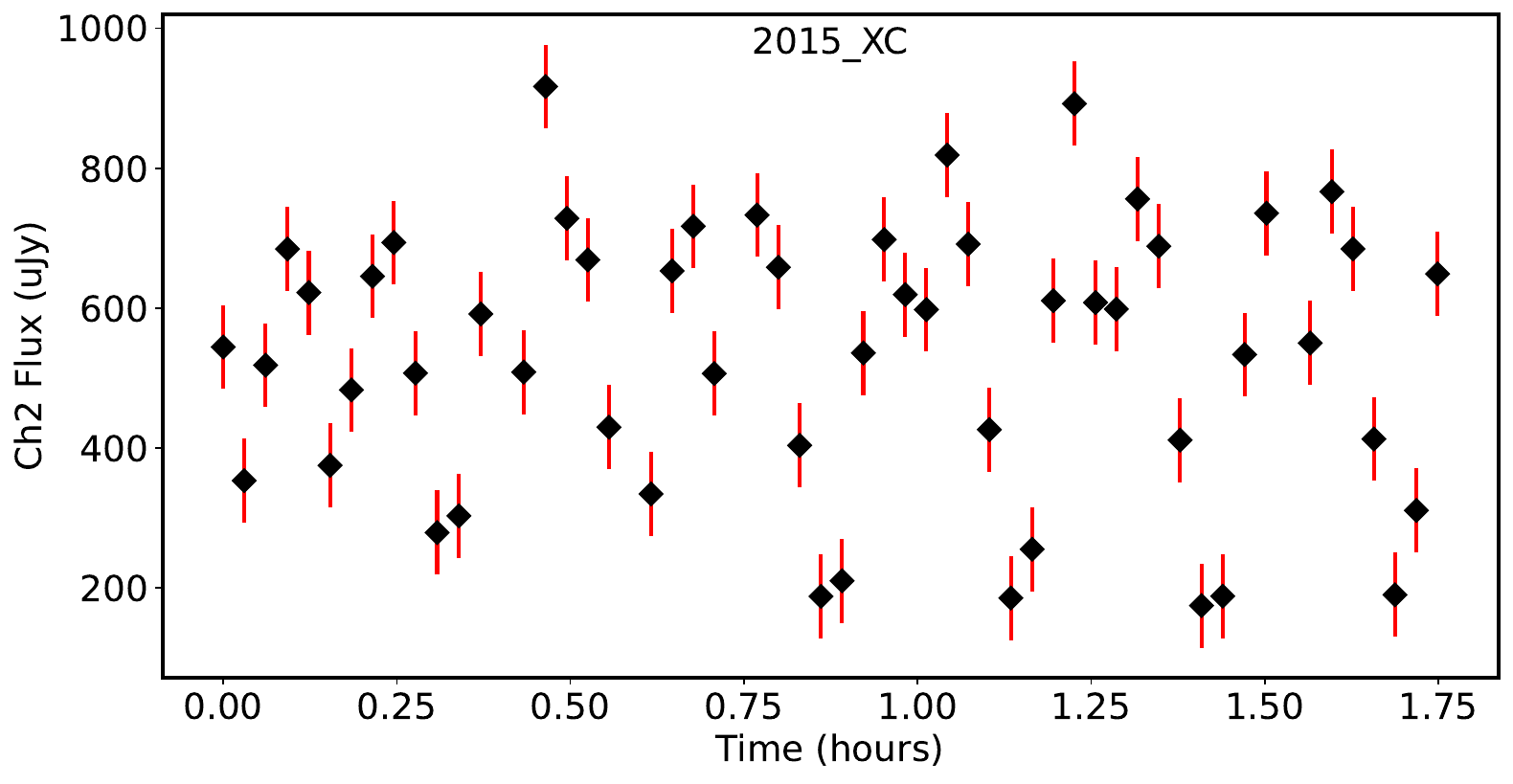}
    \includegraphics[width=0.495\linewidth]{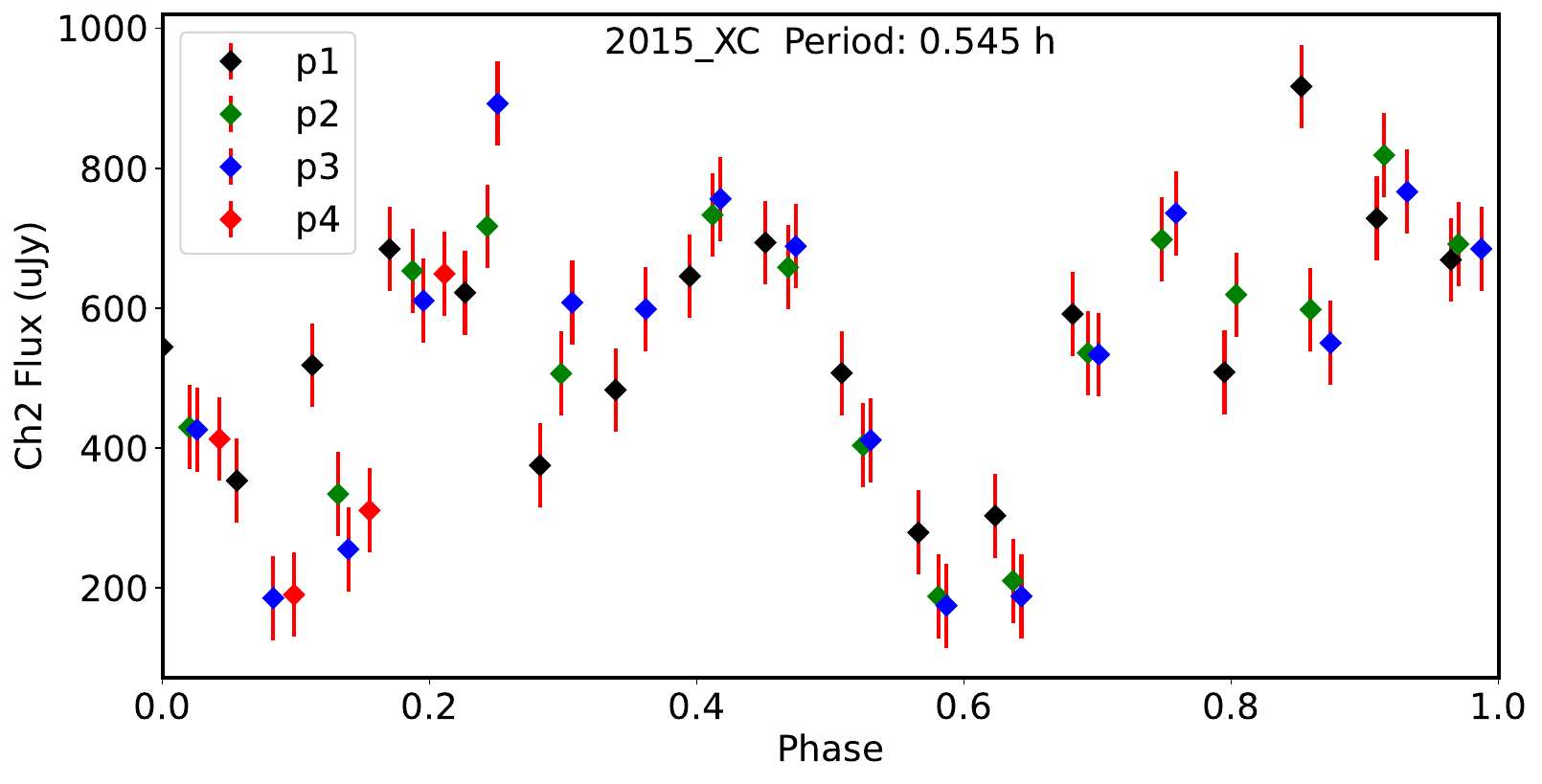}    \caption{Lightcurves (left column) and phased lightcurves (right column) for sources with one or more periods sampled and periods determined. The periods are plotted with different colors in the plots on the right.}
    \label{fig:lc8}
\end{figure*}

Figures~\ref{fig:lc1} -- \ref{fig:lc8} show the lightcurves and phase curves for the NEOs with $\sim$1 or more full phases sampled. The periods are shown in the phase plots and are listed in Table~\ref{tab:fullperiods}.

\end{document}